\documentclass[preprint,showpacs,preprintnumbers,amsmath,amssymb]{revtex4}
\usepackage{amsmath}

\usepackage{graphicx}
\usepackage{dcolumn}
\usepackage{bm}
\usepackage{amsmath} 
\usepackage{subfig}

\newcommand{\qed}{\nobreak \ifvmode \relax \else
      \ifdim\lastskip<1.5em \hskip-\lastskip
      \hskip1.5em plus0em minus0.5em \fi \nobreak
      \vrule height0.75em width0.5em depth0.25em\fi}

\begin{document}

\preprint{}

\title{Bloch Radii Repulsion in Separable Two-Qubit Systems}
\author{Paul B. Slater}
 \email{slater@kitp.ucsb.edu}
\affiliation{%
Kavli Institute for Theoretical Physics, University of California, Santa Barbara, CA 93106-4030\\
}
\date{\today}
            
\begin{abstract}
Milz and Strunz recently reported substantial evidence to  further support the previously conjectured 
separability probability of $\frac{8}{33}$
for two-qubit systems ($\rho$) endowed with Hilbert-Schmidt measure. Additionally, they found that along the radius ($r$) of the Bloch ball representing either of the two single-qubit subsystems, this value appeared constant (but jumping to unity at the locus of the pure states, $r=1$). Further, they also observed (personal communication) such separability probability $r$-invariance, when using, more broadly, random induced measure ($K=3,4,5,\ldots$), with $K=4$ corresponding to the (symmetric) Hilbert-Schmidt case.
Among the findings here is that this invariance is maintained even after splitting the separability probabilities into those parts arising from the determinantal inequality $|\rho^{PT}| >|\rho|$ and those from 
$|\rho| > |\rho^{PT}| >0$, where the partial transpose is indicated. The nine-dimensional set of generic two-re[al]bit states endowed with Hilbert-Schmidt measure is also examined, with similar $r$-invariance conclusions. Contrastingly, two-qubit separability probabilities  based on the Bures (minimal monotone) measure  {\it diminish} with $r$.    Moreover, we study the forms that the separability probabilities take as  joint (bivariate) functions of the radii ($r_A, r_B$) of the Bloch balls of {\it both} single-qubit subsystems. Here, a form of Bloch radii {\it repulsion} for separable two-qubit systems emerges in {\it all} our several analyses. Separability probabilities tend to be  smaller when the lengths of the two radii are closer. In Appendix A, we report certain  companion analytic results for the much-investigated,  
more amenable (7-dimensional) $X$-states model.
\end{abstract}

\keywords{$2 \cdot 2$ quantum systems,  Bloch sphere, Bloch ball, Peres-Horodecki conditions,  partial transpose, partial trace, reduced density matrix, determinant of partial transpose, two qubits,  Hilbert-Schmidt measure,  Bures measure,  separability probabilities,  determinantal moments, $X-states, random induced measure, two-qubits}

\maketitle

\tableofcontents

\section{Introduction}
A considerable body of diverse evidence--though yet no formal proof--has been adduced strongly indicating that the probability that a generic two-qubit system is separable/unentangled is $\frac{8}{33}$ \cite{slater833,MomentBased,slaterJModPhys,FeiJoynt,WholeHalf,Dubna} 
\cite[sec. VII]{Fonseca-Romero} \cite[sec. 4]{Shang}.
The probability is computed with respect to the Hilbert-Schmidt (flat/Euclidean) measure \cite{szHS,ingemarkarol} on the 15-dimensional convex set of $4 \times 4$ density matrices ($\rho$). Milz and Strunz  have recently conducted an analysis further supportive of this conjecture, while  injecting an interesting  new element \cite{milzstrunz}. They found that the probability of $\frac{8}{33}$ appears to hold {\it constant} in the radial direction ($r_A$) of the Bloch ball/sphere parameterization ($r_A \in [0,1]$, $\theta_A \in [0, 2 \pi)$, $\phi_A \in [0, \pi]$)
\begin{equation}
\rho_A = \mbox{Tr}_{B} \rho = 
\end{equation}
\begin{displaymath}
\frac{1}{2}  \left(
\begin{array}{cc}
 \cos \left(\phi _A\right) r_A+1 & \cos \left(\theta _A\right) \sin \left(\phi
   _A\right) r_A-i \sin \left(\theta _A\right) \sin \left(\phi _A\right) r_A \\
 \cos \left(\theta _A\right) \sin \left(\phi _A\right) r_A+i \sin \left(\theta
   _A\right) \sin \left(\phi _A\right) r_A & 1-\cos \left(\phi _A\right) r_A \\
\end{array}
\right)
\end{displaymath}
of either of  the qubit subsystems ($A,B$) of $\rho$, obtained by the partial tracing over
$\rho$ of the complementary subsystem--with a singularity occurring at the pure state boundary,  $r_A=1$ (cf. Fig.~\ref{fig:HSconfidence} below). 
(At times below, we use the symbol $r$, to denote interchangeably, $r_A$ or $r_B$. 
``The Bloch sphere provides a simple representation for the state space of the most primitive quantum unit--the qubit--resulting in geometric intuitions that are invaluable in countless fundamental information-processing scenarios'' \cite{Jevtic}.)

This same $r$-invariance phenomenon appeared to hold, in general they found, for $2 \times m$ [qubit-qudit] systems. (For $m>3$, the probability of having a positive partial transpose--which is now a necessary but not sufficient condition for separability--was employed \cite[Fig. 5]{milzstrunz}.) Further, Milz indicated in a personal communication that for the $2 \times 2$ qubit-qubit systems endowed with
random induced measure (a function of the dimension $K$ of the ancillary space) \cite{Induced,aubrun2}, $r$-invariance of these separability probabilities--the values of which can now be directly obtained from equations (3)-(5) in 
\cite{LatestCollaboration}--also seems to hold. The (symmetric) case $K=4$ is equivalent to the Hilbert-Schmidt one.

The work of Milz and Strunz is rather similar in motivation with earlier efforts in which it was sought to describe separability probabilities also as functions of single variables (but other than the Bloch radius)--namely, the ``cross-product ratio'' 
(suggested by work of Bloore \cite{bloore}),  
$\frac{\rho_{11} \rho_{44}}{\rho_{22} \rho_{33}}$  \cite{slaterPRA2,slater833,slaterJGP2}, and the maximal concurrence over spectral orbits \cite{ratios} (cf. \cite{roland2}), and the participation ratio and von Neumann-Renyi entropies \cite[Figs. 2b, 4]{ZHSL} (cf. App. B below and \cite[Figs. 1, 2]{Dubna}). None of the univariate separability functions constructed was of the highly intriguing {\it constant} form, however. (So, in a sense here, the Bloch radius serves as the extreme antithesis of an entanglement indicator, being entanglement-insensitive apparently.)

In this study, we seek to broaden the investigation of Milz and Strunz by examining the nature of the  {\it joint} (bivariate) distribution of the 
Hilbert-Schmidt separability probability over the radii ($r_A, r_B$) of  {\it both} qubit subsystems of  $\rho$ (sec.~\ref{HSsection}) (cf. \cite{mosseri,mosseri2}).  Further, we examine the use of random induced measure for the Hilbert-Schmidt ($K=4$)-``neighboring'' cases of $K=3$ 
and 5 ({sec.~\ref{Induced}). We will, similarly, examine the separability probability with respect to the another measure of substantial  interest, the  Bures \cite{szBures,ingemarkarol,BuresCauchy}  (sec.~\ref{Bures}), concluding that--contrastingly--the Bures separability probability is  not constant over $r$, but diminishes with the Bloch radius. Two-re[al]bit \cite{carl,batle2} Hilbert-Schmidt analyses are included in sec.~\ref{TwoRebits}, and an apparent related ``Dyson-index'' effect 
\cite{MatrixModels} in sec.~\ref{DysonIndexEffect}. In sec.~\ref{Division}, we examine how the various separability probabilities change as a function of $r$ when they are subdivided into that part arising from the determinantal inequality $|\rho^{PT}| >|\rho|$ and that from $|\rho| > |\rho^{PT}| >0$ 
(cf. \cite{WholeHalf,HyperDiff}). In that setting too, $r$-invariance  appears to hold. Further, we attempted companion analyses for two-quater[nionic]bit systems 
\cite{MomentBased,FeiJoynt,batle2}, but encountered certain conceptual/computational issues we have yet to successfully resolve. 

Appendix A develops upon the $X$-states analyses of Milz and Strunz \cite[Apps. A, B]{milzstrunz}, finding similar results to those reported in the main body of this paper, but now also deriving certain exact formulas. In Appendix B we report
some results (not directly pertaining to Bloch radii) based on an entanglement measure of Holik and Plastino \cite[eq. (9)]{HolikPlastino}.
\section{Background}
Milz and Strunz found numerically-based evidence that  the Hilbert-Schmidt {\it volumes} of the two-qubit systems and of their separable subsystems  were {\it both} proportional to $(1-r_A^2)^6$  \cite[eqs. (23), (30),(31)]{milzstrunz}, with the consequent {\it constant} ratio of the two (simply proportional) volume functions being the aforementioned separability probability of $\frac{8}{33}$. 

For the recently much-investigated ``toy'' model of $X$-states \cite{Xstates2,YuEberly,AliRauAbler} \cite[App. B]{Ellinas}, occupying a seven-dimensional subspace of the full fifteen-dimensional space, it was possible for them to {\it formally} demonstrate that the counterpart volume functions, somewhat similarly, were both  again proportional, but now to $(1-r_A^2)^3$ (the square root of
the higher-dimensional result). The corresponding (constant, but at $r_A=1$) separability probability was greater than $\frac{8}{33}$, that is 
$\frac{2}{5}$ \cite[Apps. A, B]{milzstrunz}. This  $\frac{2}{5}$ result was also subsequently 
proven  in \cite{LatestCollaboration}, along with companion $X$-states findings for the broader class of random induced measures \cite{Induced,aubrun2}.

However, the joint distributions over the {\it two} radii in which we are expressly interested here in discerning (either in the $X$-states and/or full model), did not seem readily derivable, even in the analytical frameworks of those two $X$-states studies (cf. \cite{BalakrishnanLai}). So recourse to numerical methods seemed indicated. The two {\it marginal} univariate distributions of the desired joint bivariate {\it volume} distributions should, of course, be proportional to $(1-r_A^2)^6$
and  $(1-r_A^2)^3$ in the full and $X$-states models, respectively.
(In Appendix A, we do succeed in constructing the desired bivariate $X$-states 
total and separable volume functions and, hence, the associated separability probability function.)
\section{Hilbert-Schmidt Analysis} \label{HSsection}
We generated 2,548,000,000 two-qubit density matrices, randomly  with respect to Hilbert-Schmidt  measure, using the simple (Ginibre ensemble) algorithm outlined in \cite[eq. (1)]{osipov}.
(That is: ``(a) prepare a square complex random matrix A of size N pertaining to the Ginibre ensemble (with real and imaginary parts of each element being independent normal random variables);
(b) compute the matrix H = AA?/(tr AA?), (generically positive definite)'' \cite{Induced}.)
For each such matrix, we 
found the values of $r_A$ and $r_B$, as well as performed the well-known Peres-Horodecki (determinantal-based \cite{Demianowicz}) test for separability \cite{asher,michal} on the partial transpose of $\rho$. (That is, a two-qubit state is separable if and only if this determinant is nonnegative.) Then, we discretized/binned the values of the $r_A$'s and $r_B$'s obtained to lie in intervals of 
length $\frac{1}{100}$. Thus, we obtained two  $100 \times 100$ matrices of counts (which we symmetrize for added stability).
In Figs.~\ref{fig:TotalCounts} and \ref{fig:SeparabilityCounts}, we show the histograms of these two sets of counts (cf. \cite[Fig. 3]{milzstrunz}). (Of the 10,000 bins, 9,364 of the total and 9,199 of the separable ones are occupied.)

The first ({\it total} counts) plot appears to be somewhat broader in nature than the second ({\it separable} counts) plot, while appearing qualitatively rather similar. 

A natural null (product/independence) 
hypothesis to adopt to explain the nature of Figs.~\ref{fig:TotalCounts} and \ref{fig:SeparabilityCounts}--in light of the assertions of Milz and 
Strunz \cite{milzstrunz}--is
that both these surfaces are proportional (taking into account the spherical area formula) to 
\begin{equation}
Q(r_A,r_B)=16 \pi^2 r_A^2 r_B^2 (1-r_A^2)^6  (1-r_B^2)^6. 
\end{equation}
As (visual) tests of these hypotheses, we show the residuals (Figs.~\ref{fig:TotalResiduals} and \ref{fig:SeparabilityResiduals})--that is, the discrepancies (the yet unexplained amount) from the first two figures--based on such predictions. $Q(r_A,r_B)$ was normalized so as to minimize the sum of squares of the residuals. We note, interestingly, that both sets of residuals are bimodal in nature--as opposed to seemingly random in nature--so a null product/independence 
hypothesis does not seem suitable, and further analysis is required. (We can somewhat improve the fits [more so, in the total count case] by employing instead $Q(r_A,r_B)=16 \pi^2 r_A^2 r_B^2 \frac{(1-r_A^2)^8  (1-r_B^2)^8}{(1-r_A^2 r_B^2)^{13}}$.)

In Fig.~\ref{fig:SepProbRatios}, we show the estimated {\it joint} separability probabilities (the ratio of the surface in Fig.~\ref{fig:SeparabilityCounts} to that in Fig.~\ref{fig:TotalCounts})--which are clearly now, in contrast to the univariate case--{\it not} uniform over their (unit square) domain of definition. The initial motivation for this study
was to discern the functional nature of this derived (separability probability) surface.

In Fig.~\ref{fig:Rotate}, for ease of visualization purposes, we perform a $\frac{\pi}{4}$ rotation of Fig.~\ref{fig:SepProbRatios} (cf. \cite[eq. (7)]{BSM}).

In Fig.~\ref{fig:DiagonalSepProbs} we present the $r_A=r_B$ cross-section of 
Fig.~\ref{fig:SepProbRatios}, and indicate a closely-fitting model for it. 
(Let us observe--certainly quite consistently with this figure--that the four Bell states are themselves unpolarized, that is $r_A=r_B=0$, {\it and} maximally entangled (cf. \cite{vanik}).)
For this curve, the total volume--forming the denominator of the separability probability curve--appears to be proportional to $(1-r)^8 (1+8 r)$, and the numerator comprised of the separable volume to be proportional  $(1-r)^9   (1+\frac{17}{2} r+ 29 r^2)$, leading to a factor of $(1-r)$ in the equation of the curve itself. The sample separability probability for those states for which $r_A=r_B$, was recorded as 0.22753675.

Quite contrastingly, in Fig.~\ref{fig:ConstantSumSepProbs}, we show the $r_A+r_B=1$ (U-shaped) cross-section of the two-dimensional separability
probability surface (Fig.~\ref{fig:SepProbRatios}). A joint plot of these last two 
($r_A=r_B$ and $r_A+r_B =1$) curves is given in Fig.~\ref{fig:JointPlot}--the first indication we have here of the titular ``Bloch radii repulsion'' effect. 

The estimated separability (marginal) probabilities over either one of the Bloch radii are shown
in Fig.~\ref{fig:HSconfidence}, along with a $95\%$ confidence interval (based on the suitable large-sample normal distribution approximation to the binomial distribution) about the conjectured value of $\frac{8}{33} \approx 0.242424$.  (The sample estimate of this probability that we obtained here was 0.242425003.) This plot helps to confirm the chief ($r$-invariance) $2 \times 2$ findings of Milz and Strunz \cite[Fig. 4]{milzstrunz}. 
\section{Random Induced Measure Analyses} \label{Induced}
We explore the questions raised above, but now in the broader context of random induced measure \cite{Induced,ingemarkarol,aubrun2}, involving the use of the natural, rotationally invariant measure on the set of all pure states of a $4 \times K$ composite system (with $K=4$ yielding the Hilbert-Schmidt measure). By tracing over the $K$-dimensional ancillary system, one obtains the two-qubit states that we will analyze.
We generate random matrices with respect to these measures using the algorithm specified in \cite{Miszczak} (cf. \cite{Miszczak2}).
\subsection{The case K=3} \label{RIK3}
Setting $k = K -4=-1$, in equation (2) of the recent study \cite{LatestCollaboration},
\begin{equation} \label{ComplexRule}
P^{qubit}_k=1-\frac{3\ 4^{k+3} (2 k (k+7)+25) \Gamma \left(k+\frac{7}{2}\right) \Gamma (2k+9)}{\sqrt{\pi } \Gamma (3 k+13)},
\end{equation}
we obtain that the associated separability probability for this scenario is $\frac{1}{14} \approx 0.0714285$.
The related figures--based on 764,000,000 randomly generated $4 \times 4$ density matrices--are Figs.~\ref{fig:TotalCountsInducedK3} to 
\ref{fig:K3confidence} (Fig. [11] total counts histogram; [12] separable counts histogram; [13] joint separability probability estimates;
[14] $r_A=r_B$ curve; [15] $r_A+r_B=1$ curve; [16] joint plot of $r_A=r_B$ and $r_A+r_B=1$ curves; and [17] separability probability estimates over Bloch radius). The sample separability probability estimate is 0.0714333, quite close to the theoretically-predicted value. 

For the $r_A= r_B$ curve 
(Fig.~\ref{fig:DiagonalSepProbsInducedK3}), the total volume--forming the denominator of the separability probability curve--appears to be proportional to $(1-r)^5$, and the numerator comprised of the separable volume to contain a factor $(1-r)^6$, leading--again, as in the $K=4$ case--to a factor of $(1-r)$ in the equation of the curve.
The sample separability probability for those states for which $r_A=r_B$ was recorded as $\frac{3396373}{53960055} \approx 0.0629424$.
\subsection{The case K=5} \label{Kfive}
Now, inserting $k=1$ into the two-qubit random-induced formula (\ref{ComplexRule}) above, we obtain the associated separability probability for this scenario, 
$\frac{61}{143} \approx 0.426573$.
The related figures--parallel to those (Figs.~\ref{fig:TotalCountsInducedK3}-\ref{fig:K3confidence}) for the $K=3$ analysis--are now Figs.~\ref{fig:TotalCountsInducedK5} to \ref{fig:K5confidence} (Fig. [18] total counts histogram; [19] separable counts histogram; [20] joint separability probability estimates;
[21] $r_A=r_B$ curve; [22] $r_A+r_B=1$ curve; [23] joint plot of $r_A=r_B$ and $r_A+r_B=1$ curves; and [24] separability probability estimates over Bloch radius). These are based on 1,267,000,000 randomly generated  density matrices. The estimated separability probability is 0.426549.

For the $r_A=r_B$ curve (Fig.~\ref{fig:DiagonalSepProbsInducedK5}), the total volume--forming the denominator of the separability probability curve--appears to be proportional to $(1-r)^{11} (1+ 11 r +40 r^2)$, and the numerator comprised of the separable volume to contain a factor $(1-r)^{12}$, leading, again, 
as with $K=3, 4$, to an apparent factor of $(1-r)$ in the equation of the curve.

\section{Bures Analysis} \label{Bures}
Our analyses now are based on 424,000,000 randomly generated $4 \times 4$ density matrices with respect to the Bures (minimal monotone) measure 
\cite{ingemarkarol,szBures}, using the algorithm given in \cite[eq. (4)]{osipov}.
Paralleling those figures given above for the $K=3$ and $K=5$ random induced  measures, we present a corresponding Bures-based series of figures (Figs.~\ref{fig:TotalCountsBures}--\ref{fig:BuresConfidence})  (Fig. [25] total counts histogram; [26] separable counts histogram; [27] joint separability probability estimates;
[28] $r_A=r_B$ curve; [29] $r_A+r_B=1$ curve; [30] joint plot of $r_A=r_B$ and $r_A+r_B=1$ curves; and [31] separability probability estimates over Bloch radius).

Fig.~\ref{fig:BuresConfidence} indicates that the separability probability in the radial direction of either reduced qubit subsystem is not constant, but diminishes with $r$, in strong contrast to the cases analyzed above. The estimate of the Bures separability probability \cite{slaterC,slaterJGP,Dubna} itself that we obtain is 0.0733096.
A ``silver mean'' conjecture for this last value, that is $\frac{1680 \sigma_{Ag}}{\pi^8} \approx 0.0733389$, where the silver mean  is defined as $\sigma_{Ag}=\sqrt{2}-1$, had been advanced
in \cite[eq. (16)]{slaterJGP}. Clearly, however, the supporting case for this decade-old Bures 
two-qubit separability probability conjecture (based on quasi-Monte Carlo sampling, and not the more recent algorithm \cite{osipov}) is not nearly as strong as is the multifaceted case that has been accumulating in the past few years 
for the corresponding Hilbert-Schmidt conjecture of $\frac{8}{33}$ \cite{slater833,MomentBased,slaterJModPhys,FeiJoynt,WholeHalf}. 
\section{Two-Rebit Hilbert-Schmidt Analysis} \label{TwoRebits}
We have been noting that a remarkably strong, diverse  body 
of evidence  \cite{slater833,MomentBased,slaterJModPhys,FeiJoynt,
WholeHalf,Dubna}--though yet no 
formal proof (cf. \cite{LatestCollaboration})--has been 
accumulating in the past several years for the proposition that the Hilbert-Schmidt
separability probability of generic (15-dimensional) two-qubit states is $\frac{8}{33} 
\approx 0.242424$. Accompanying these results has also been evidence that the
Hilbert-Schmidt separability probability of generic (9-dimensional) two-re[al]bit states
\cite{carl,batle2} is $\frac{29}{64} \approx 0.453125$.

In Figs.~\ref{fig:RealTotalCounts}-\ref{fig:Realconfidence} we present a parallel set of figures to those above, now  for $4 \times 4$
density matrices with real entries with respect to Hilbert-Schmidt measure 
(Fig. [32] total counts histogram; [33] separable counts histogram; [34] joint separability probability estimates;
[35] $r_A=r_B$ curve; [36] $r_A+r_B=1$ curve; [37] joint plot of $r_A=r_B$ and $r_A+r_B=1$ curves; and [38] separability probability estimates over Bloch radius).
(We follow the prescription given in \cite[p. 7]{osipov} regarding the generation of such random
matrices, of which we generate 2,751,000,000.) The separability probability estimate that we obtain is 0.453115. In this case, each Bloch radius has a two-dimensional, rather than a three-dimensional character (nor one-dimensional aspect, as will be the case with the $X$-states).

The two $100 \times 100$ matrices of total counts in the Hilbert-Schmidt two-rebit analysis (Fig.~\ref{fig:RealTotalCounts}) 
and in the Hilbert-Schmidt two-qubit analysis 
(Fig.~\ref{fig:TotalCounts}) have 8,325 of the 10,000 cells both of size at least 100. The correlation between the estimated separability probabilities for those two sets of 8,325 cells was 0.987954. On the other hand, if we first square the values of the two-rebit separability probabilities--as random matrix (Dyson-index) theory \cite{MatrixModels} might suggest--the correlation is slightly higher, 0.989462. 
\subsection{Dyson-index relations between rebit and qubit $r_A=r_B$ formulas} \label{DysonIndexEffect}

Further pursuing this Dyson-index {\it ansatz}, we plot in Fig.~\ref{fig:QubitRebit2}, the ratio of the 
estimated two-qubit $r_A=r_B$ probability distribution to the {\it square} of its two-rebit counterpart. We see that this ratio appears to hold quite constant at roughly $\frac{5}{4}$. This ratio is also, clearly, formally a ratio ($R=\frac{R_{sep}}{R_{tot}}$) of two auxiliary quantities. The numerator $R_{sep}$ is itself the ratio of the two-qubit separable volume to the square of the two-rebit separable volume, and the denominator $R_{tot}$, likewise in terms of  total volumes. Our well-fitting estimates of $R_{sep}$ 
were $0.0023976 r+0.019464$ and of $R_{tot}$, 
$0.0023868 r+0.0171397$, respectively. So, the slopes of the two lines are extremely close, yielding--{\it via} the equivalence $\frac{R_{sep}}{R_{tot}}$--the near flatness of 
Fig.~\ref{fig:QubitRebit2}.

Along such ``Dyson-index'' lines of thought, we have been investigating--with some numerical encouragement--the possibility that in the $r_A=r_B$ scenarios in the two-rebit case, the volume functions might have factors of $(1-r)^4$ and $(1-r)^{\frac{9}{2}}$ in the total and separable cases, respectively. 
Such an investigation stemmed from the apparent occurrences, noted above, of factors $(1-r)^8$ and $(1-r)^9$ in the two-qubit counterpart volume functions--which appear to be, in full, proportional to 
$(1-r)^8 (1+8 r)$ and $(1-r)^9 (1+\frac{17}{2} r +29 r^2)$.

We would anticipate that related Dyson-index patterns would continue to hold in the case of two-quater[nionic]bits.
\section{Division of separability probabilities between $|\rho^{PT}|>
|\rho|$ and $|\rho| > |\rho^{PT}| >0$} \label{Division}
It now appears, based on the above analyses, that for the two-qubit states endowed with random induced measure, the separability probabilities are constant along either Bloch radius 
(except at the isolated point $r=1$) of the reduced single-qubit states (Figs.~\ref{fig:HSconfidence}, \ref{fig:K3confidence}, \ref{fig:K5confidence}).
An interesting supplementary question which we will now investigate is what are the contributions to the separability probabilities arising
from the determinantal inequality $|\rho^{PT}|>
|\rho|$ and, complementarily, from $|\rho| > |\rho^{PT}| >0$.
\subsection{Hilbert-Schmidt ($K=4$) case}
We know from preceding work \cite[Table IV]{WholeHalf} \cite{HyperDiff} that in the Hilbert-Schmidt case, the conjectured probability of $\frac{8}{33}$ appears to be evenly divided, with each inequality contributing $\frac{4}{33}$. In a further analysis conducted here,  based on
1,419,000,000 random matrices, it appears that this amount of $\frac{4}{33} \approx 0.121212$  (the sample estimate being 0.121208) 
is--paralleling the Milz-Strunz finding for the total (undivided, that is  $|\rho^{PT}| >0$) separability probability--{\it constant} along the Bloch radius of either reduced single-qubit system (Fig.~\ref{fig:HSconfidenceDivision}).
\subsection{Random Induced $K=3$ case}
For the $K=3$ $(k=-1)$ case, the entire separability probability of $\frac{1}{14}$ is associated with the inequality $|\rho^{PT}| >|\rho|$, so no similar {\it nontrivial} splitting can take place. (In this scenario, $\rho$ {\it must} possess at least one zero eigenvalue, and hence $|\rho|=0$, thus explaining this unique phenomenon.)
\subsection{Random Induced $K=5$ case}
For the $K=5$ $(k=1)$ case, the total separability probability appears--as already noted (sec.~\ref{Kfive})--to be
$\frac{61}{143} \approx 0.426573$. Table IV of \cite{WholeHalf} asserts that the {\it proportion} of this associated with $|\rho^{PT}| > |\rho|$ is 
$\frac{45}{122}$, yielding then $(\frac{61}{143})(\frac{45}{122})=\frac{45}{286} \approx 0.157343$. In Fig.~\ref{fig:K5confidenceDivision}, we show--based on 1,267,000,000 random matrices--an associated flat-like plot, conforming closely to this value (the overall sample estimate being 0.157323). Thus, it appears that this separability (sub-)probability--and, of course, 
its complementary value of $\frac{61}{143} -\frac{45}{286} =\frac{7}{26}$--is constant along either Bloch radius.
So, these analyses serve as an expansion--and a type of further validation--of the Milz-Strunz findings \cite{milzstrunz}.
\subsection{Two-rebit case}
In Fig.~\ref{fig:RealconfidenceDivision} we present the Hilbert-Schmidt two-rebit counterpart of these several figures, with the sample estimate of the overall separability probability of $\frac{1}{2} (\frac{29}{64}) = \frac{29}{128} \approx 0.226563$, being 0.226554. (Here, the 2,751,000,000 random density 
matrices employed in sec.~\ref{TwoRebits} were again employed.)
\subsection{Bivariate Extension}
Additionally, if we similarly split the three {\it bivariate} separability probability plots (Figs.~\ref{fig:SepProbRatios}, \ref{fig:SepProbRatiosInducedK5}, \ref{fig:RealSepProbRatios}) for the $K=4$, $K=5$ two-qubit and $K=4$ two-rebit cases, in accordance with the two determinantal inequalities,
$|\rho^{PT}| >|\rho|$ and $|\rho| > |\rho^{PT}| >0$, the resultant plots appear alike in shape to the parent plots. So, we can certainly conjecture a similar equal splitting of probability phenomenon in that higher-dimensional domain.
\section{Concluding Remarks}
All the bivariate separability probability estimates presented 
(Figs.~\ref{fig:SepProbRatios}, \ref{fig:SepProbRatiosInducedK3}, \ref{fig:SepProbRatiosInducedK5}, \ref{fig:SepProbRatiosBures}, \ref{fig:RealSepProbRatios}) appear to have a saddle point at $(\frac{1}{2},\frac{1}{2})$, or somewhere in the neighborhood thereof, with the $r_A=r_B$ curves (Figs. \ref{fig:DiagonalSepProbs}, \ref{fig:DiagonalSepProbsInducedK3}, \ref{fig:DiagonalSepProbsInducedK5}, \ref{fig:DiagonalSepProbsBures}, \ref{fig:RealDiagonalSepProbs}), possibly achieving their maxima at $r_A=r_B=\frac{1}{2}$ and the $r_A+r_B=1$ curves (Figs.~\ref{fig:ConstantSumSepProbs}, \ref{fig:ConstantSumSepProbsInducedK3}, \ref{fig:ConstantSumSepProbsInducedK5}, \ref{fig:ConstantSumSepProbsBures}, \ref{fig:RealConstantSumSepProbs}) attaining their minima there. A simple probability distribution over $[0,1]^2$ with such a saddle point property, that,  in addition, has the required marginal univariate {\it uniform} distributions over $r_A$ or $r_B$,  is
\begin{equation} \label{tung}
p(r_A,r_B)= 2 r_A +2 r_B -4 r_A r_B.
\end{equation}
(This functional form was suggested by Brian Tung in response to a Math Stack Exchange 
question https://math.stackexchange.com/questions/1271549/bivariate-probability-distributions-over-unit-square-uniform-marginals-midpo.)
In Fig.~\ref{fig:MathStackResiduals} we show the residuals/discrepancies obtained by subtracting $\frac{8 p(r_A,r_B)}{33}$ from
the estimated two-qubit Hilbert-Schmidt separability probabilities of Fig.~\ref{fig:SepProbRatios}.

We were able to obtain a somewhat superior fit (also satisfying the marginal constraints) to this one--as measured by the sum of absolute values of residuals from the fit--using a higher-degree form of probability distribution over the unit square, namely
\begin{equation}
p'(r_A,r_B)=\frac{3}{2} r_A^2 r_B+\frac{3}{2} r_A r_B^2-6 r_A r_B-\frac{3 r_A^2}{4}+\frac{5
   r_A}{2}-\frac{3 r_B^2}{4}+\frac{5 r_B}{2}.
\end{equation}

From Figs.~\ref{fig:JointPlot}, \ref{fig:K3JointPlot}, \ref{fig:K5JointPlot},  \ref{fig:BuresJointPlot} and \ref{fig:RealJointPlot} we see a form of Bloch radii {\it repulsion}. That is, separability probabilities tend to increase as the gap in value between the lengths of the two radii increase.

At this point, we have not yet achieved our motivating goal in undertaking this study, that is, to determine the precise  nature of the bivariate distributions
over the pair of Bloch radii.

A remaining related case that is still not successfully analyzed is that of the 27-dimensional set of generic two-quat[ernionic]bits \cite{batle2}, for which the Hilbert-Schmidt separability probability appears to be
$\frac{26}{323} \approx 0.0804954$ \cite{MomentBased,FeiJoynt}.
An interesting question here is how to determine the corresponding ``Bloch radii'' for randomly generated 
two-quaterbit states (cf. \cite{BG}). Further, we have not yet developed a computationally feasible (Mathematica-implemented) algorithm for the random generation of such matrices (cf. \cite[Fig. 1]{MatrixModels} \cite{osipov,Miszczak}).

Let us note that the 
two reduced qubit systems of a {\it pure} two-qubit system must have their
Bloch radii equal (that is, totally ``non-repulsive''). The separable pure two-qubit systems form a four-dimensional submanifold of the six-dimensional manifold of pure two-qubit systems 
\cite[p. 368]{ingemarkarol}, and thus are of relative measure/probability zero. (These observations, it would seem, at least in an informal qualitative manner, are not inconsistent with our general set of results.) In terms of the ``pseudo-pure'' two-qubit states, that is those having only two distinct nondegenerate eigenvalues, Scutaru has shown that "the Bloch vectors of the corresponding qubits are related by a rotation" \cite{scutaru}.

In Table~\ref{table:1} 
we present the results of an auxiliary set of analyses. Five million random density matrices were generated for each scenario indicated, and the correlation computed between the lengths of the corresponding Bloch radii, both for all the density matrices generated, and also just for the subset of separable density matrices. The consistently smaller correlations for the separable states are a manifestation of the repulsion effect we have documented in this study. (We note, however, that none of these correlations is negative. 
So, perhaps rather than the term ``repulsion'', the use of ``relative repulsion'' or ``diminished attraction'' might be more strictly appropriate.)
Obviously, the correlations in the table based on the Bures measure are exceptionally large.
\begin{table}[h!]
\centering\begin{tabular}{ c || c | c} \label{tab:CorrelationTable}
scenario & all states & separable states \\
\hline
Random Induced $K=5$ & 0.145496 & 0.0968024 \\
Hilbert-Schmidt two-qubits & 0.183026 & 0.107762 \\
Hilbert-Schmidt two-rebits & 0.176898& 0.118049 \\
Random Induced $K=3$ & 0.248993 & 0.125835 \\
Bures two-qubits & 0.388250 & 0.210838 \\
\hline
\end{tabular}
\caption{Correlations between pairs of Bloch radii for all states and all separable states for differing scenarios. The lower
correlations for the separable states are consistent with a relative repulsion effect.}
\label{table:1}
\end{table}
\newpage
Evidence adduced by Milz and Strunz indicated that both the Hilbert-Schmidt total and separable univariate volume functions of two-qubit states were simply proportional to $(1-r^2)^6$ \cite[eq. (23)]{milzstrunz}. Quite supportingly, we ourselves fit a function of the form 
$c (1-r^2)^p$ to the large sample of such states employed above (sec.~\ref{HSsection}), and obtained estimates of 5.99965 and 5.99926, respectively, of this exponent for the two volumes.
Similarly, for the two-rebit set of analyses (sec.~\ref{TwoRebits}), the estimates were 6.00439 and 
6.00447. For random-induced measure with $K=3$ (sec.~\ref{RIK3}),  3.99973 and 
4.00015 were obtained, while for the case $K=5$ (sec.~\ref{Kfive}), the corresponding results were 7.99923 and 7.99917. (A parallel exercise based on the Bures measure [sec.~\ref{Bures}] yielded the rather proximate--but certainly far from integral--results of 3.48845 and 3.58319, respectively.)

Following the work of Braga, Souza and Mizrahi \cite[eq. (7)]{BSM}, it might prove advantageous in our quest to model the various bivariate total and separable volume and probability functions discussed above, to employ transformed variables of the form
$u_+ = \frac{r_A + r_B}{2}$ and $u_-= \frac{r_A -r_B}{2}$. In fact, this appears to be the case in the following appendix devoted to $X$-states analyses \cite{YuEberly,AliRauAbler,Ellinas}.
\section{Appendix A: $X$-states analyses}
\subsection{$X$-states bivariate formulas}
We employed the $X$-states parametrization and transformations indicated by Braga, Souza and Mizrahi \cite[eqs. (6), (7)]{BSM}. Based on these, we were formally able to reproduce the Hilbert-Schmidt univariate volume result of Milz and Strunz 
\cite[eq. (20), Fig. 1]{milzstrunz},
\begin{equation}
V^{(X)}_{HS}(r) =\frac{\pi^2}{2304} (1-r^2)^3,
\end{equation}
as the marginal distribution (over either $r_A$ or $r_B$) of the {\it bivariate} distribution (Fig.~\ref{fig:Xbivariate})
\begin{equation} \label{Xtotal} 
_{{tot}}V^{(X)}_{HS}(r_A,r_B) =
\end{equation}
\begin{displaymath}
\begin{cases}
 -\frac{1}{960} \pi ^2 \left(r_A-1\right){}^3 \left(r_A \left(r_A+3\right)-5
   r_B^2+1\right) & r_A>r_B \\
 -\frac{1}{960} \pi ^2 \left(r_B-1\right){}^3 \left(-5 r_A^2+r_B
   \left(r_B+3\right)+1\right) & r_A<r_B.
\end{cases}
\end{displaymath}
To now obtain the desired $X$-states bivariate separability probability distribution (perhaps--and hopefully--suggestive of the full 15-dimensional counterpart), we find the separable volume counterpart 
(Fig.~\ref{fig:XbivariateSep}) to (\ref{Xtotal}) 
\begin{equation} \label{Xsep} 
_{{sep}}V^{(X)}_{HS}(r_A,r_B) =
\end{equation}
\begin{displaymath}
\begin{cases}
 -\frac{\pi ^2 \left(r_A-1\right){}^3 \left(5 \left(r_A+3\right) r_B^4-10 \left(3
   r_A+1\right) r_B^2+8 r_A^2+9 r_A+3\right)}{7680} & r_A>r_B \\
 -\frac{\pi ^2 \left(r_B-1\right){}^3 \left(5 r_A^4 \left(r_B+3\right)-10 r_A^2 \left(3
   r_B+1\right)+r_B \left(8 r_B+9\right)+3\right)}{7680} & r_A<r_B,
\end{cases}
\end{displaymath}
and take their ratio (Fig.~\ref{fig:XbivariateSepProb}) (noting the cancellation of the $(r-1)^3$-type factors), 
obtaining thereby the $X$-states bivariate separability probability formula,
\begin{equation} \label{BivSepProb}
p^{X-states}(r_A,r_B)=
\end{equation}
\begin{displaymath}
\begin{cases}
 \frac{5 \left(r_A+3\right) r_B^4-10 \left(3 r_A+1\right) r_B^2+8 r_A^2+9 r_A+3}{8
   \left(r_A \left(r_A+3\right)-5 r_B^2+1\right)} & r_A>r_B \\
 \frac{5 r_A^4 \left(r_B+3\right)-10 r_A^2 \left(3 r_B+1\right)+r_B \left(8
   r_B+9\right)+3}{8 \left(-5 r_A^2+r_B \left(r_B+3\right)+1\right)} & r_A<r_B
\end{cases}.
\end{displaymath}
(Numerical integration of this function over $[0,1]^2$ yielded 0.381678.) 
Also, 
Fig.~\ref{fig:XbivariateUpperLower} shows the (lower) $r_A=r_B$
and (upper) $r_A +r_B =1$ cross-sections of 
Fig.~\ref{fig:XbivariateSepProb}. (We note in the next section immediately below an exception to the  
``repulsion'' phenomenon, which we have repeatedly observed above. We computed the correlation between the pair of Bloch radii to be $1-\frac{11206656}{37748736-10080 \pi ^2+\pi ^4} \approx 0.702341$ for all states and only slightly less, $1-\frac{74649600}{235929600-25200 \pi ^2+\pi ^4} \approx 0.68326$, for the separable $X$-states.)

In light of the $X$-states results (\ref{Xtotal}) and (\ref{Xsep}), we might speculate that the counterpart bivariate total and separable volumes for the 15-dimensional set of two-qubits states will both consist of the product of 
$(1-r)^6$ and certain polynomials. The corresponding separability probability function (cf. Fig.~\ref{fig:SepProbRatios}) would then be  a rational one.
\subsection{Certain univariate $X$-states separability probability {\it conditional} distributions}
The analytic form of the $r_A=r_B$ separability probability curve  for the $X$-states is
\begin{equation} \label{Lower}
p^{\{X-states\}}(r_A=r_B) =-\frac{(r-1) (5 r (r (r+5)+3)+3)}{32 r+8}.
\end{equation}

The value of this $X$-states separability probability univariate function at 
$(\frac{1}{2},\frac{1}{2})$ is 
$\frac{139}{384}=
\frac{139}{2^7 \cdot 3}$, at $(0,0)$ it is $\frac{3}{8}$, and at (1,1)  it is 0. The maximum of the  $r_A=r_B$ 
curve is achieved at the positive root ($r \approx 0.2722700792$) of the cubic
equation $3 r^3+9 r^2+r-1=0$, its value there ($\approx 0.393558399$) being the positive 
root of the cubic equation $54 r^3 + 108 r^2 - 28 r - 9=0$. 
On the other hand, the minimum ($\frac{139}{384} 
\approx 0.361979$) of the $r_A+r_B=1$ curve 
\begin{equation} \label{Upper}
p^{\{X-states\}}(r_A=1-r_B) = 
\begin{cases}
 -\frac{(r_A-2) r_A \left(5 r_A \left(r_A^2+r_A-10\right)+28\right)+8}{8 (r_A (4 r_A-13)+4)} & 2 r_A>1 \\
 \frac{r_A (r_A (5 r_A ((r_A-4) r_A-6)+32)+25)-20}{8 (r_A (4 r_A+5)-5)} & 2 r_A<1
\end{cases}
\end{equation}
is attained more simply at $r_A = r_B =\frac{1}{2}$. The maximum of $\frac{1}{2}$ is situated at $r_A=r_B=0$ or 1. So, at least in this model there does not seem to be a corresponding {\it minimax} result.
(Let us very interestingly note that in the interval $r \in [0.40182804, \frac{1}{2}]$, the $p^{\{X-states\}}(r_A=1-r_B)$ curve is dominated by the $p^{\{X-states\}}(r_A=r_B)$ curve. The maximum gap of 0.0056796160 is attained at $r=0.4564893379$. So, at least in the $X$-states setting, the "repulsion phenomenon" does not fully hold.)

Further, setting $r_B=\frac{1}{2}$, we have
\begin{equation} \label{OneHalf}
p^{\{X-states\}}(r_B=\frac{1}{2}) =
\begin{cases}
 \frac{r_A \left(128 r_A+29\right)+23}{32 \left(4 r_A \left(r_A+3\right)-1\right)} &
   \frac{1}{2}<r_A<1 \\
 \frac{35 r_A^4-50 r_A^2+19}{44-80 r_A^2} & 0<r_A<\frac{1}{2}
\end{cases}.
\end{equation}
Putting $r_B =0$, we obtain
\begin{equation} \label{Zero}
p^{\{X-states\}}(r_B=0) =\frac{r_A \left(8 r_A+9\right)+3}{8 \left(r_A \left(r_A+3\right)+1\right)},
\end{equation}
and with $r_B =1$,
\begin{equation}
p^{\{X-states\}}(r_B=1) =  0.
\end{equation}
It certainly appears that this last result is in direct contradiction with certain assertions of Milz and Strunz: ``The latter fact that $p_{sep}^{(X)}(1)=1$ is clear: a pure reduced state $(r=1)$ can only be realized by a product and thus, a separable total state'' \cite[p. 8]{milzstrunz} (see also the discussion prior to their eq. (23)). We anticipate an eventual clarification of this apparent conflict, possibly in terms of differing dimensionalities of the measures employed.

If we restrict the $X$-states to those for which $r_A=r_B$, then the separability probability for this continuum $r_A \in [0,1]$ of states is $\frac{8}{21} \approx 0.3809524$. For those for which $r_A =1-r_B$, the corresponding separability probability is slightly higher, $\frac{58}{147} \approx 0.3945578$.
\subsection{Use of Fano correlation parameter $c_{33}=M_{zz}$}
In their $X$-state studies, both Milz and Strunz \cite{milzstrunz} and Braga, Souza and Mizrahi \cite{BSM} employ the well-known Fano parameterization of two-qubit systems \cite{fano}.
Milz and Strunz denote the Fano correlation parameter in the 
(conventionally denoted) $z-$ or $x_3$-direction by $c_{33}$, while Braga, Souza and Mizrahi employ the notation $M_{zz}$. (The alignments of the Bloch radii--$r_A$ and $r_B$--are along this same direction in the 
$X$-states model, we interestingly note.) Focusing on this parameter  yields a number of analytic results, such as the associated $X$-states separability probability 
(Fig.~\ref{fig:MzzSepProb}). (Numerical integration of this function over [-1,1] yielded 0.416283.)
\subsection{Random Induced $K=5$ case}
We have found here--introducing a factor of $|\rho|$ into the integrations in the previously-conducted Hilbert-Schmidt $X$-states analyses--that the total volume bivariate distribution for the induced measure case of $K=5$ equals 
\begin{equation} \label{Xinduced} 
_{{tot}}V^{(X)}_{K=5}(r_A,r_B) =
\end{equation}
\begin{displaymath} 
\begin{cases}
 -\frac{\pi ^2 \left(r_A-1\right){}^5 \left(-6 r_A \left(r_A+5\right) r_B^2+r_A
   \left(r_A+1\right) \left(r_A \left(r_A+4\right)+5\right)+21 r_B^4-6
   r_B^2+1\right)}{1290240} & r_A>r_B\\
 -\frac{\pi ^2 \left(r_B-1\right){}^5 \left(-6 r_A^2 \left(r_B
   \left(r_B+5\right)+1\right)+21 r_A^4+r_B \left(r_B+1\right) \left(r_B
   \left(r_B+4\right)+5\right)+1\right)}{1290240} & r_A<r_B
\end{cases}
.
\end{displaymath}
The total volume itself is $\frac{\pi ^2}{9979200} \approx 9.8901759 \cdot 10^{-7}$.  
The marginal distributions of the total volume bivariate distribution (\ref{Xinduced}) are of the (again, proportional to $\pi^2 (1-r^2)^n$) form 
\begin{equation}
V^{(X)}_{K=5}(r) =\frac{\pi ^2 (1-r^2)^5}{3686400} =\frac{\pi ^2 (1-r^2)^5}{2^{14} \cdot 3^2 \cdot 5^2}.
\end{equation}
The separable volume is given by
\begin{equation} \label{XinducedSep} 
_{{sep}}V^{(X)}_{K=5}(r_A,r_B) =
\end{equation}
\begin{displaymath}
\begin{cases}
 -\frac{\pi ^2 \left(r_A-1\right){}^5 \left(-21 \left(r_A+5\right) r_B^6+63 \left(5
   r_A+1\right) r_B^4-27 r_A \left(8 r_A+5\right) r_B^2+r_A \left(8 r_A
   \left(r_A+2\right) \left(r_A+3\right)+25\right)-27 r_B^2+5\right)}{10321920} &
   r_A>r_B \\
 -\frac{\pi ^2 \left(r_B-1\right){}^5 \left(-21 r_A^6 \left(r_B+5\right)+63 r_A^4
   \left(5 r_B+1\right)-27 r_A^2 \left(r_B \left(8 r_B+5\right)+1\right)+25 r_B+8 r_B^2
   \left(r_B+2\right) \left(r_B+3\right)+5\right)}{10321920} & r_A<r_B 
\end{cases}.
\end{displaymath}
Its marginal distributions are of the form
\begin{equation}
_{sep}V^{(X)}_{K=5}(r) =\frac{\pi ^2 (1-r^2)^5}{5734400} =\frac{\pi ^2 (1-r^2)^5}{2^{15} \cdot 5^2 \cdot 7}.
\end{equation}
The separability probability we found was $\frac{9}{14} \approx 0.642857$--a result also derivable from a formula \cite[p.13]{LatestCollaboration2},
\begin{equation} \label{CFDformula}
\mbox{Pr} \{|\rho^{PT}| > 0\} = 1-\frac{2 \Gamma(2 k+4)^2}{\Gamma(k+2) \Gamma(3 k +6)},
\end{equation}
inserting $k=1$.
The separability probability function (Fig.~\ref{fig:XstatesK5}) is 
\begin{equation} \label{BivSepProbK5}
p^{X-states}_{K=5}(r_A,r_B)=
\end{equation}
\begin{displaymath}
\begin{cases}
 \frac{-21 \left(r_A+5\right) r_B^6+63 \left(5 r_A+1\right) r_B^4-27 r_A \left(8
   r_A+5\right) r_B^2+r_A \left(8 r_A \left(r_A+2\right) \left(r_A+3\right)+25\right)-27
   r_B^2+5}{8 \left(-6 r_A \left(r_A+5\right) r_B^2+r_A \left(r_A+1\right) \left(r_A
   \left(r_A+4\right)+5\right)+21 r_B^4-6 r_B^2+1\right)} & r_A>r_B \\
 \frac{-21 r_A^6 \left(r_B+5\right)+63 r_A^4 \left(5 r_B+1\right)-27 r_A^2 \left(r_B
   \left(8 r_B+5\right)+1\right)+25 r_B+8 r_B^2 \left(r_B+2\right)
   \left(r_B+3\right)+5}{8 \left(-6 r_A^2 \left(r_B \left(r_B+5\right)+1\right)+21
   r_A^4+r_B \left(r_B+1\right) \left(r_B \left(r_B+4\right)+5\right)+1\right)} &
   r_A<r_B
\end{cases}.
\end{displaymath}
In Fig.~\ref{fig:XstatesK5joint}, we show the $r_A=r_B$ and $r_A=1-r_B$ sections of this plot. The minimum of the $r_A=1-r_B$ section is once again found at $r=\frac{1}{2}$ with a value of $\frac{1261}{2176} =\frac{13 \cdot 97}{2^7 \cdot 17} \approx 0.579504$ there. For the $r_A=r_B$ section (cf. (\ref{Lower})) 
\begin{equation} \label{LowerXK5}
p^{X-states}_{K=5}(r_A,r_B)_{(r_A=r_B)} =\frac{(1-r) (r (21 r (r (r+8)+6)+40)+5)}{8 (r (16 r+7)+1)}.
\end{equation}
The maximum, 0.63964, of this curve is attained at $r=0.238465$.
\subsection{Random Induced $K=6$ case}
Here the total volume itself is $\frac{\pi ^2}{9081072000} \approx 1.0868325 \cdot 10^{-9}$.
The marginal distributions are of the form
\begin{equation}
V^{(X)}_{K=6}(r) = \frac{\pi ^2 \left(1-r^2\right)^7}{2890137620} = \frac{\pi ^2 \left(1-r^2\right)^7}{2^{18} \cdot 3^2 \cdot 5^2 \cdot 7^2}.
\end{equation}
The separability probability is $\frac{26}{33}$, given by  (\ref{CFDformula}), inserting  $k=2$.
\subsection{Random Induced $K=7$ case}
Here the total volume itself is $\frac{\pi ^2}{5866372512000} \approx 1.68240328 \cdot 10^{-12}$.
The marginal distributions are, continuing the observed pattern, of the form
\begin{equation}
V^{(X)}_{K=7}(r) = \frac{\pi ^2 \left(r^2-1\right)^9}{1664719257600} = 
\frac{\pi ^2 \left(r^2-1\right)^9}{2^{24} \cdot 3^4 \cdot 5^2 \cdot 7^2}.
\end{equation}
The separability probability is $\frac{125}{143}$, given by  (\ref{CFDformula}), inserting   $k=3$ there.
\subsection{Random Induced $K=3$ case}
The total volume is $\frac{2}{3} \pi ^2 \log ^2(2) \approx 3.16125412$.
The bivariate total volume distribution is
\begin{equation}
_{{tot}}V^{(X)}_{K=3}(r_A,r_B) =
\begin{cases}
 -2 \pi ^2 \log ^2(2) \left(r_A-1\right) & r_A>r_B\land r_A+r_B>0\land r_A<1 \\
 -2 \pi ^2 \log ^2(2) \left(r_B-1\right) & r_B>r_A\land r_A+r_B>0\land r_B<1
\end{cases}
\end{equation}
The two marginal distributions are of the form $\pi ^2 \left(1-r^2\right) \log ^2(2)$.
Here we found the separability probability to equal $\frac{1}{3}$.
We have not been able to find analytic formulas for the bivariate separability volume function and the bivariate separability probability function, but in Fig.~\ref{fig:K3XstatesPlot}, we present a numerically-based estimate of the separability probability function.

\section{Appendix B: Separability probabilities as a function of \newline $||\rho-\rho^A \otimes \rho^B||_{\it{HS}}$}
In a recent paper of Holik and Plastino, the expression
\begin{equation} \label{HP}
||\rho-\rho^A \otimes \rho^B||_{\it{HS}}
\end{equation}
is put forth as a measure of entanglement \cite[eq. (9)]{HolikPlastino},
where $||\ldots||_{\it{HS}}$ is the Hilbert-Schmidt norm
\begin{equation} \label{dagger}
||A||^2_{\it{HS}} =\mbox{tr}(A A^{\dagger}).
\end{equation}

In Fig.~\ref{fig:HolikPlastino} we show estimates of the two-qubit separability
probabilities as a function of this term for the Hilbert-Schmidt and Bures measures, based on 54,000,000  random realizations in the former case and 43,000,000  in the latter (cf. \cite[Fig. 2]{Dubna}). 
The probability diminishes as the Hilbert-Schmidt distance from product states 
($\rho^A \otimes \rho^B$) increases. (Of course, it might be of some interest to employ Bures counterparts of (\ref{HP}) and (\ref{dagger})).

\bibliography{Repulsion2016}

\begin{acknowledgments}
I would like to express appreciation to  Simon Milz and Charles Dunkl--who suggested the use of confidence bands--for helpful communications.
\end{acknowledgments}

\begin{figure}
\includegraphics{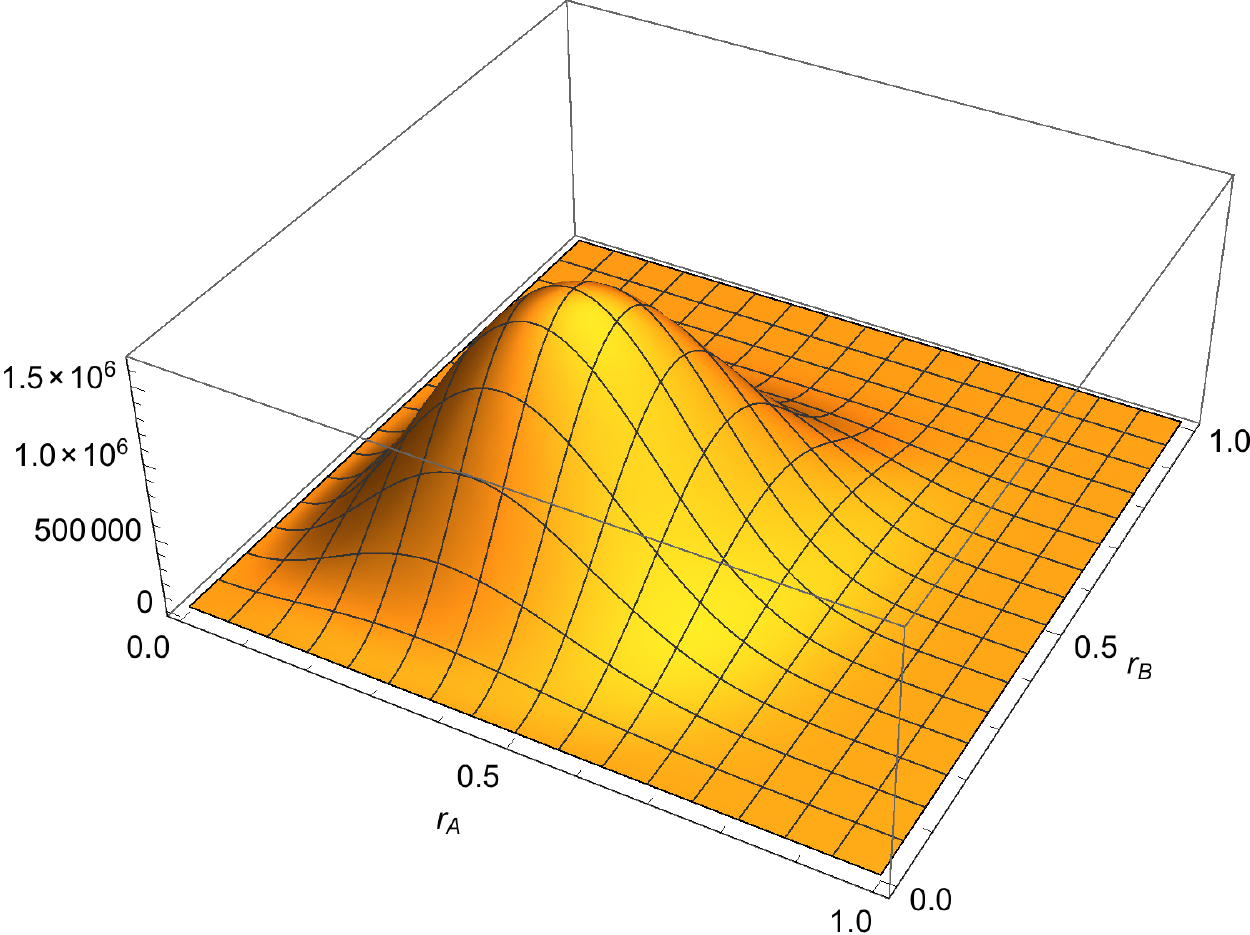}
\caption{\label{fig:TotalCounts}Histogram of Hilbert-Schmidt randomly sampled
two-qubit density matrices parameterized by $r_A$ and $r_B$}
\end{figure}
\begin{figure}
\includegraphics{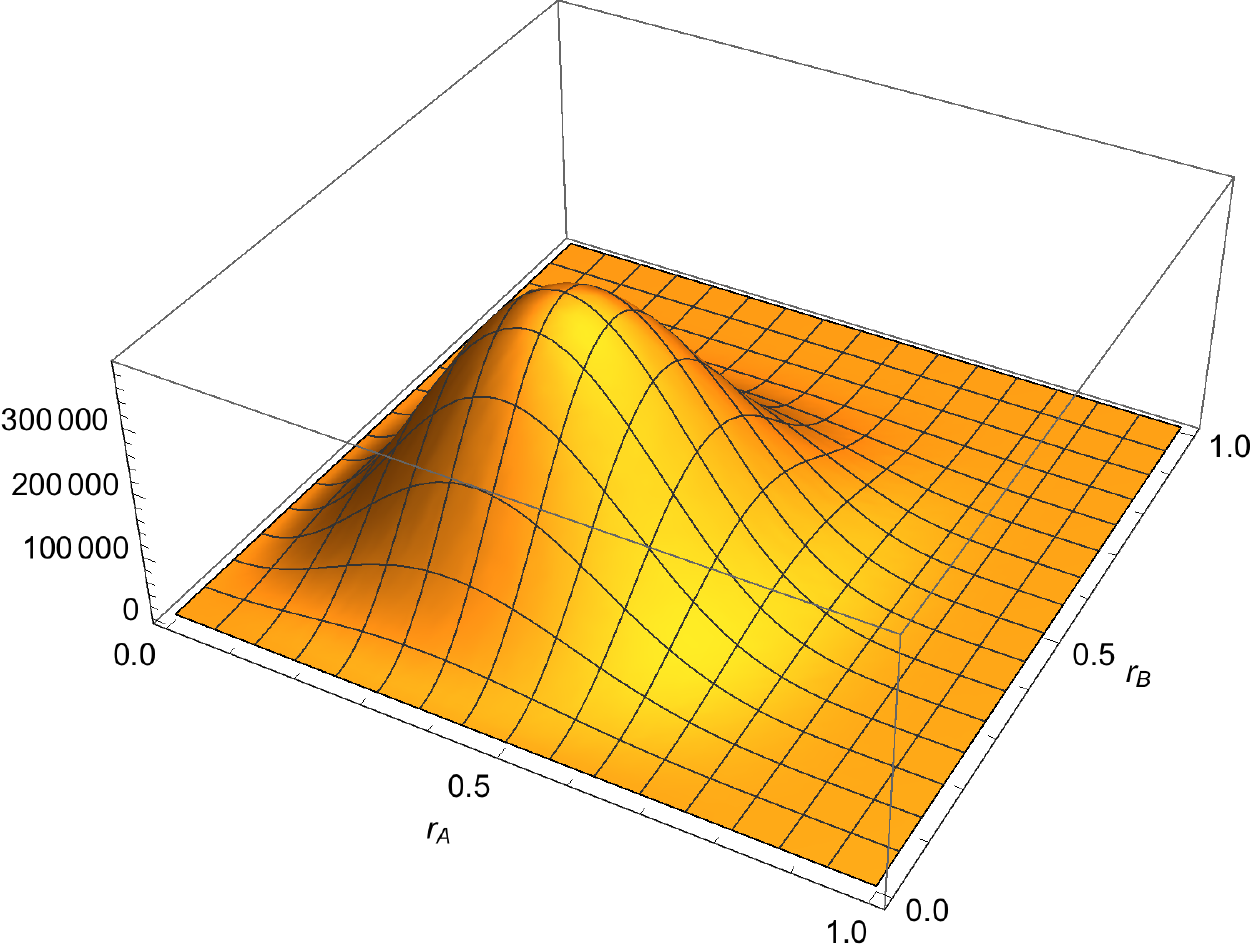}
\caption{\label{fig:SeparabilityCounts}Histogram of Hilbert-Schmidt randomly sampled
{\it separable} two-qubit density matrices parameterized by $r_A$ and $r_B$}
\end{figure}
\begin{figure}
\includegraphics{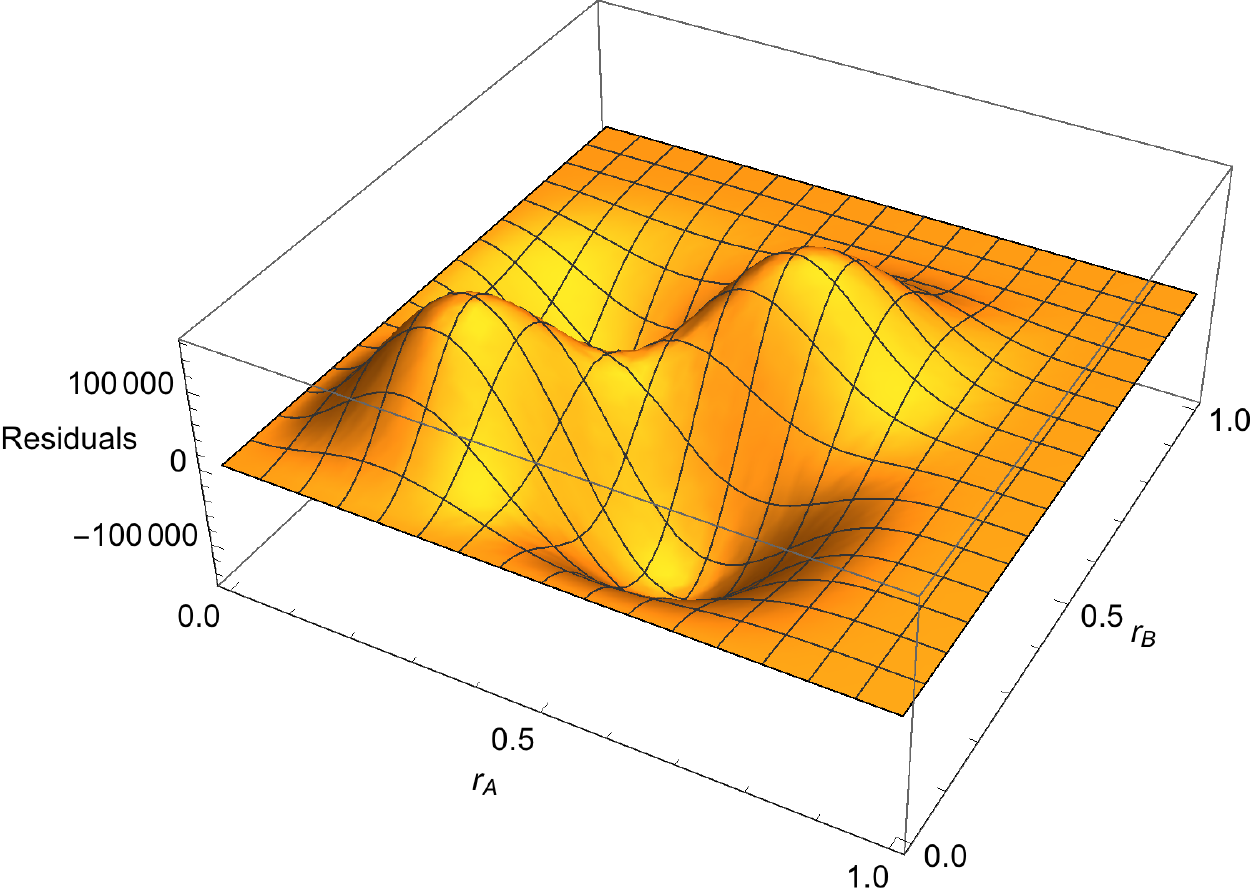}
\caption{\label{fig:TotalResiduals}Residuals (discrepancies) from a fit to the total counts (Fig.~\ref{fig:TotalCounts}) of a normalized form of 
$Q(r_A,r_B)=16 \pi^2 r_A^2 r_B^2 (1-r_A^2)^6  (1-r_B^2)^6 $}
\end{figure}
\begin{figure}
\includegraphics{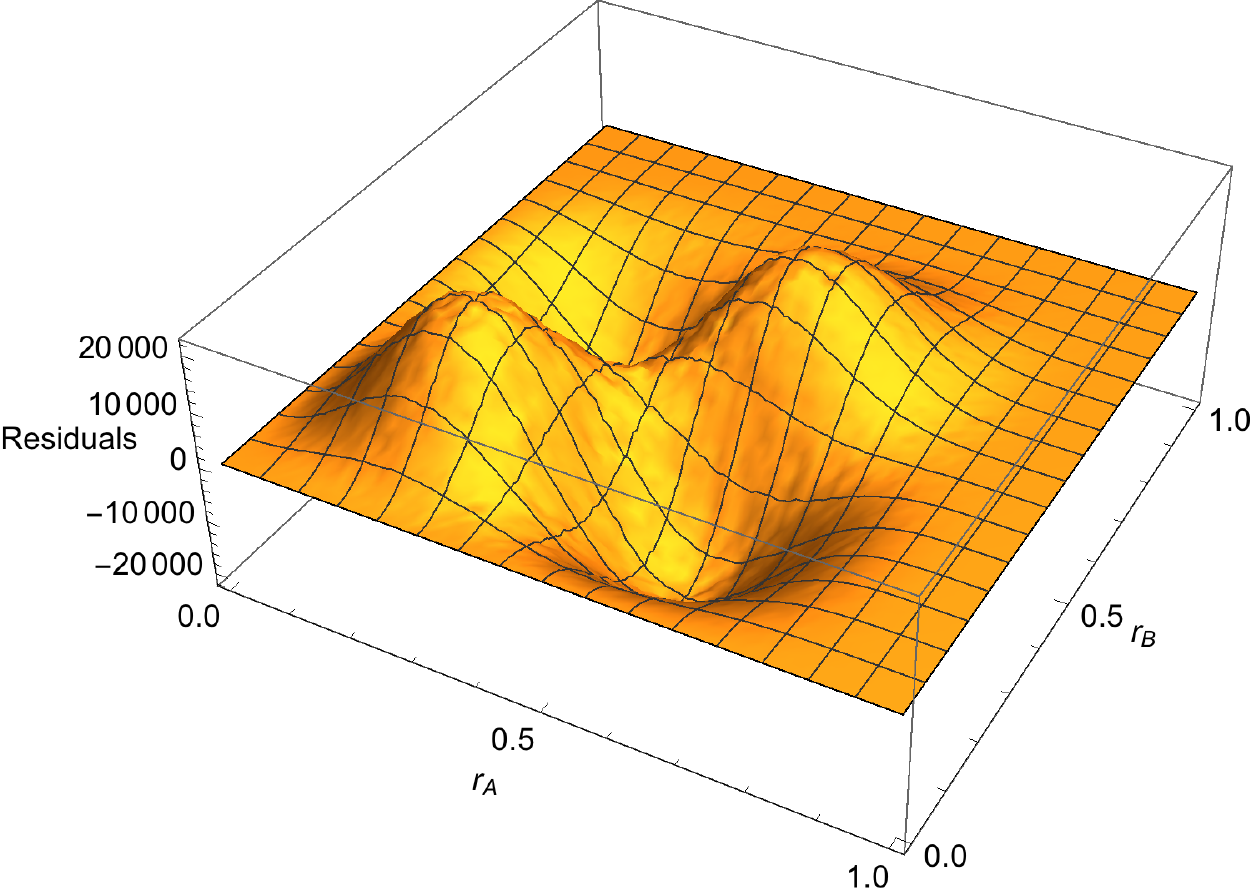}
\caption{\label{fig:SeparabilityResiduals}Residuals (discrepancies) from a fit to the separable counts (Fig.~\ref{fig:SeparabilityCounts}) of a normalized form of $Q(r_A,r_B)=16 \pi^2 r_A^2 r_B^2 (1-r_A^2)^6  (1-r_B^2)^6 $}
\end{figure}
\clearpage
\begin{figure}
\includegraphics{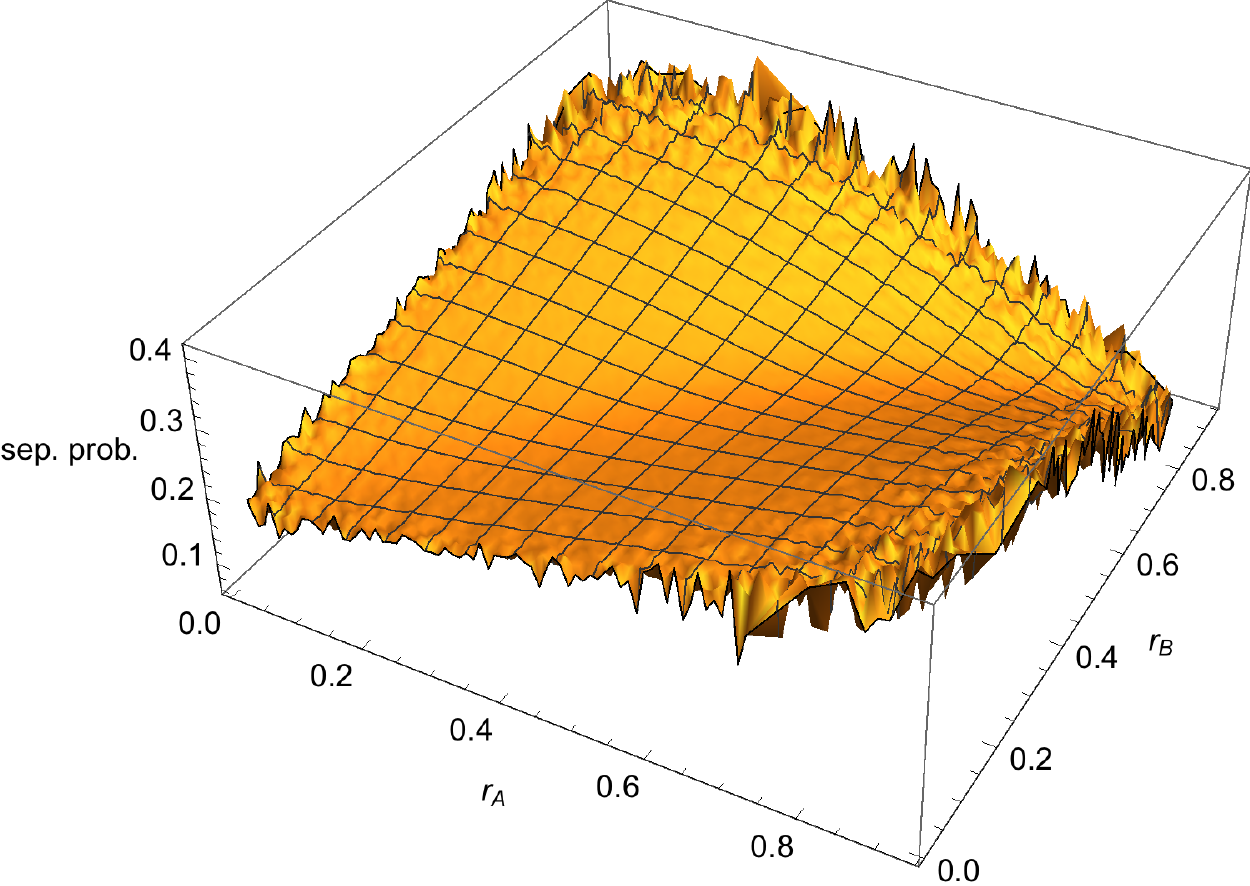}
\caption{\label{fig:SepProbRatios}Estimated joint Hilbert-Schmidt two-qubit separability probabilities--the ratio of 
the separable counts in Fig.~\ref{fig:SeparabilityCounts} to the total counts in Fig.~\ref{fig:TotalCounts}}
\end{figure}
\newpage
\clearpage
\begin{figure}
\includegraphics{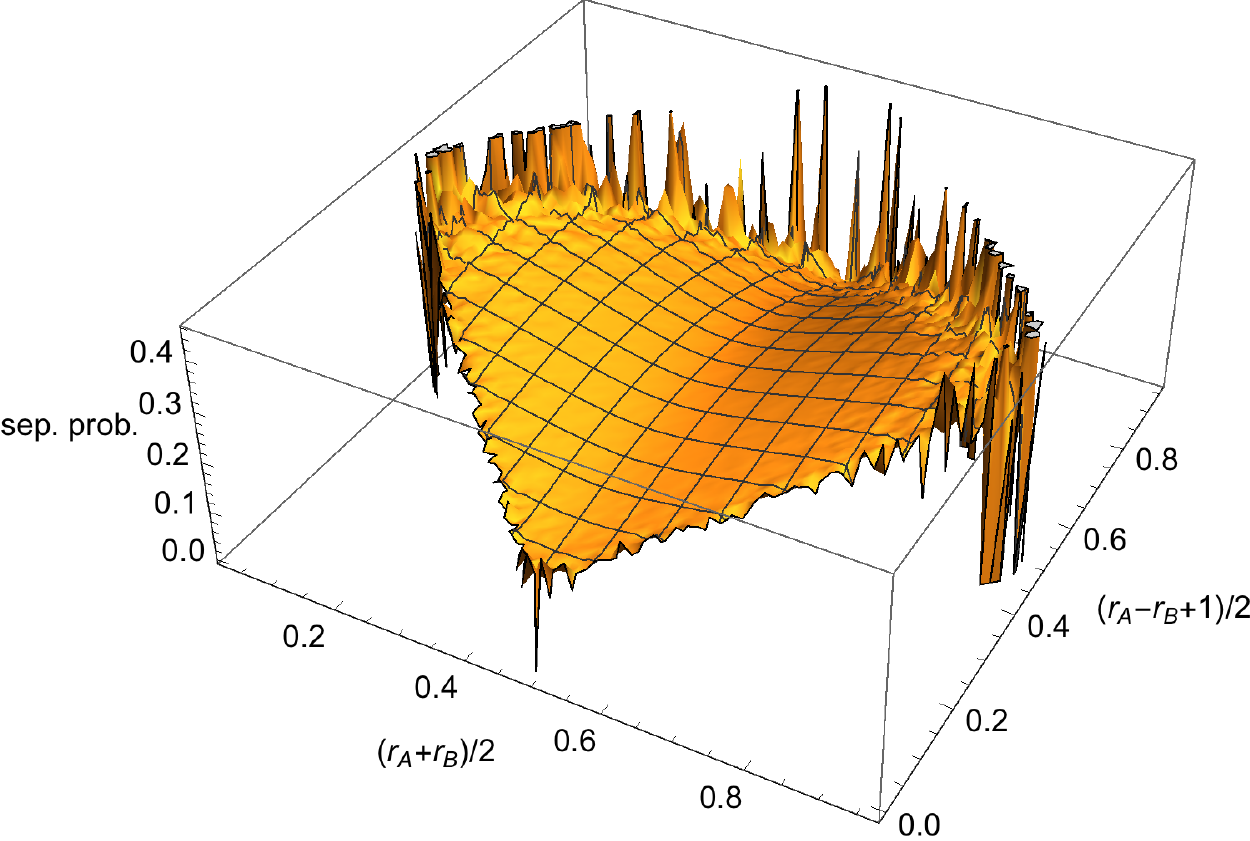}
\caption{\label{fig:Rotate}$\frac{\pi}{4}$-rotation of estimated joint Hilbert-Schmidt two-qubit separability probabilities (Fig.~\ref{fig:SepProbRatios})}
\end{figure}
\begin{figure}
\includegraphics{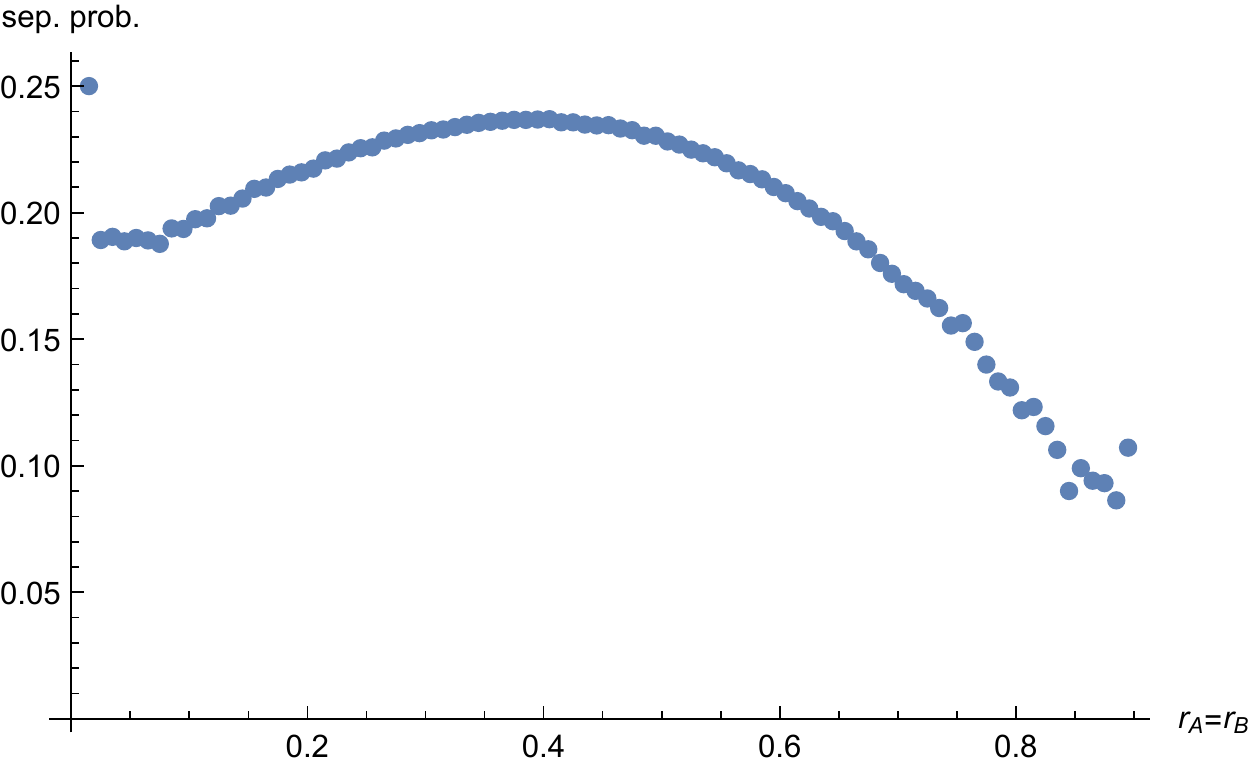}
\caption{\label{fig:DiagonalSepProbs}Estimated Hilbert-Schmidt  separability probabilities for $r_A=r_B$. A closely-fitting model (ratio of apparent separable and total volumes) for this curve is $ \frac{375 (1-r) \left(58 r^2+17 r+2\right)}{4096 (8 r+1)}$.}
\end{figure}
\begin{figure}
\includegraphics{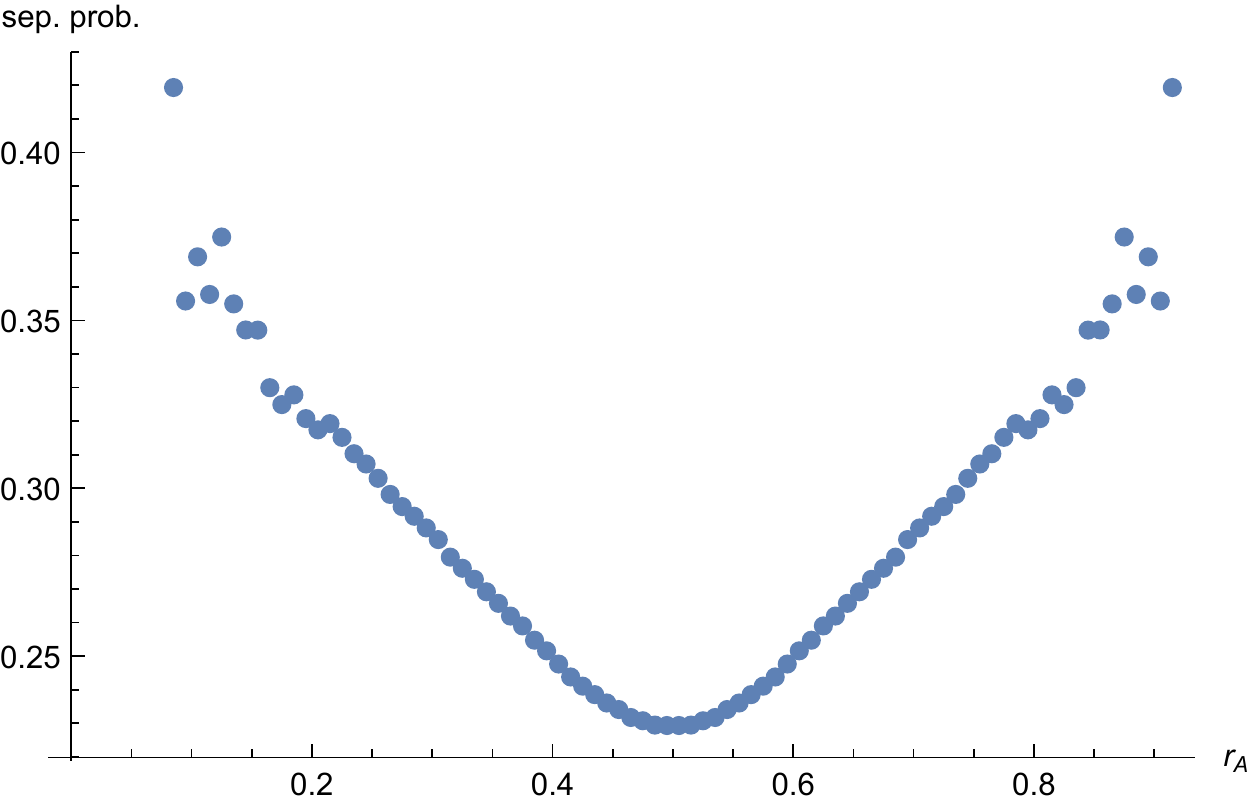}
\caption{\label{fig:ConstantSumSepProbs}Estimated Hilbert-Schmidt  two-qubit separability probabilities for $r_A+r_B=1$. A closely-fitting model for this curve is 
$ \frac{r+4}{24 r+8}$ $(0<r<\frac{1}{2})$ and $\frac{r-5}{24 r-32}$ $(\frac{1}{2}<r<1)$.}
\end{figure}
\begin{figure}
\includegraphics{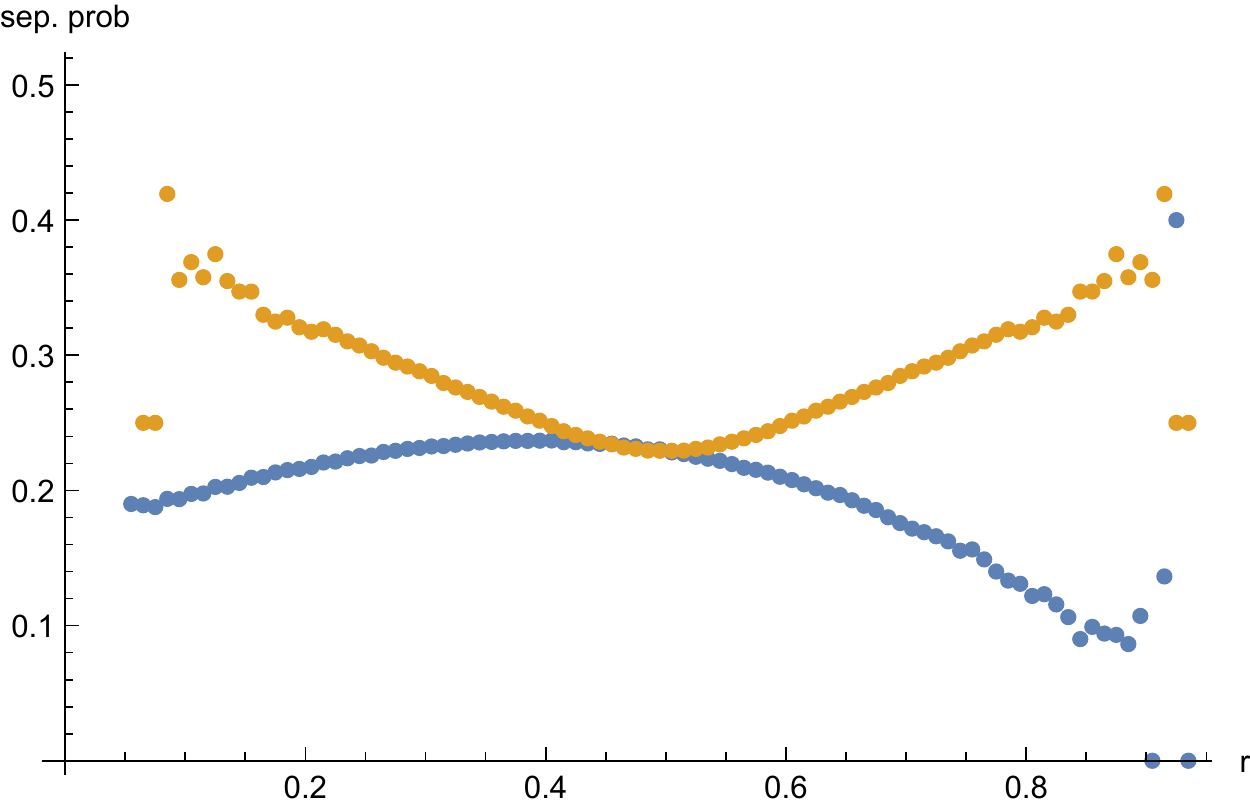}
\caption{\label{fig:JointPlot}Joint plot of the last two figures}
\end{figure}
\begin{figure}
\includegraphics{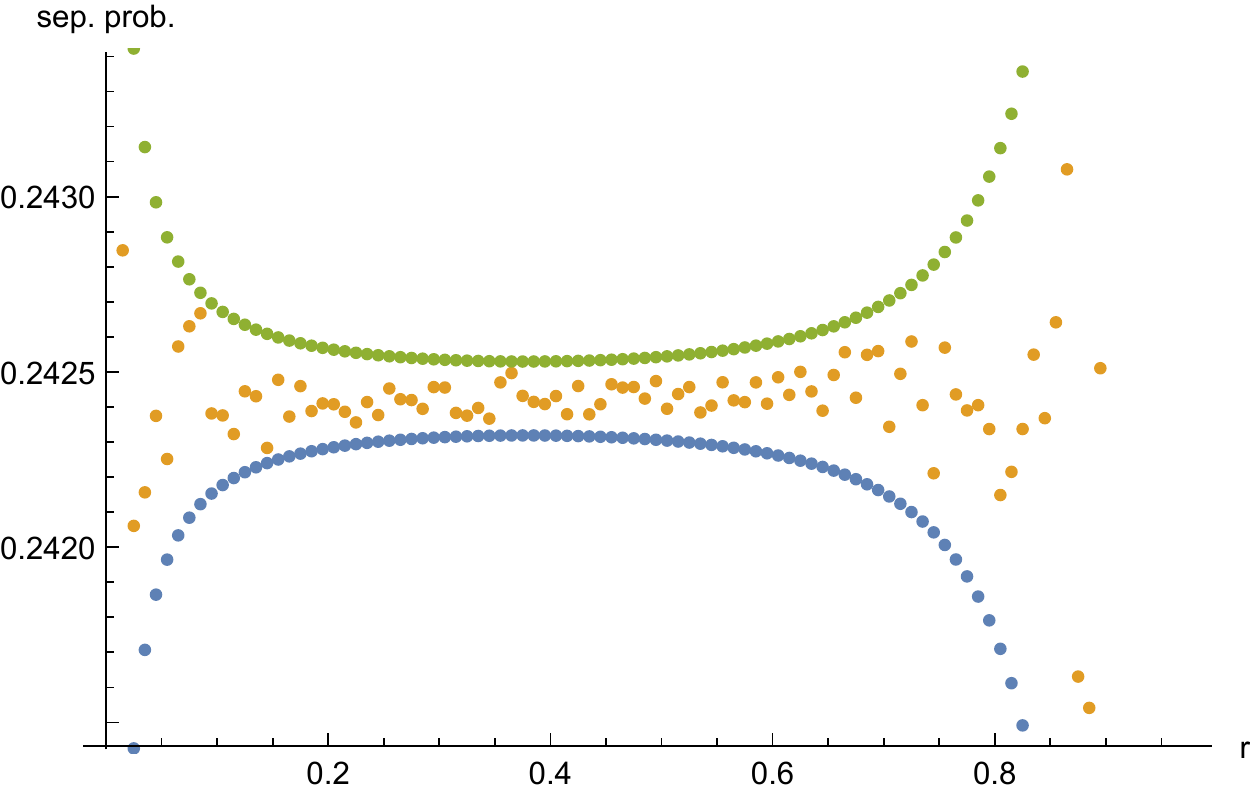}
\caption{\label{fig:HSconfidence}Estimated (marginal) Hilbert-Schmidt  two-qubit separability probabilities along either Bloch radius, together with 95$\%$ confidence limits about the conjectured value of $\frac{8}{33} \approx 0.242424$}
\end{figure}
\begin{figure}
\includegraphics{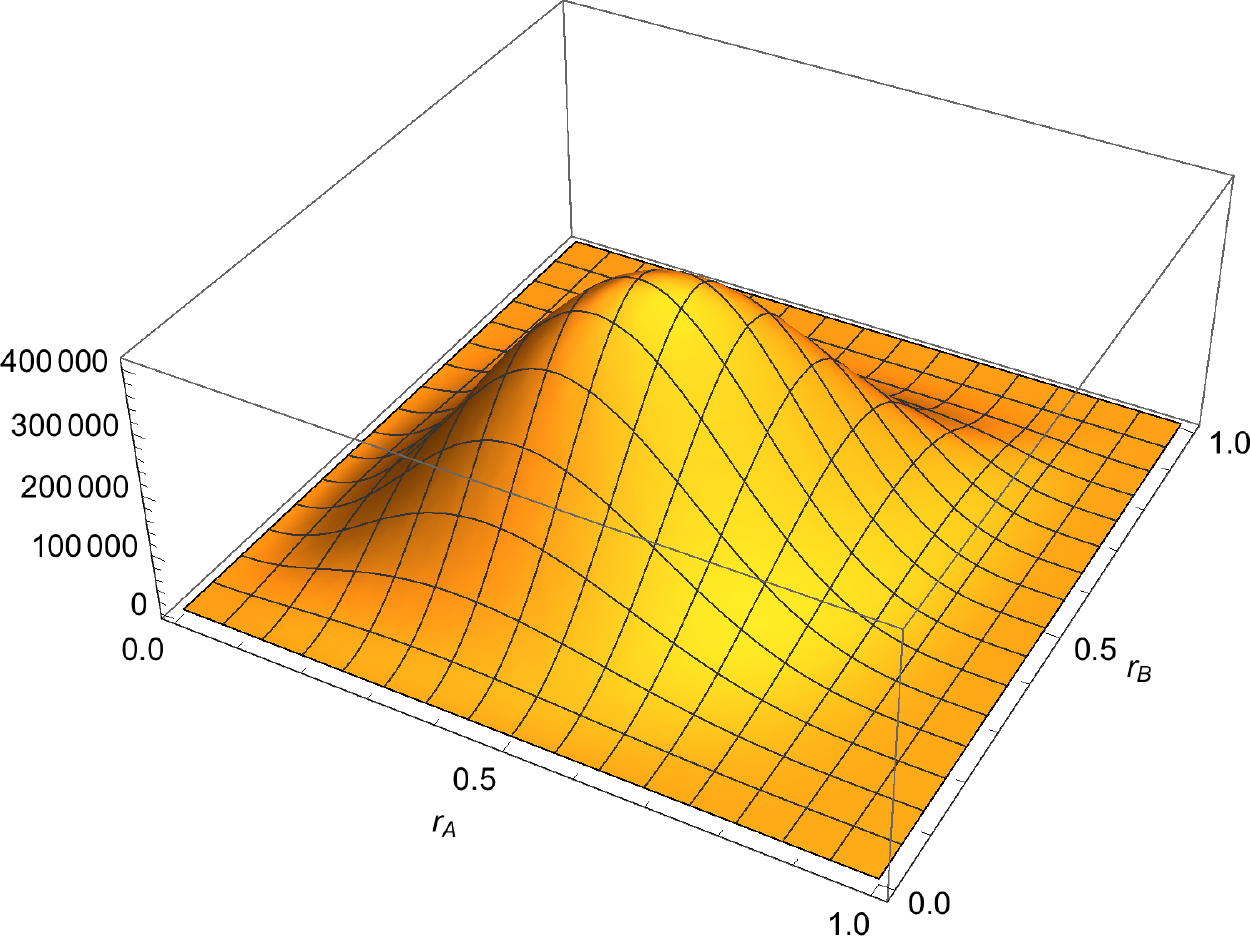}
\caption{\label{fig:TotalCountsInducedK3}Histogram of randomly sampled (with respect to the random induced $K=3$ measure) 
two-qubit density matrices}
\end{figure}
\begin{figure}
\includegraphics{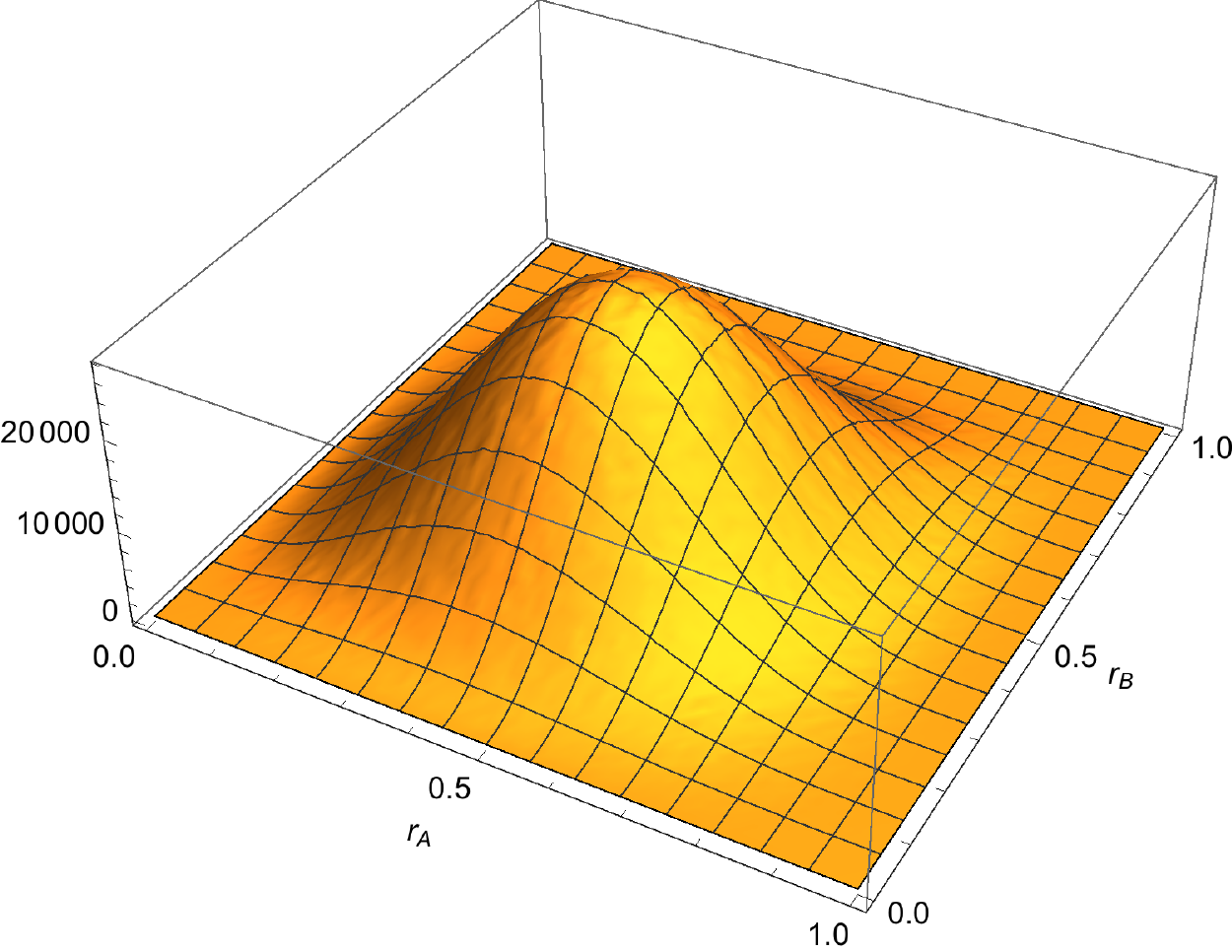}
\caption{\label{fig:SeparabilityCountsInducedK3}Histogram of randomly sampled (with respect to the random induced $K=3$ measure) 
{\it separable} two-qubit density matrices}
\end{figure}
\begin{figure}
\includegraphics{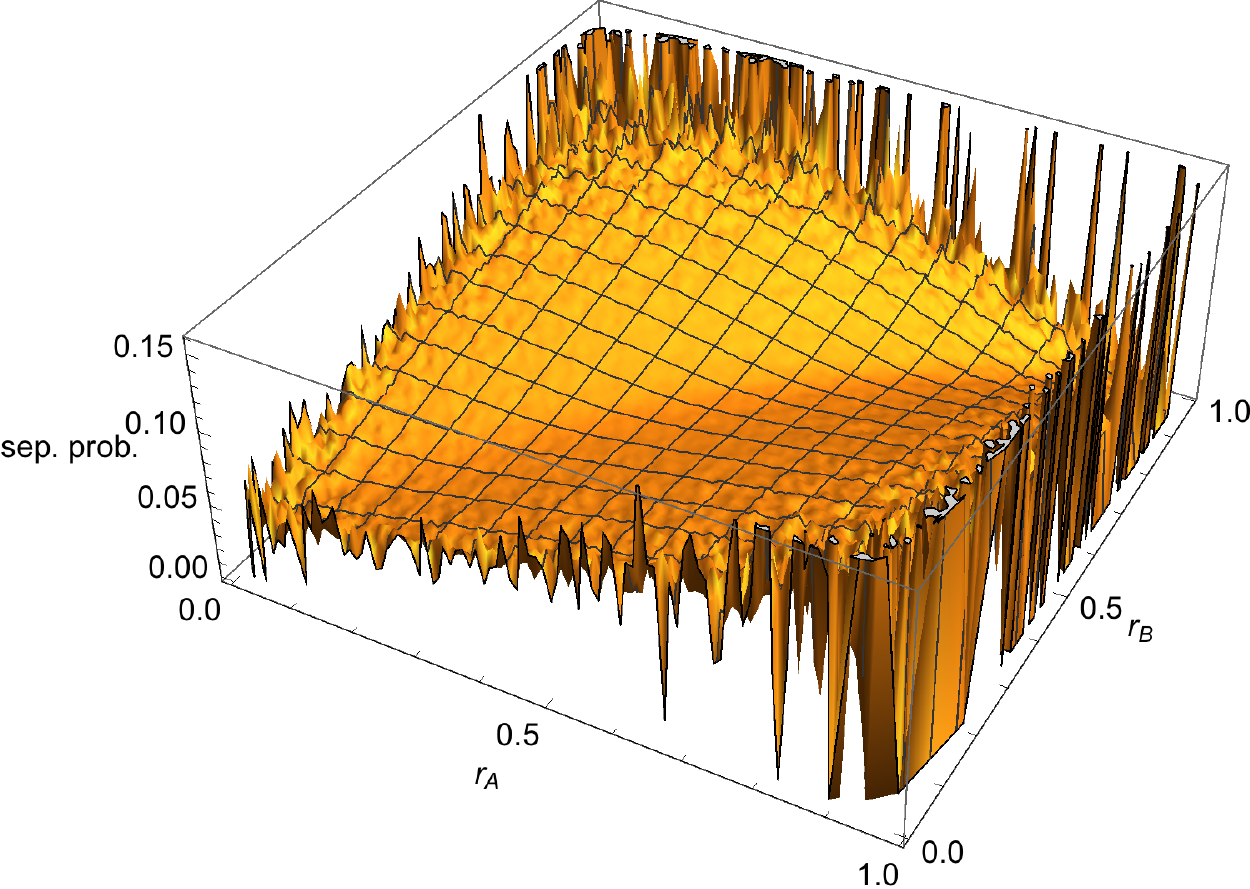}
\caption{\label{fig:SepProbRatiosInducedK3}Estimated joint random induced ($K=3$) separability probabilities--the ratio of Fig.~\ref{fig:SeparabilityCountsInducedK3} to Fig.~\ref{fig:TotalCountsInducedK3}}
\end{figure}
\begin{figure}
\includegraphics{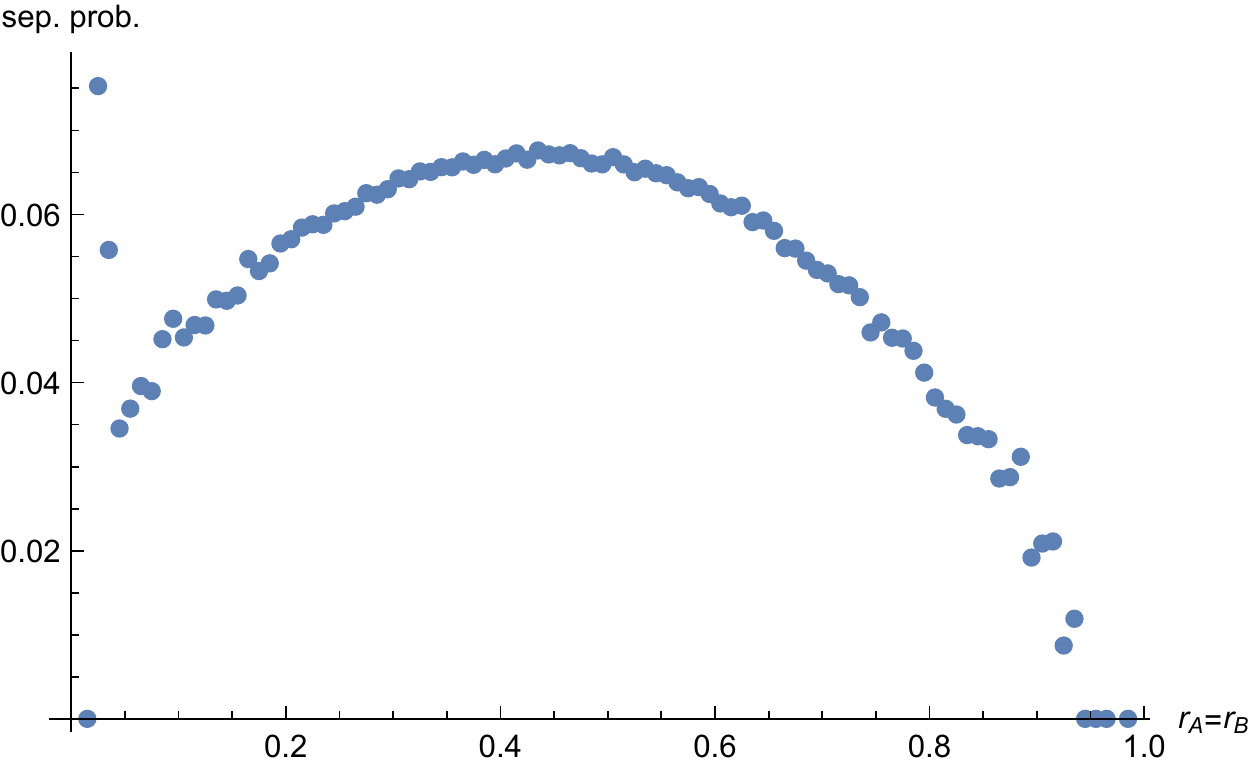}
\caption{\label{fig:DiagonalSepProbsInducedK3}Estimated random induced ($K=3$) two-qubit separability probabilities for $r_A=r_B$. A closely-fitting model (ratio of apparent separable and total volumes) for this curve is $\frac{1}{32} (1-r) \left(r^2+6 r+1\right)$, giving a total separability probability for the continuum of such states of $\frac{9}{143} \approx 0.0629371$. The maximum of the curve $\frac{1}{27} (5 \sqrt{10}-14) \approx 0.0670885$ occurs at $r= \frac{1}{3} (2 \sqrt{10}-5) \approx  0.441518$}
\end{figure}
\begin{figure}
\includegraphics{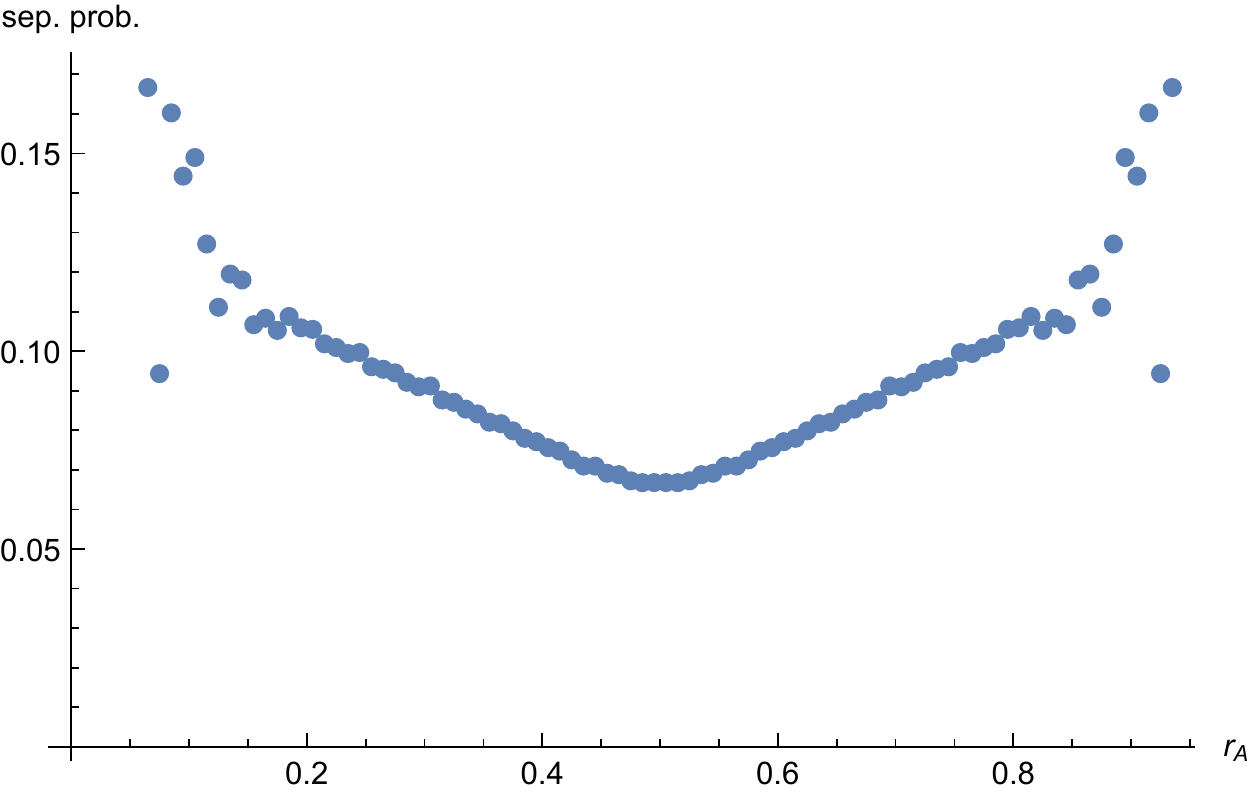}
\caption{\label{fig:ConstantSumSepProbsInducedK3}Estimated random induced ($K=3$) two-qubit separability probabilities for $r_A+r_B=1$.}
\end{figure}
\begin{figure}
\includegraphics{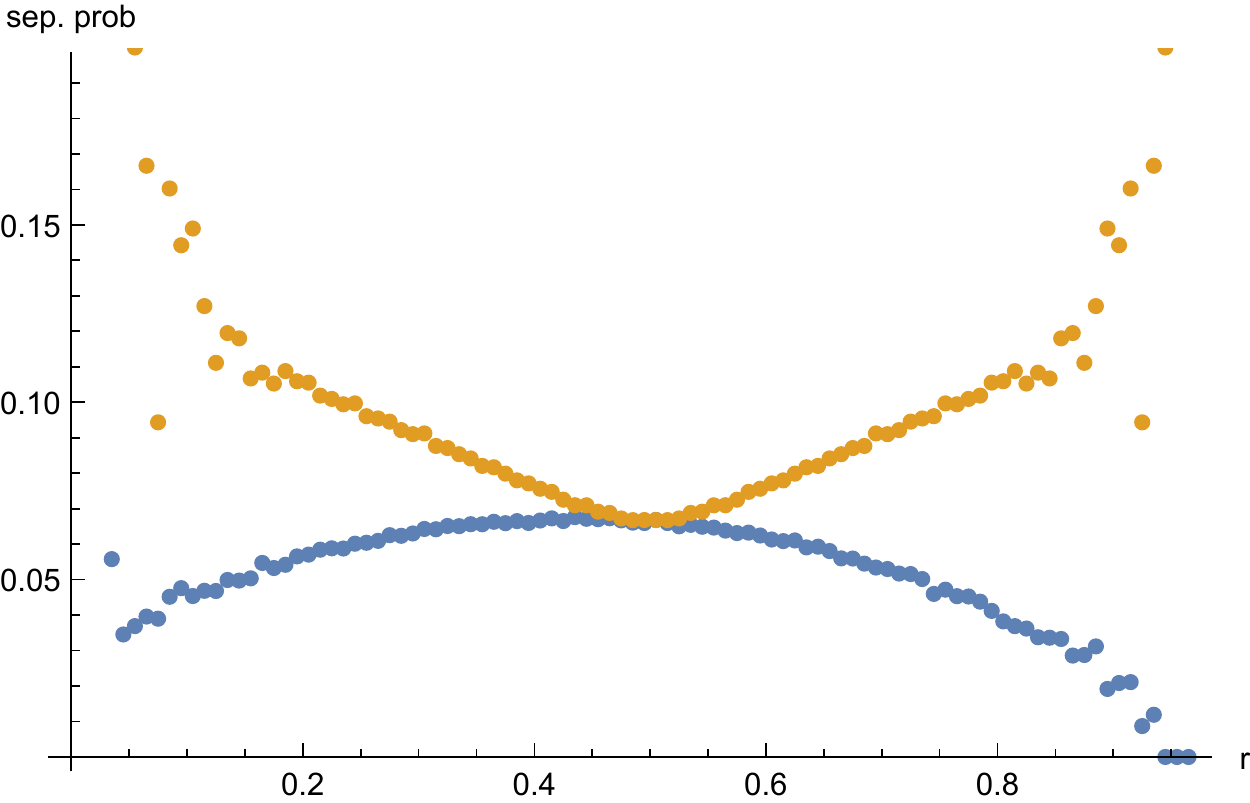}
\caption{\label{fig:K3JointPlot}Joint plot of the last two figures}
\end{figure}
\begin{figure}
\includegraphics{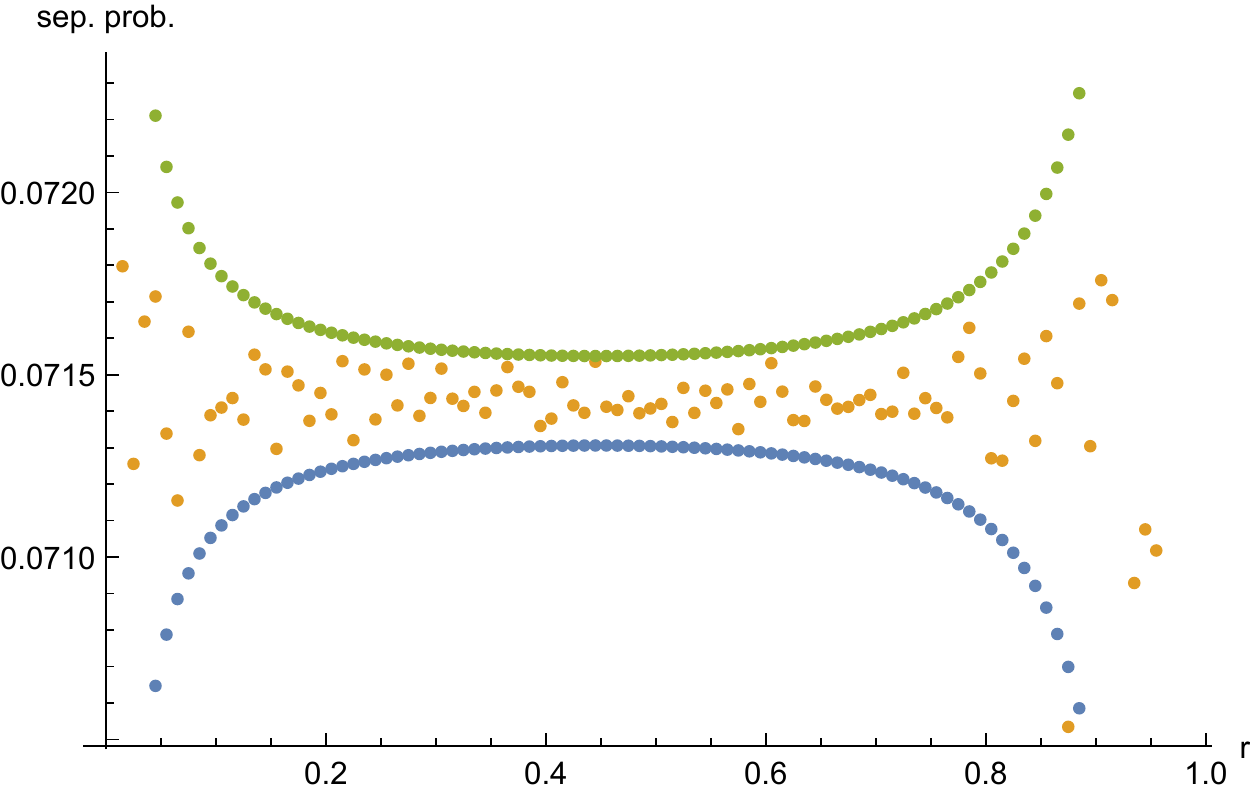}
\caption{\label{fig:K3confidence}Estimated random induced ($K=3$) two-qubit separability probabilities over either one of the Bloch radii, along with 95$\%$ confidence limits about the conjectured value of $\frac{1}{14} \approx 0.0714285$}
\end{figure}
\begin{figure}
\includegraphics{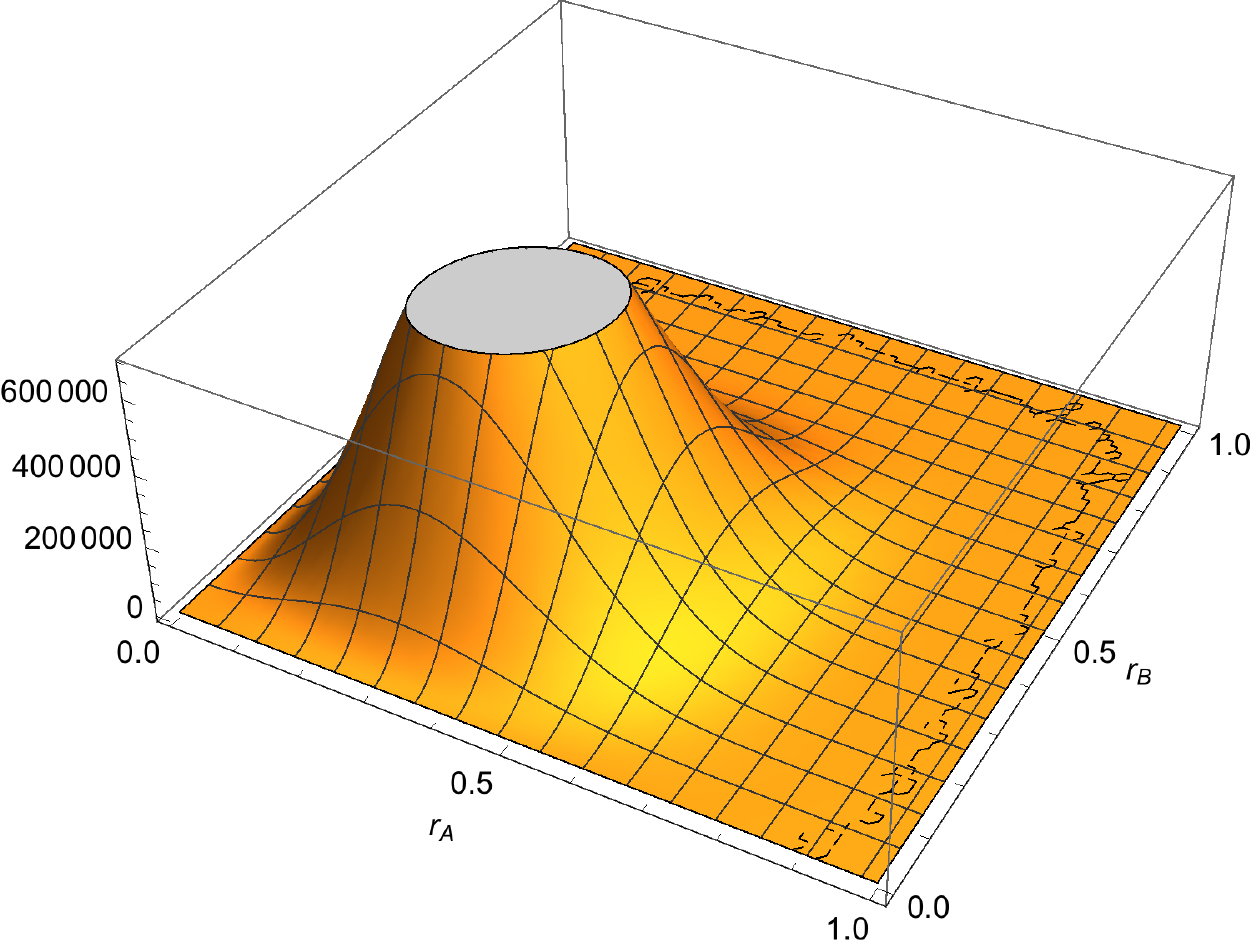}
\caption{\label{fig:TotalCountsInducedK5}Histogram of randomly sampled (with respect to the random induced $K=5$ measure) 
two-qubit density matrices}
\end{figure}
\begin{figure}
\includegraphics{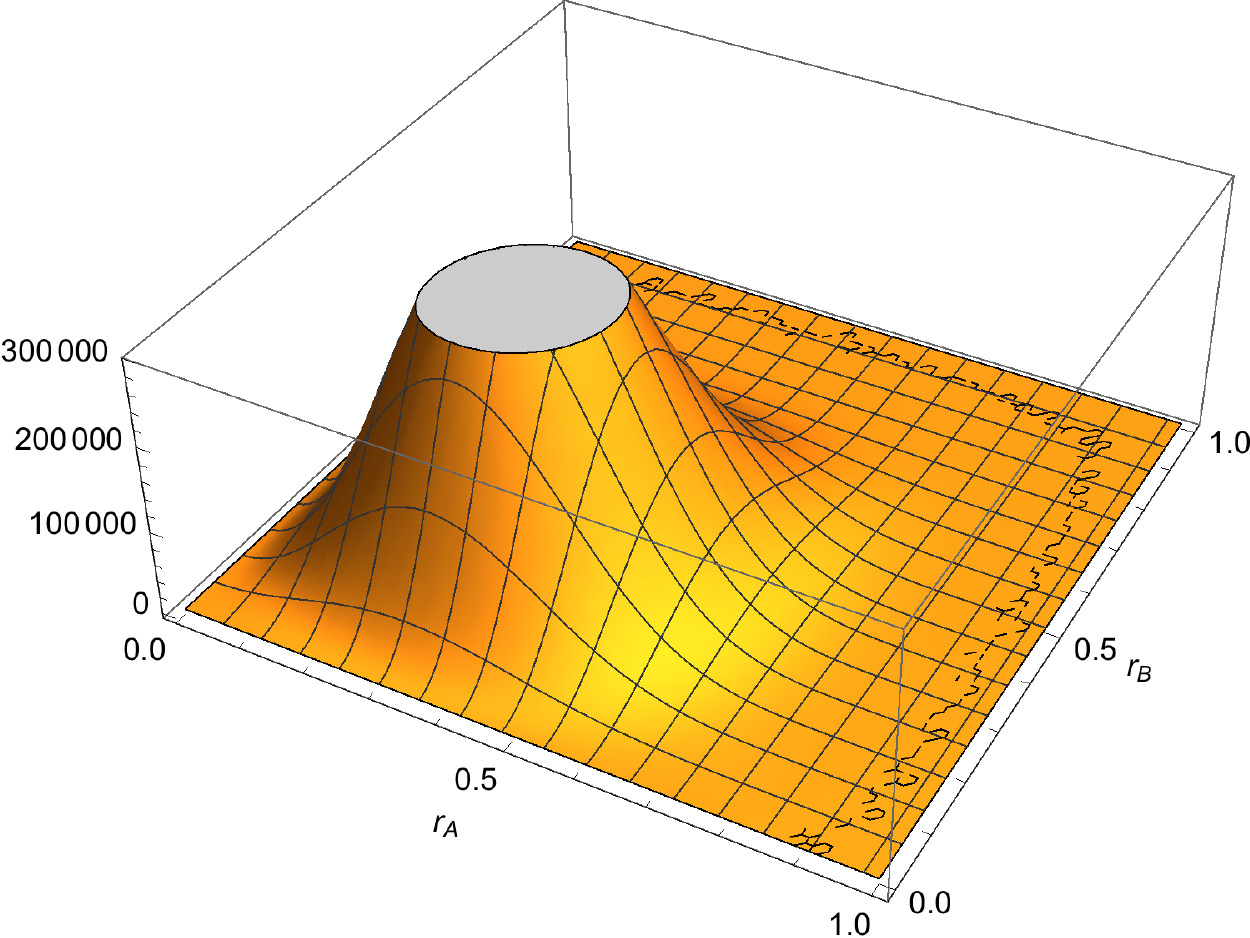}
\caption{\label{fig:SeparabilityCountsInducedK5}Histogram of randomly sampled (with respect to the random induced $K=5$ measure) 
{\it separable} two-qubit density matrices}
\end{figure}
\newpage
\clearpage
\begin{figure}
\includegraphics{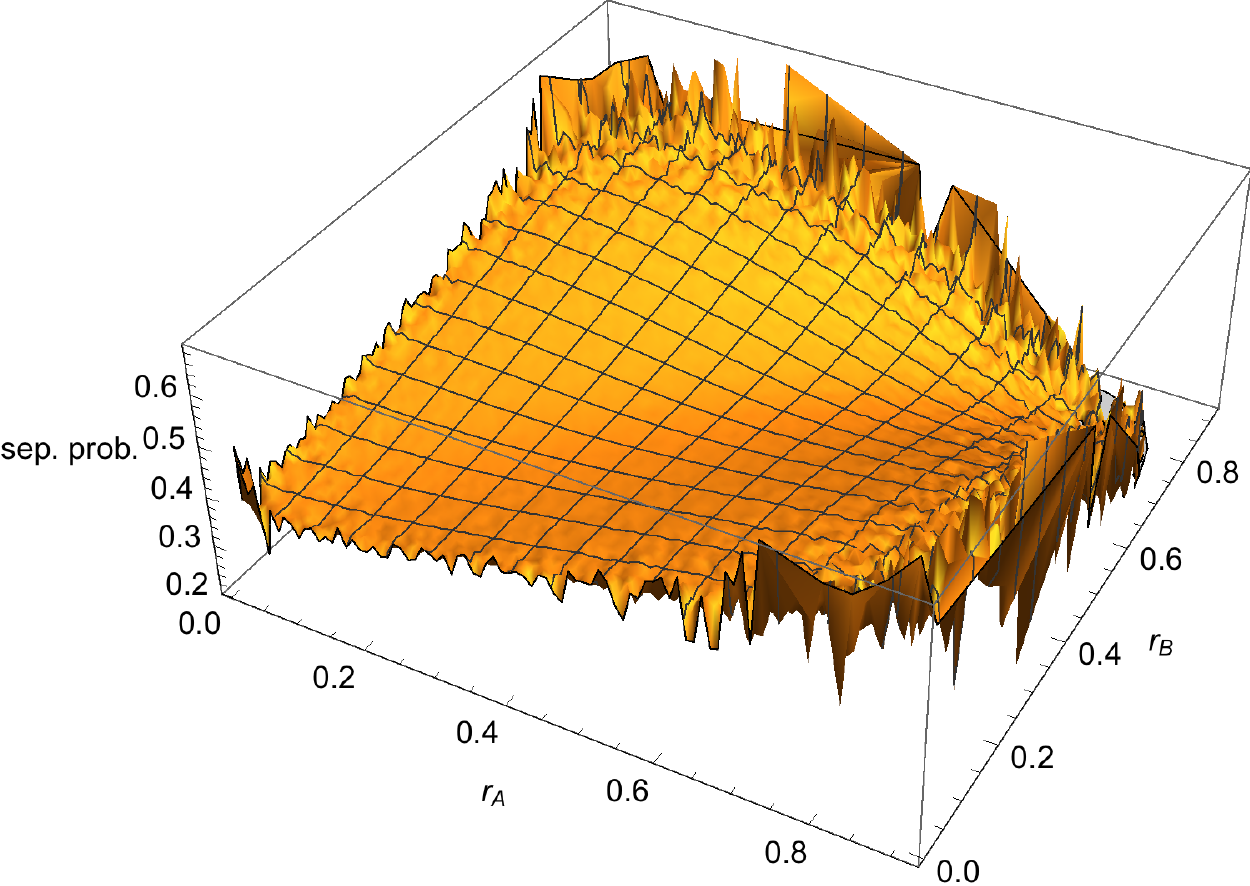}
\caption{\label{fig:SepProbRatiosInducedK5}Estimated joint random induced ($K=5$) separability probabilities--the ratio of Fig.~\ref{fig:SeparabilityCountsInducedK5} to Fig.~\ref{fig:TotalCountsInducedK5}}
\end{figure}
\begin{figure}
\includegraphics{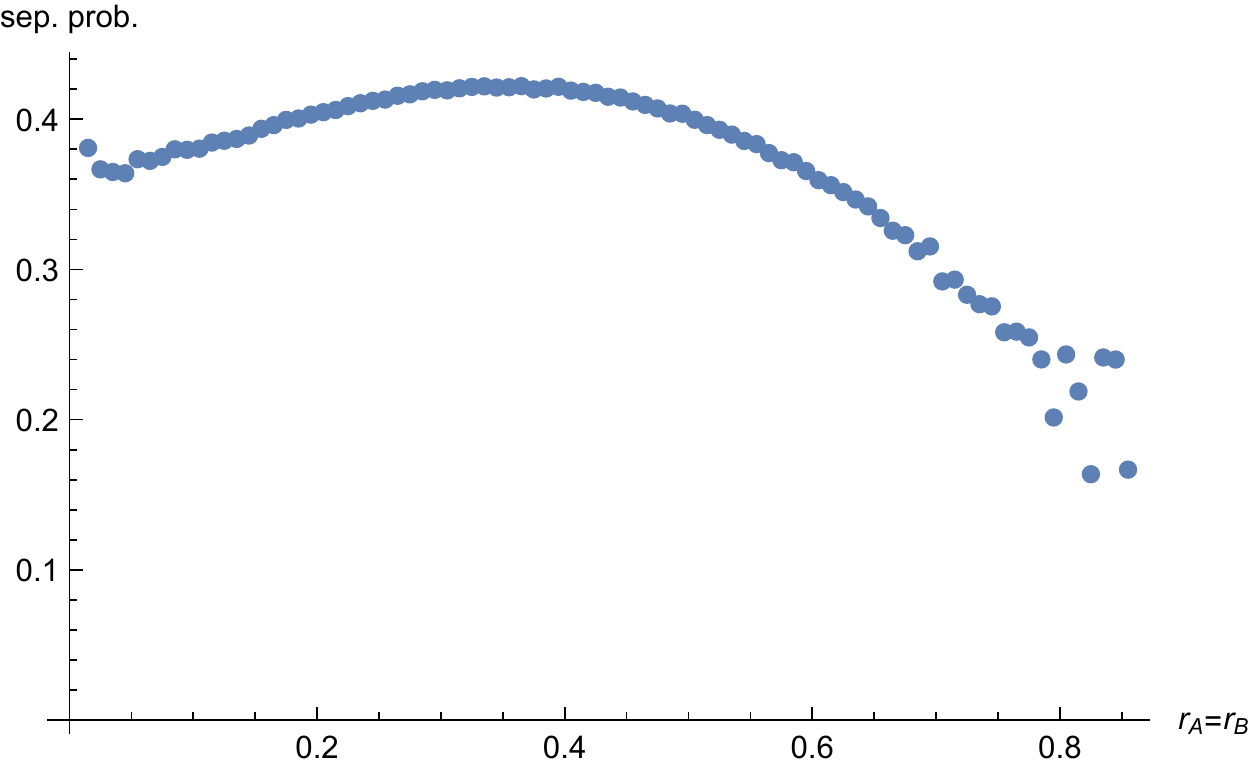}
\caption{\label{fig:DiagonalSepProbsInducedK5}Estimated random induced ($K=5$) two-qubit separability probabilities for $r_A=r_B$.  A closely-fitting model (ratio of apparent separable and total volumes) for this curve is $\frac{(1-r) \left(\frac{87 r^3}{2}+\frac{85 r^2}{4}+\frac{17
   r}{4}+\frac{1}{3}\right)}{40 r^2+11 r+1}$.}
\end{figure}
\begin{figure}
\includegraphics{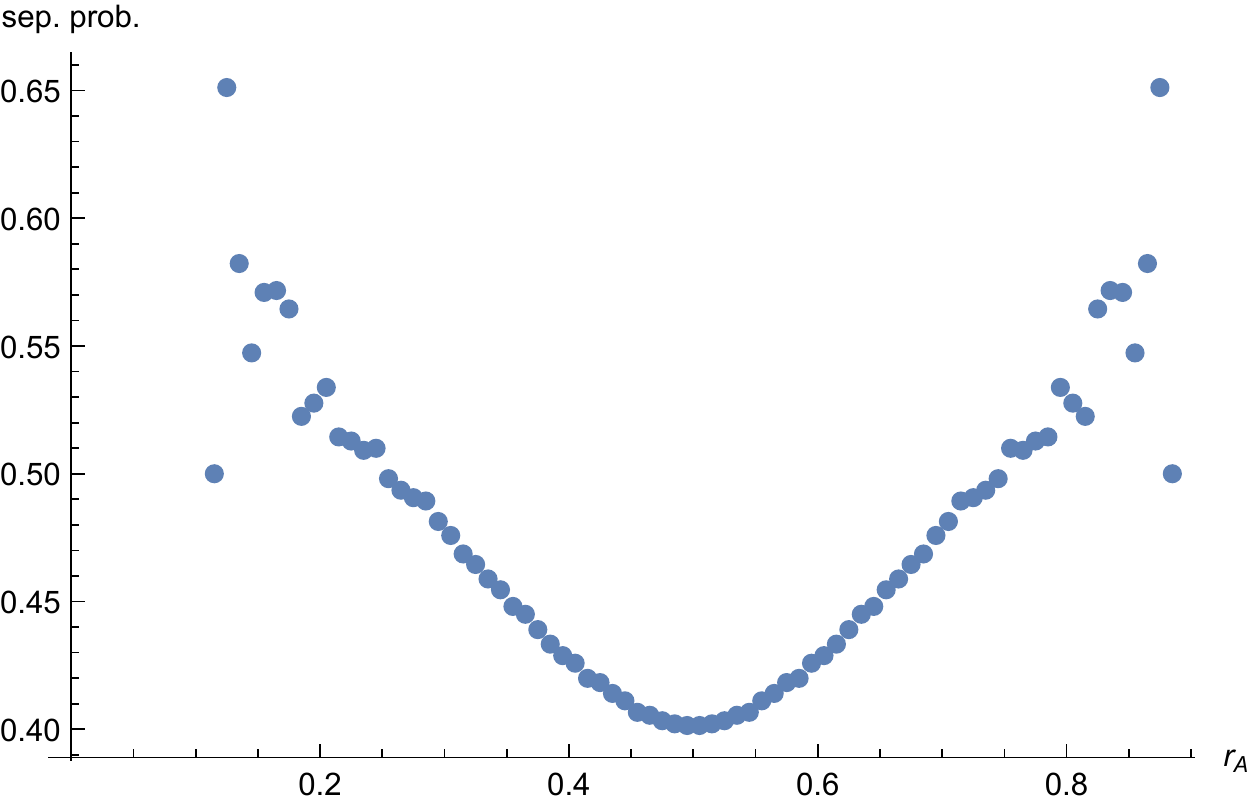}
\caption{\label{fig:ConstantSumSepProbsInducedK5}Estimated random induced ($K=5$) two-qubit separability probabilities for $r_A+r_B=1$.}
\end{figure}
\newpage
\clearpage
\begin{figure}
\includegraphics{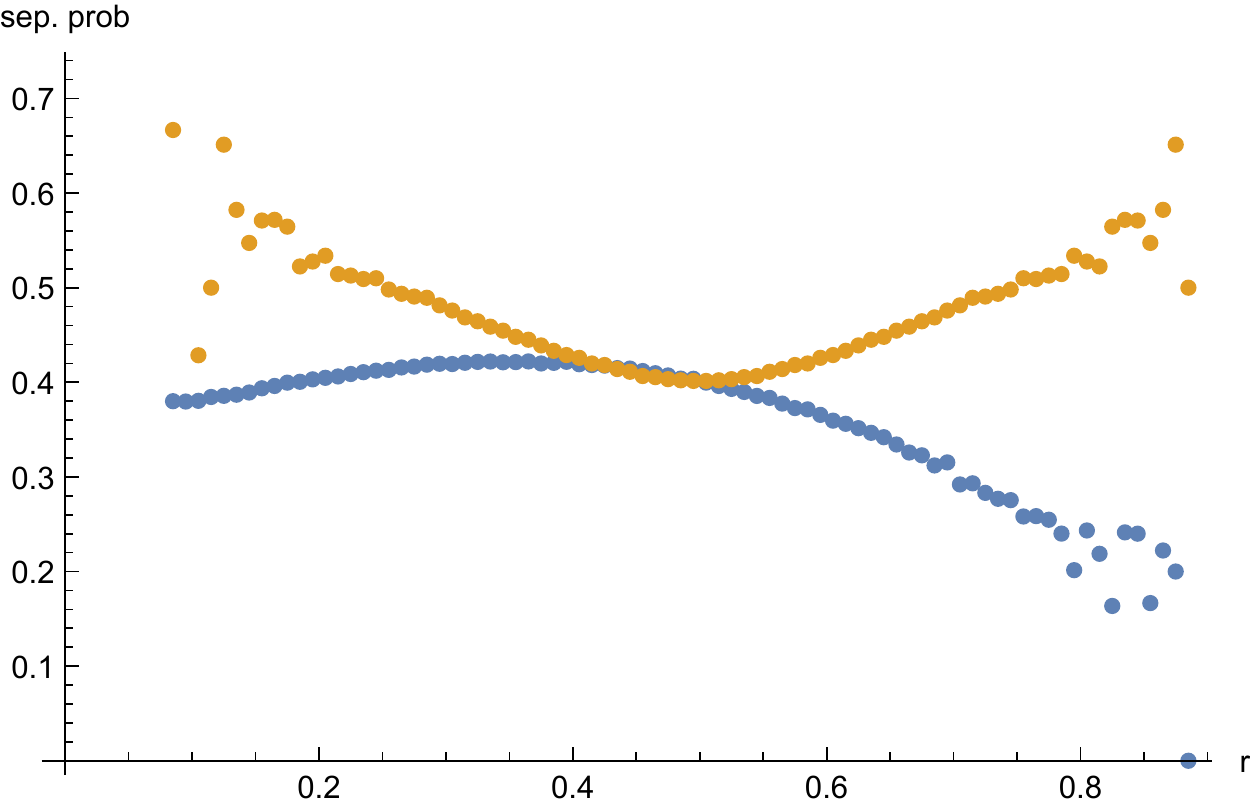}
\caption{\label{fig:K5JointPlot}Joint plot of the last two figures}
\end{figure}
\begin{figure}
\includegraphics{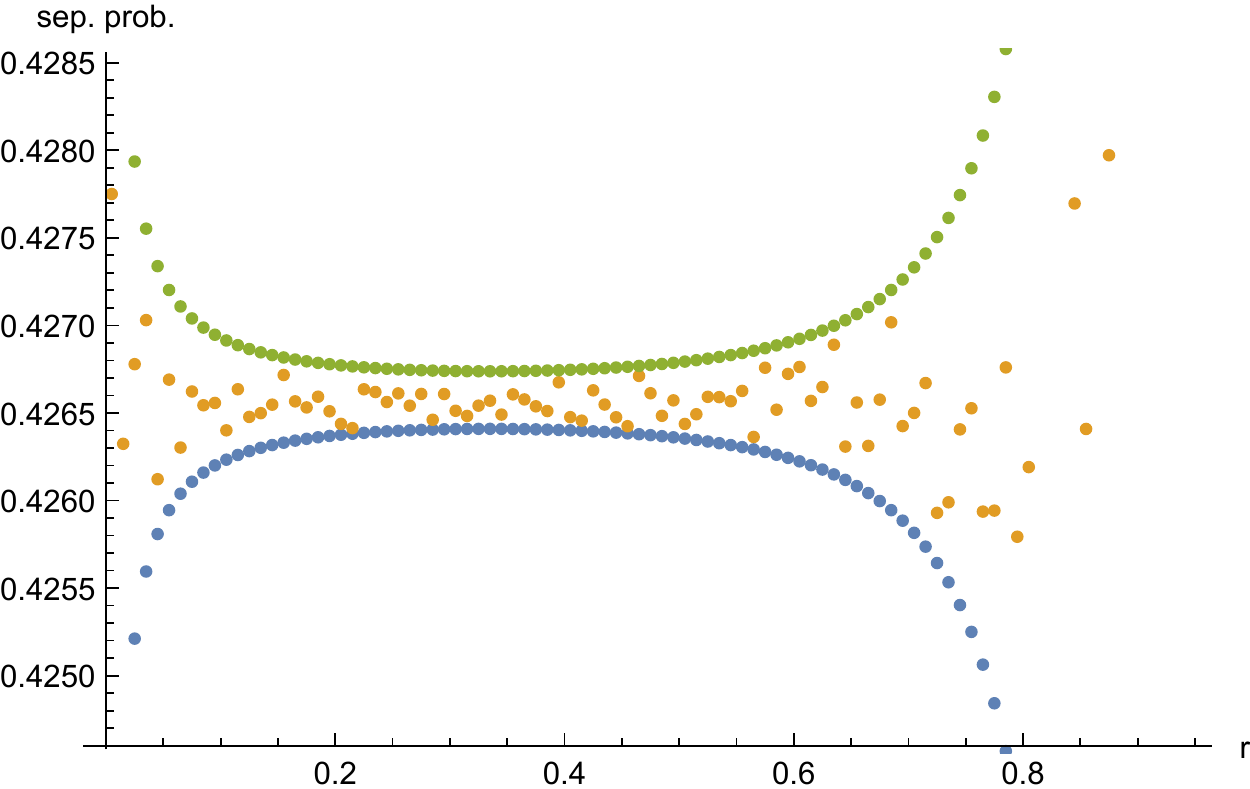}
\caption{\label{fig:K5confidence}Estimated random induced ($K=5$) two-qubit separability probabilities over either one of the Bloch radii, along with 95$\%$ confidence limits about the conjectured value of $\frac{61}{143} \approx 0.426573$}
\end{figure}
\begin{figure}
\includegraphics{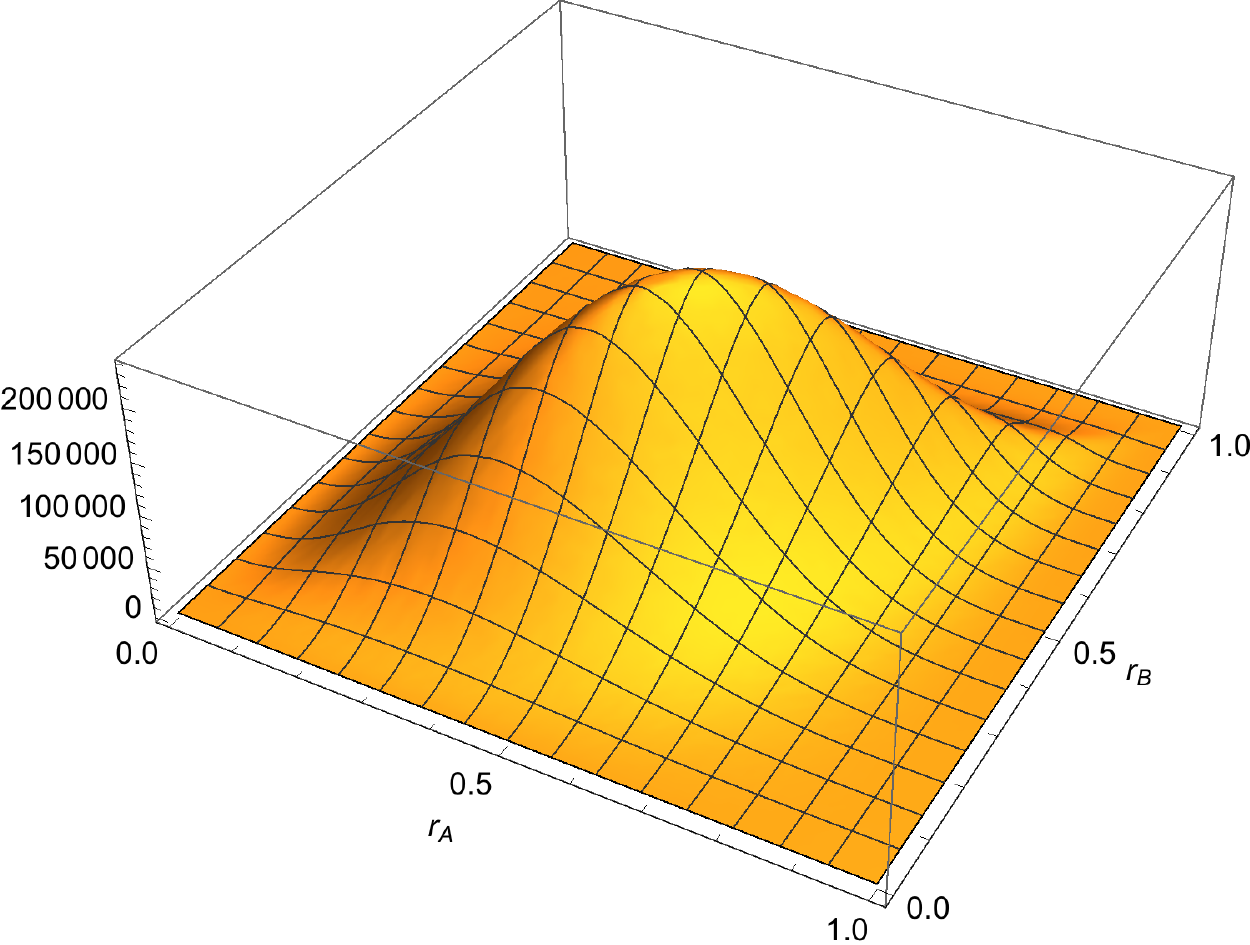}
\caption{\label{fig:TotalCountsBures}Histogram of randomly sampled (with respect to the Bures measure) 
two-qubit density matrices}
\end{figure}
\begin{figure}
\includegraphics{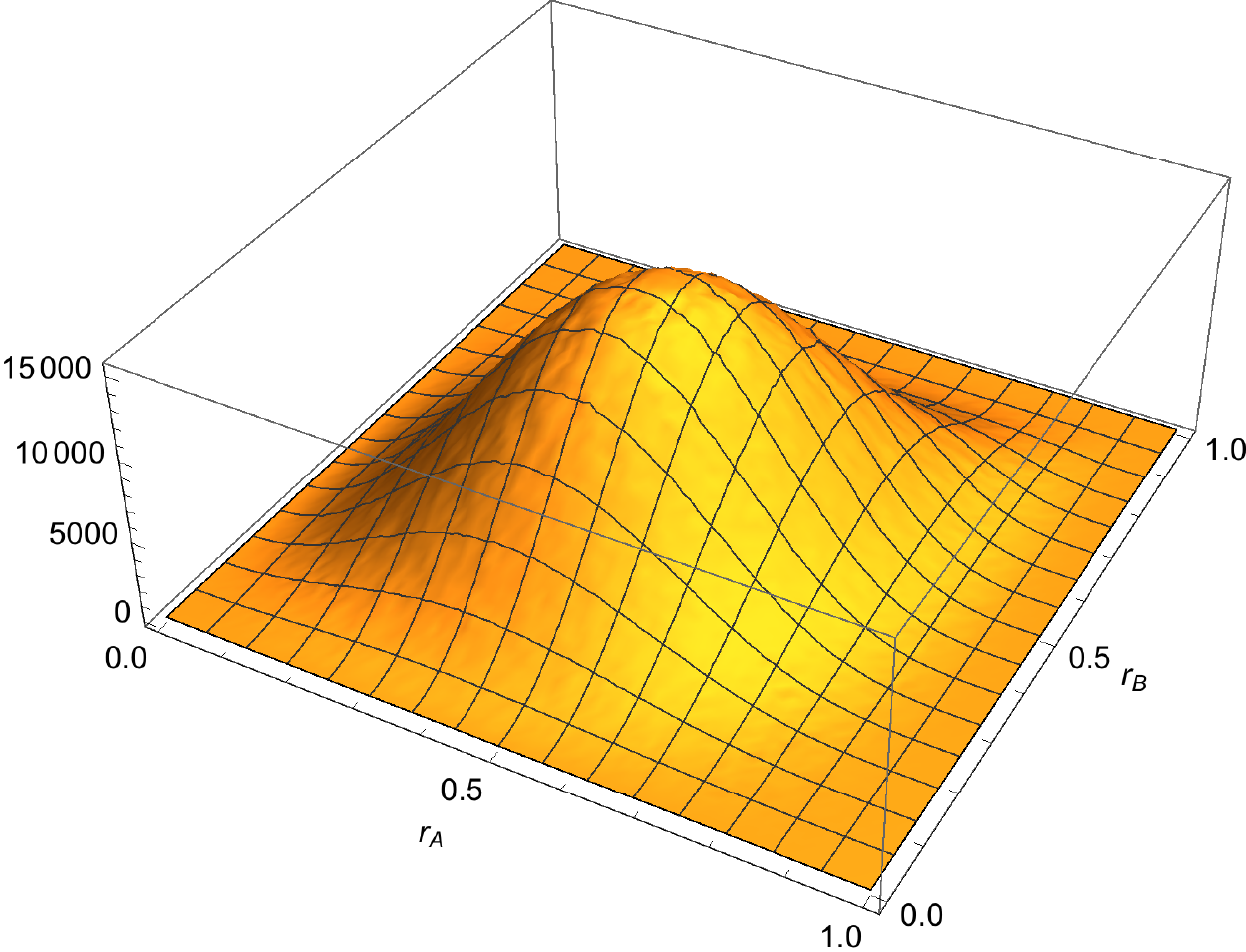}
\caption{\label{fig:SeparabilityCountsBures}Histogram of randomly sampled (with respect to the Bures measure) 
{\it separable} two-qubit density matrices}
\end{figure}
\clearpage
\newpage
\begin{figure}
\includegraphics{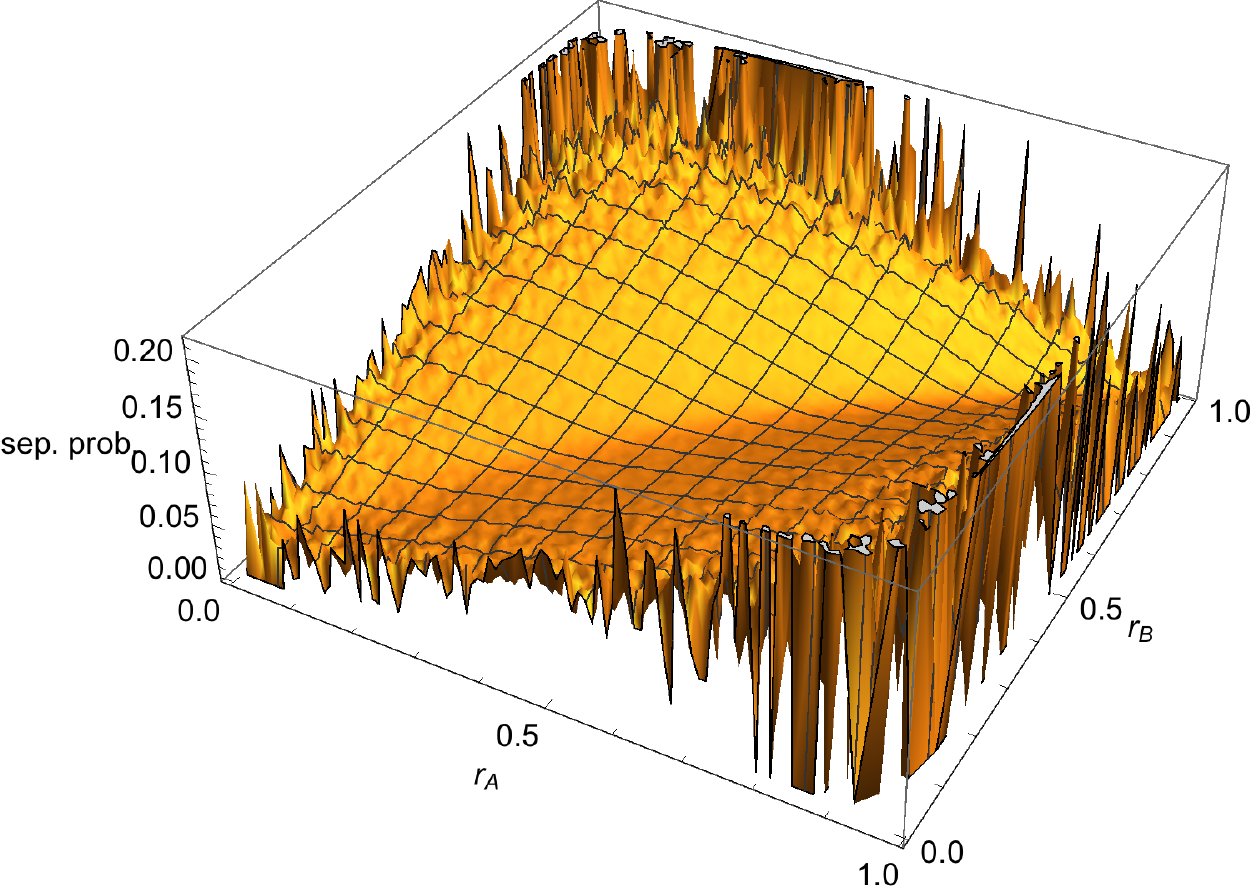}
\caption{\label{fig:SepProbRatiosBures}Estimated joint Bures two-qubit separability probabilities--the ratio of Fig.~\ref{fig:SeparabilityCountsBures} to Fig.~\ref{fig:TotalCountsBures}}
\end{figure}
\begin{figure}
\includegraphics{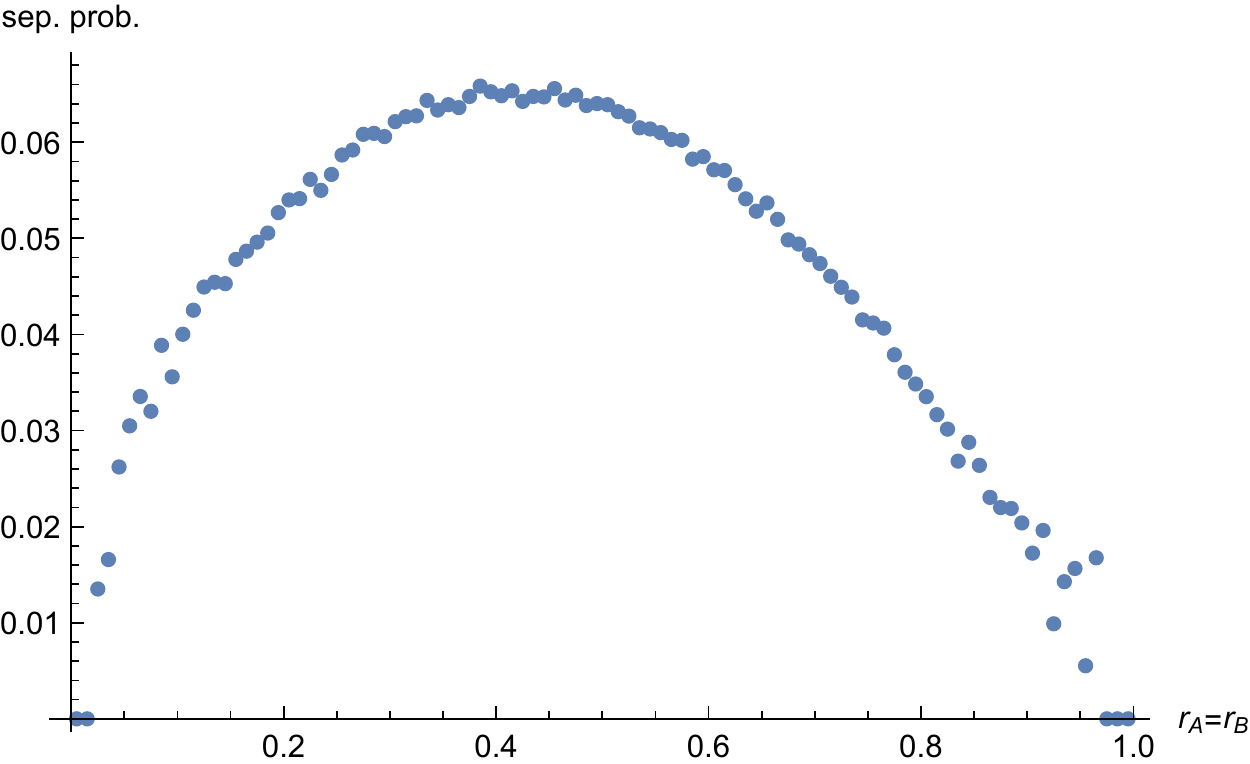}
\caption{\label{fig:DiagonalSepProbsBures}Estimated Bures two-qubit separability probabilities for $r_A=r_B$}
\end{figure}
\begin{figure}
\includegraphics{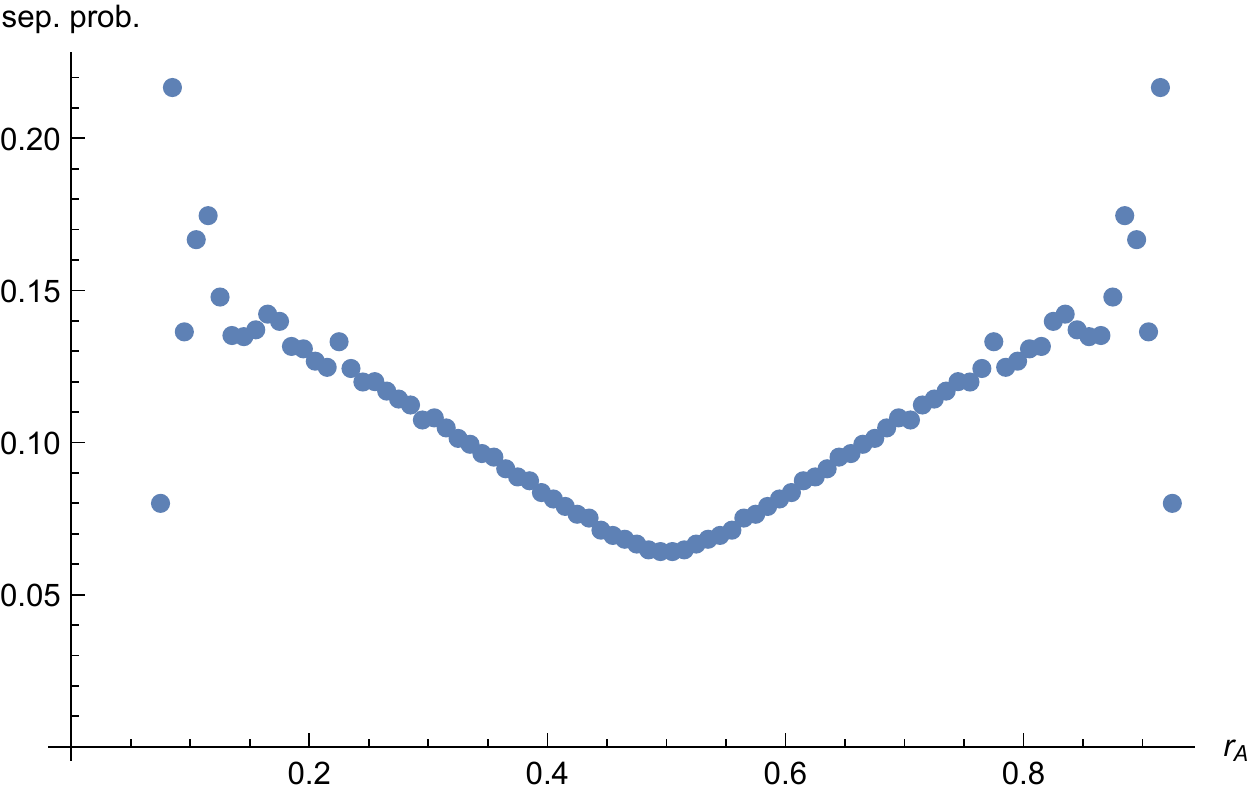}
\caption{\label{fig:ConstantSumSepProbsBures}Estimated Bures two-qubit separability probabilities for $r_A+r_B=1$}
\end{figure}
\begin{figure}
\includegraphics{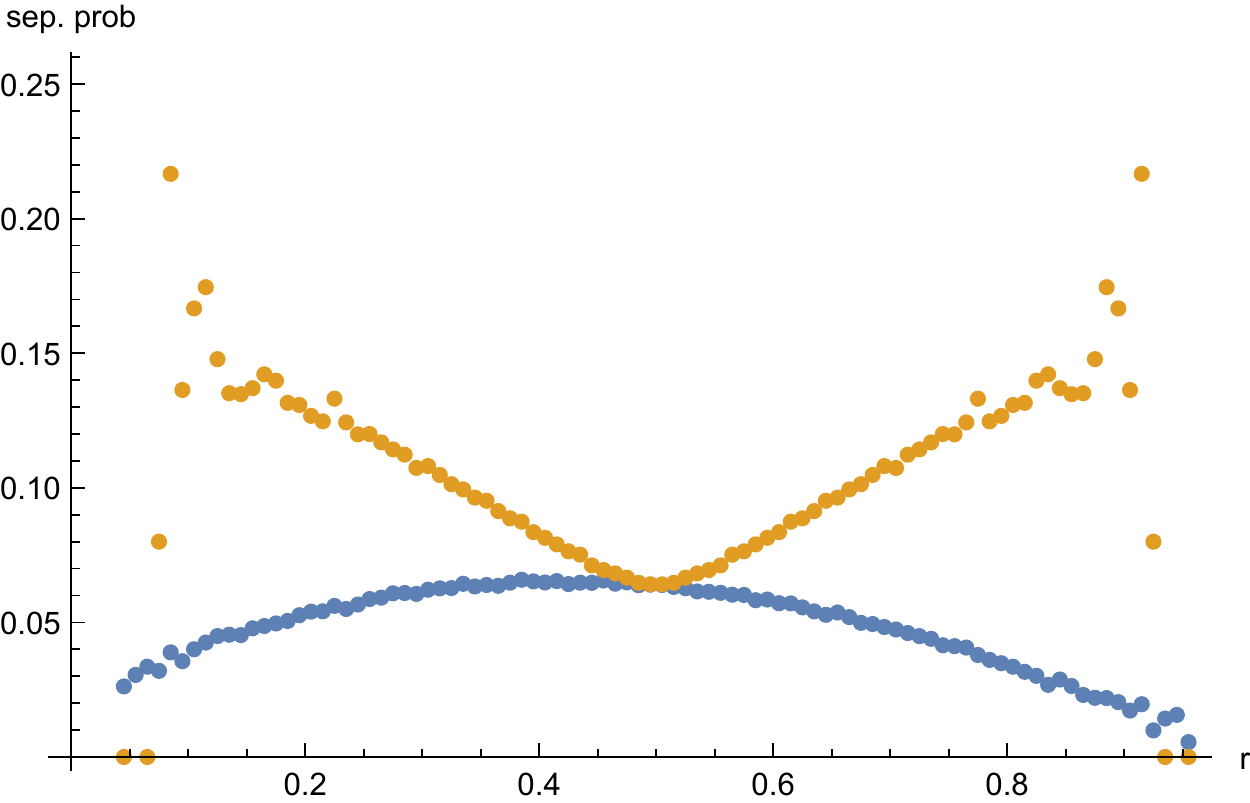}
\caption{\label{fig:BuresJointPlot}Joint plot of the last two figures}
\end{figure}
\begin{figure}
\includegraphics{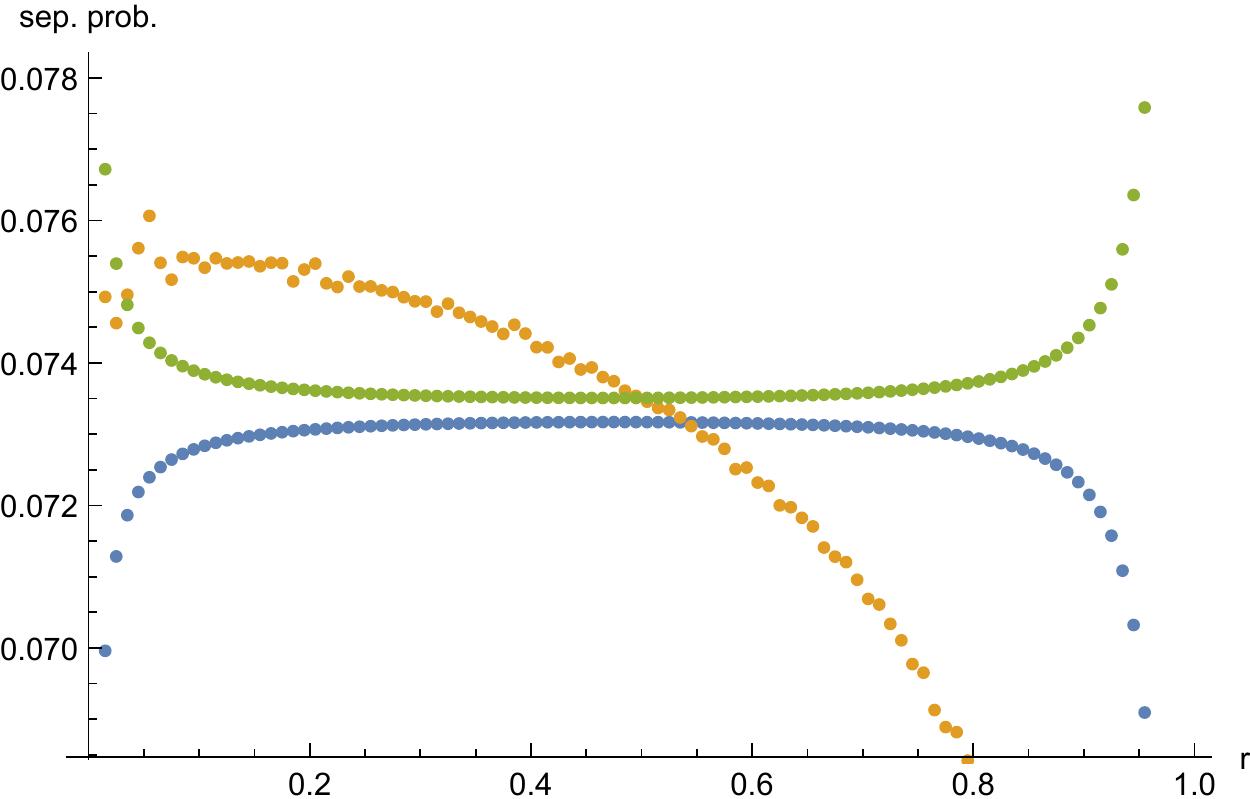}
\caption{\label{fig:BuresConfidence}Estimated Bures two-qubit separability probabilities over either one of the Bloch radii, along with ill-fitting (in contrast to Hilbert-Schmidt and random induced) 95$\%$ confidence limits about the conjectured overall probability of 
 $\frac{1680 (\sqrt{2}-1) }{\pi^8} \approx 0.0733389$.}
\end{figure}
\begin{figure}
\includegraphics{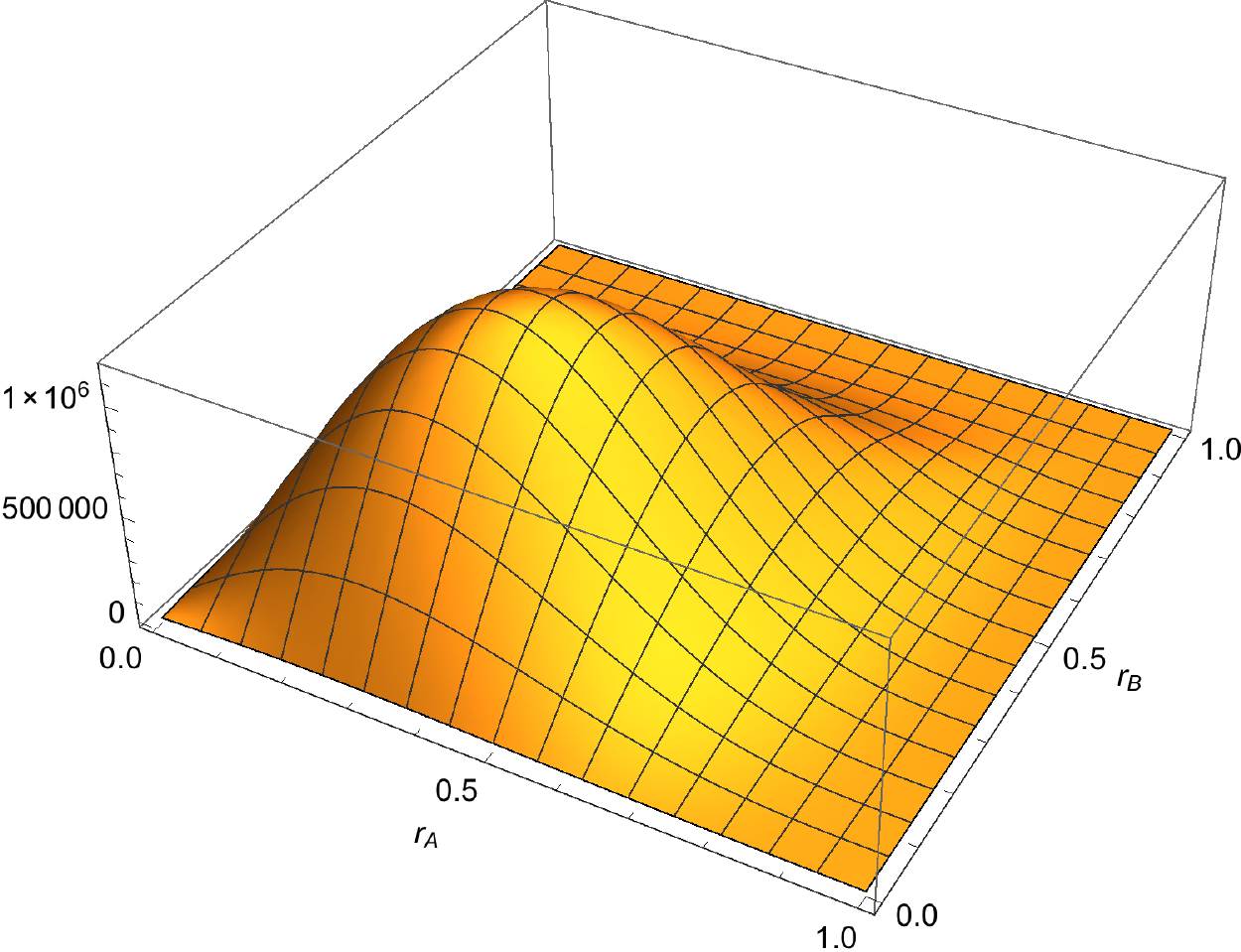}
\caption{\label{fig:RealTotalCounts}Histogram of randomly sampled
two-{\it rebit} density matrices}
\end{figure}
\begin{figure}
\includegraphics{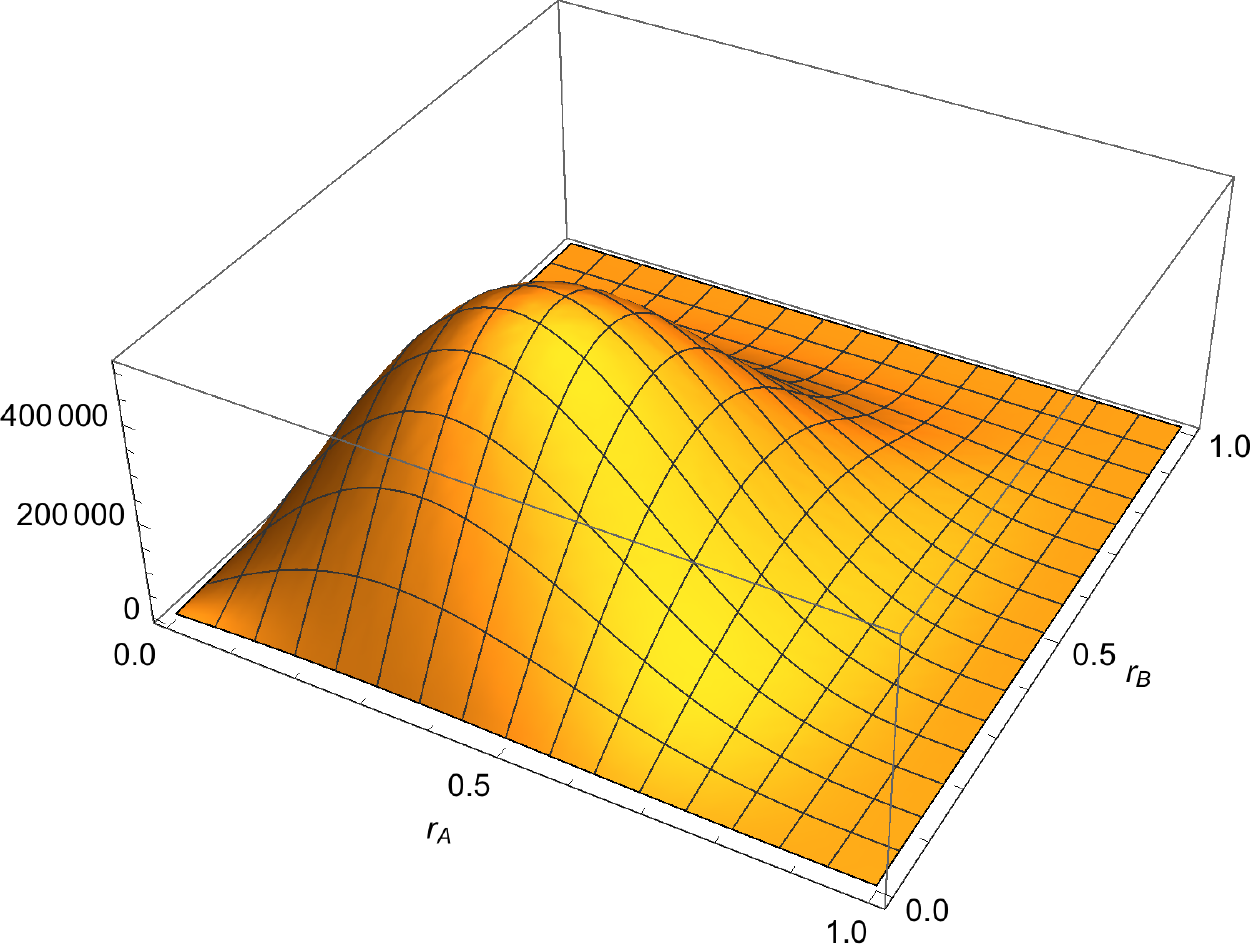}
\caption{\label{fig:RealSeparabilityCounts}Histogram of randomly sampled
{\it separable} two-rebit density matrices}
\end{figure}
\begin{figure}
\includegraphics{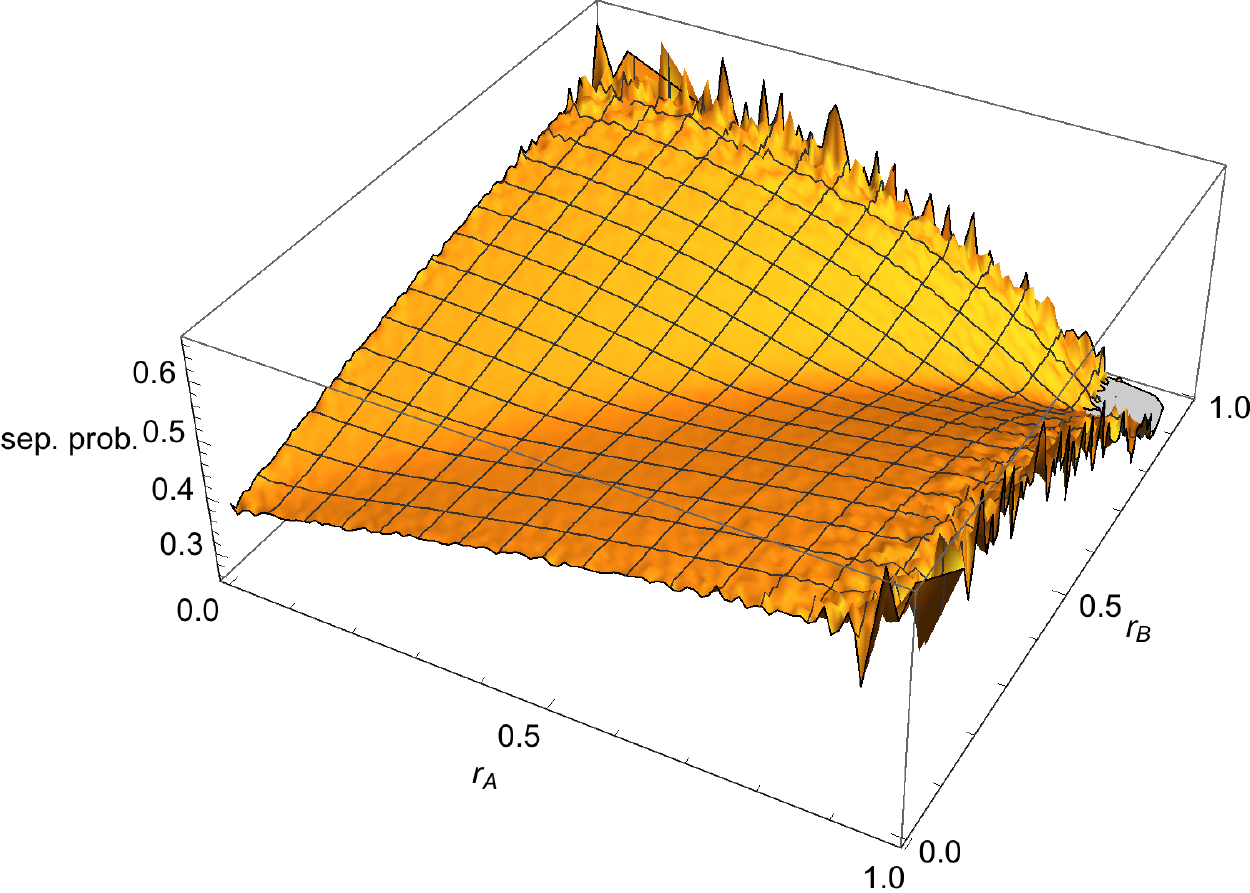}
\caption{\label{fig:RealSepProbRatios}Estimated joint Hilbert-Schmidt two-{\it rebit} separability probabilities--the ratio of Fig.~\ref{fig:RealSeparabilityCounts} to Fig.~\ref{fig:RealTotalCounts}}
\end{figure}
\begin{figure}
\includegraphics{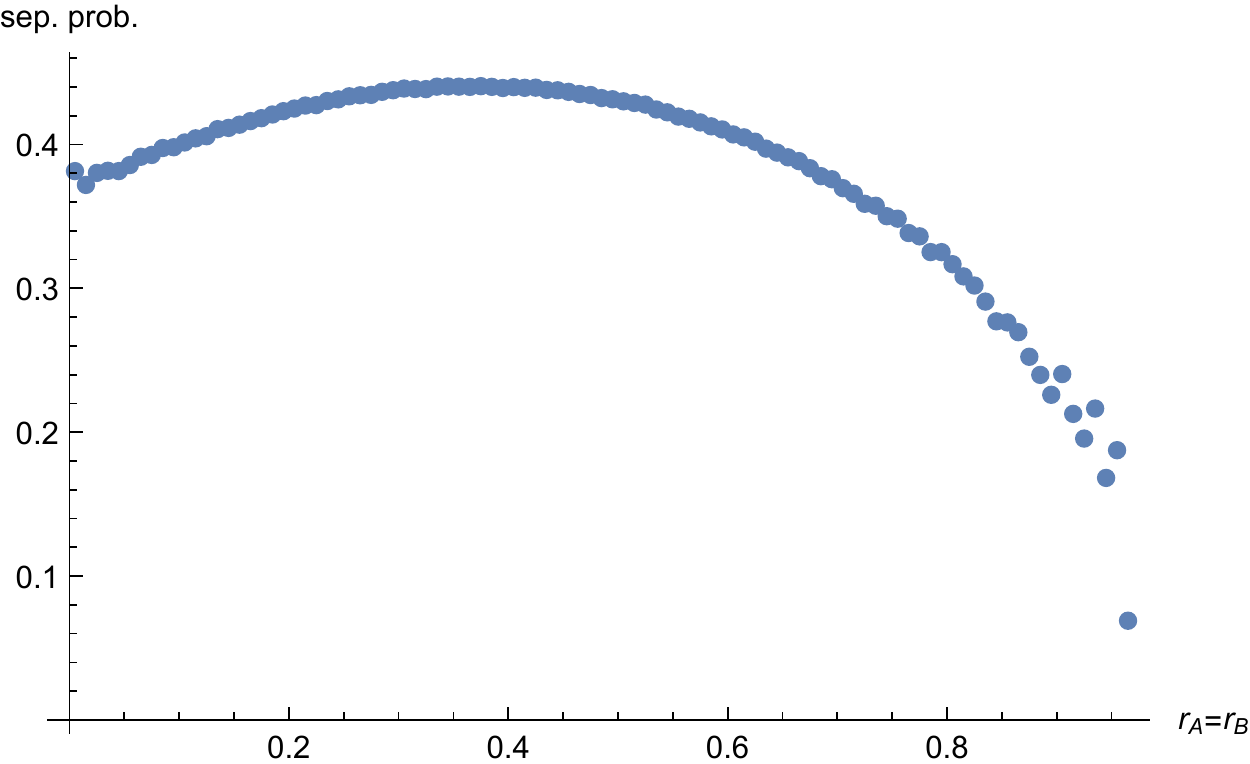}
\caption{\label{fig:RealDiagonalSepProbs}Estimated Hilbert-Schmidt  two-{\it rebit} separability probabilities for $r_A=r_B$}
\end{figure}
\newpage
\clearpage
\begin{figure}
\includegraphics{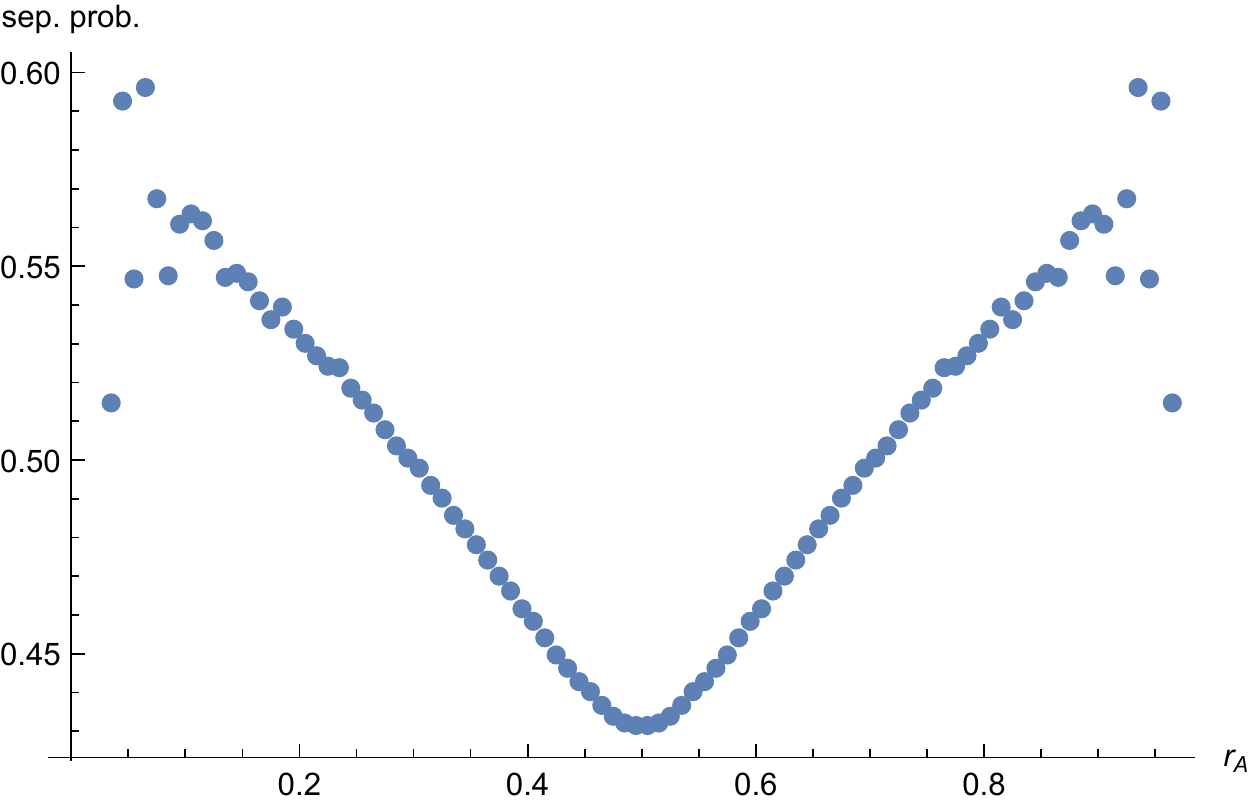}
\caption{\label{fig:RealConstantSumSepProbs}Estimated Hilbert-Schmidt two-{\it rebit} separability probabilities for $r_A+r_B=1$}
\end{figure}
\begin{figure}
\includegraphics{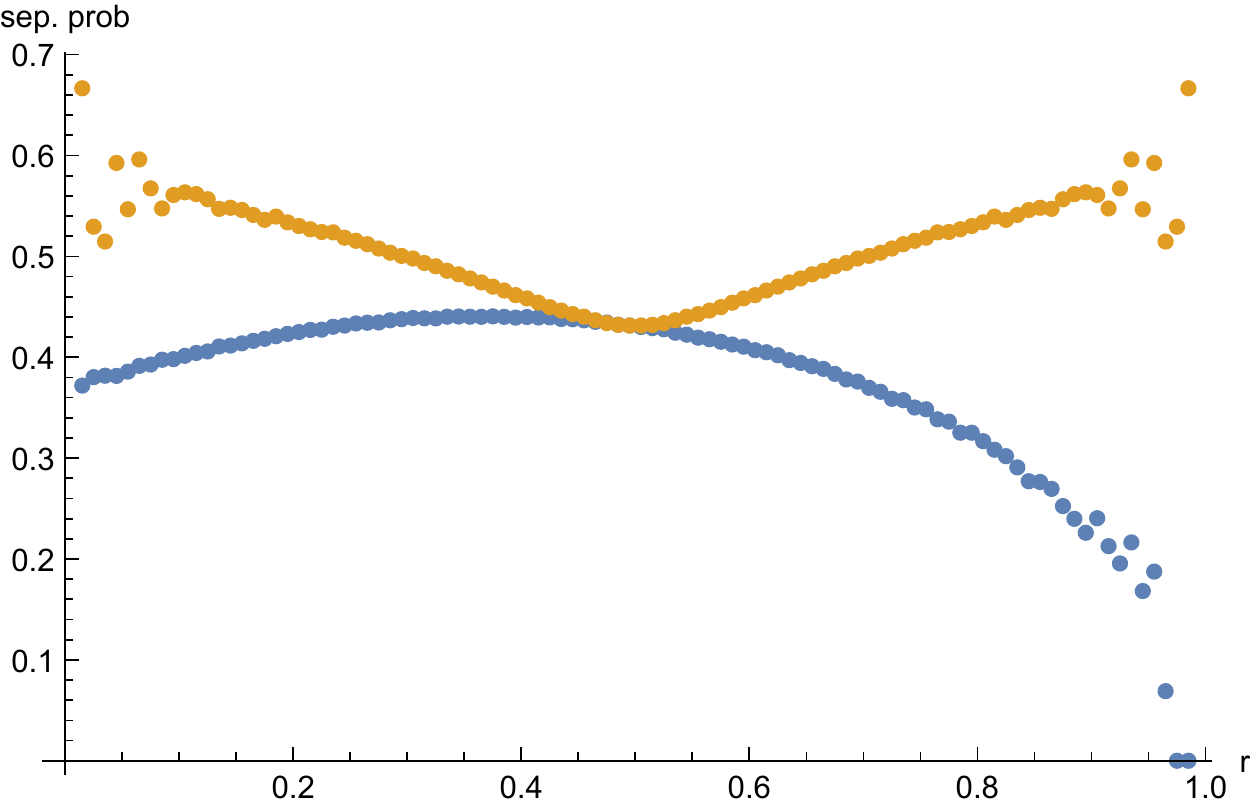}
\caption{\label{fig:RealJointPlot}Joint plot of the last two figures}
\end{figure}
\begin{figure}
\includegraphics{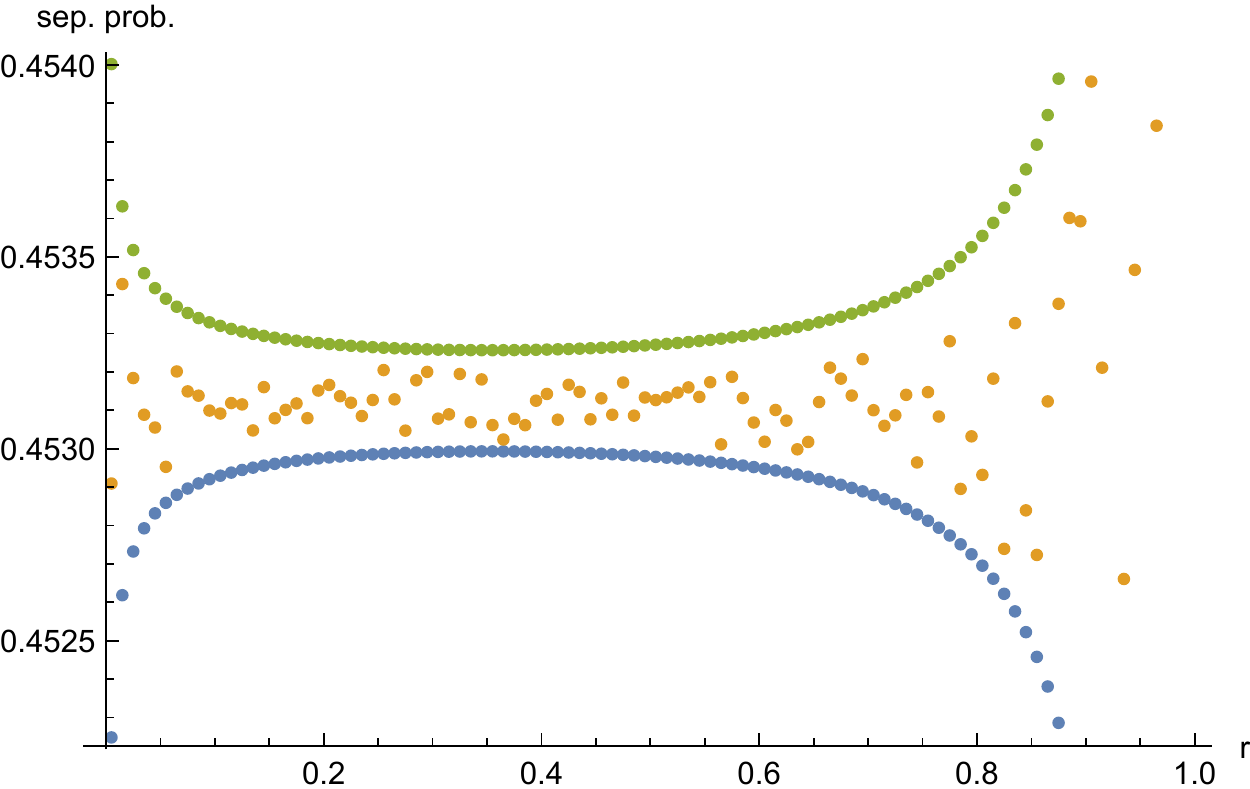}
\caption{\label{fig:Realconfidence}Estimated (marginal) Hilbert-Schmidt  
two-{\it rebit} separability 
probabilities over either one of the Bloch radii, along with 95$\%$ confidence limits about the conjectured value of $\frac{29}{64} \approx  0.4531250$}
\end{figure}
\begin{figure}
\includegraphics{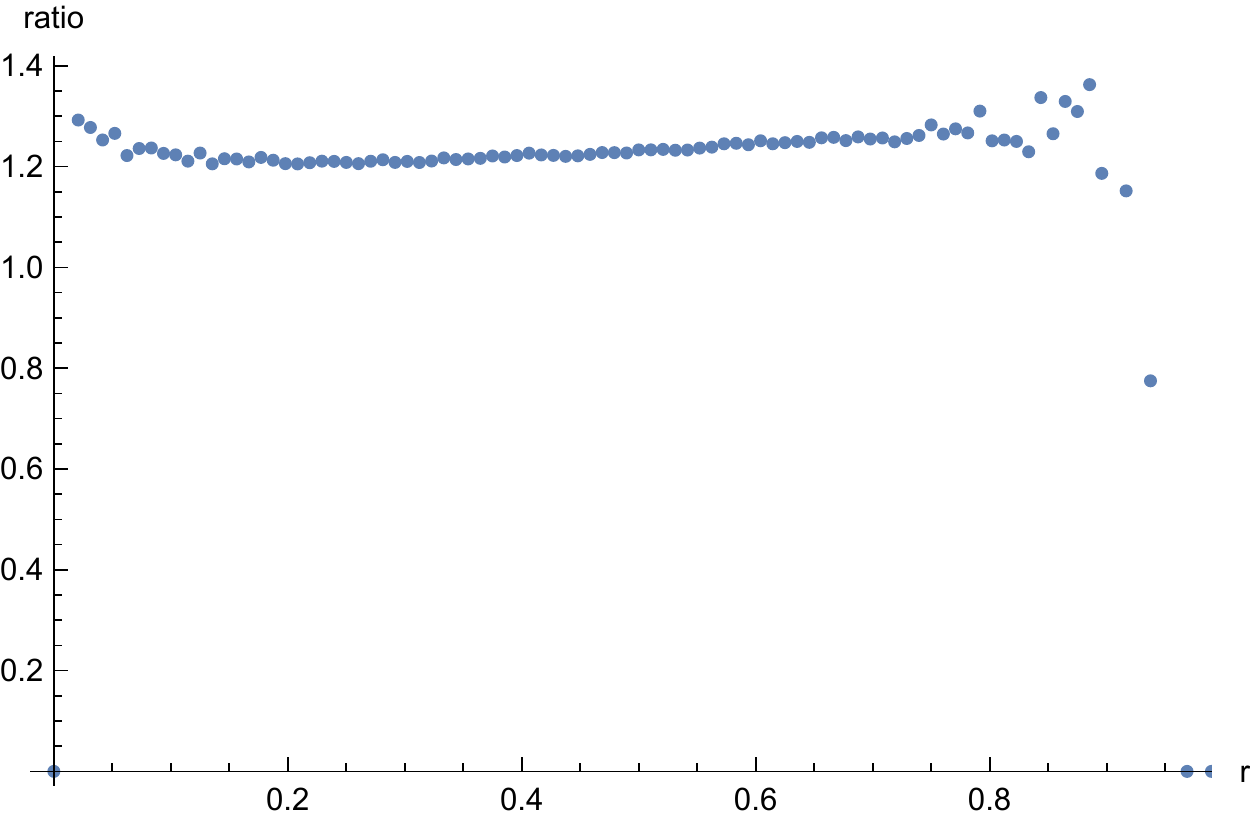}
\caption{\label{fig:QubitRebit2}Ratio of the separability probability $p^{2-qubit}(r_A=r_B)$ to the square of 
the separability probability $p^{2-rebit}(r_A=r_B)$}
\end{figure}
\begin{figure}
\includegraphics{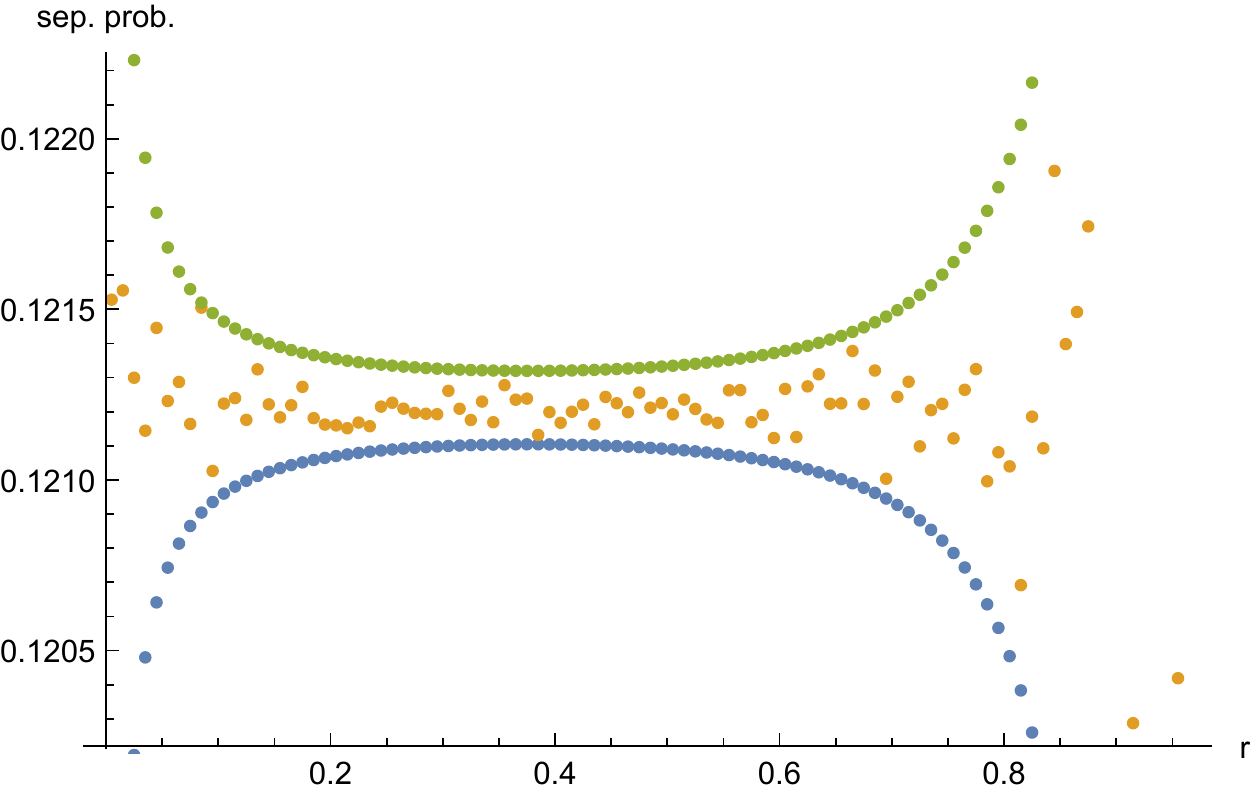}
\caption{\label{fig:HSconfidenceDivision}Estimated two-qubit separability probability for the Hilbert-Schmidt case $(k=0)$ associated with the determinantal inequality
$|\rho^{PT}| > |\rho|$, along with 95$\%$ confidence limits about  $\frac{1}{2} (\frac{8}{33}) = \frac{4}{33} \approx 0.121212$.} 
\end{figure}
\begin{figure}
\includegraphics{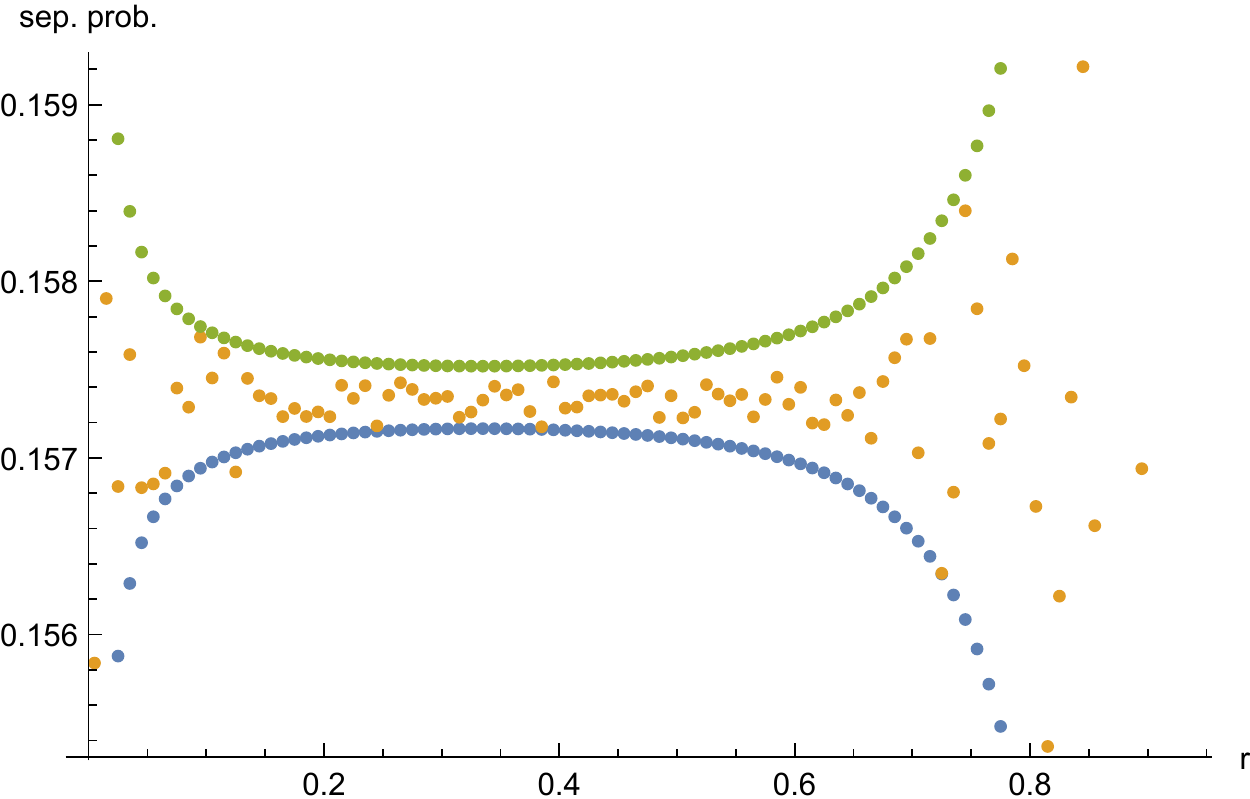}
\caption{\label{fig:K5confidenceDivision}Two-qubit separability probability for the random  induced measure case of $K=5$ $(k=1)$ associated with the inequality
$|\rho^{PT}| > |\rho|$, along with 95$\%$ confidence limits about $\frac{45}{286} \approx 0.157343$, with the complementary probability for $|\rho| > |\rho^{PT}| >0$ being $\frac{61}{143}-\frac{45}{286} =\frac{7}{26} \approx 0.269231$.} 
\end{figure}
\begin{figure}
\includegraphics{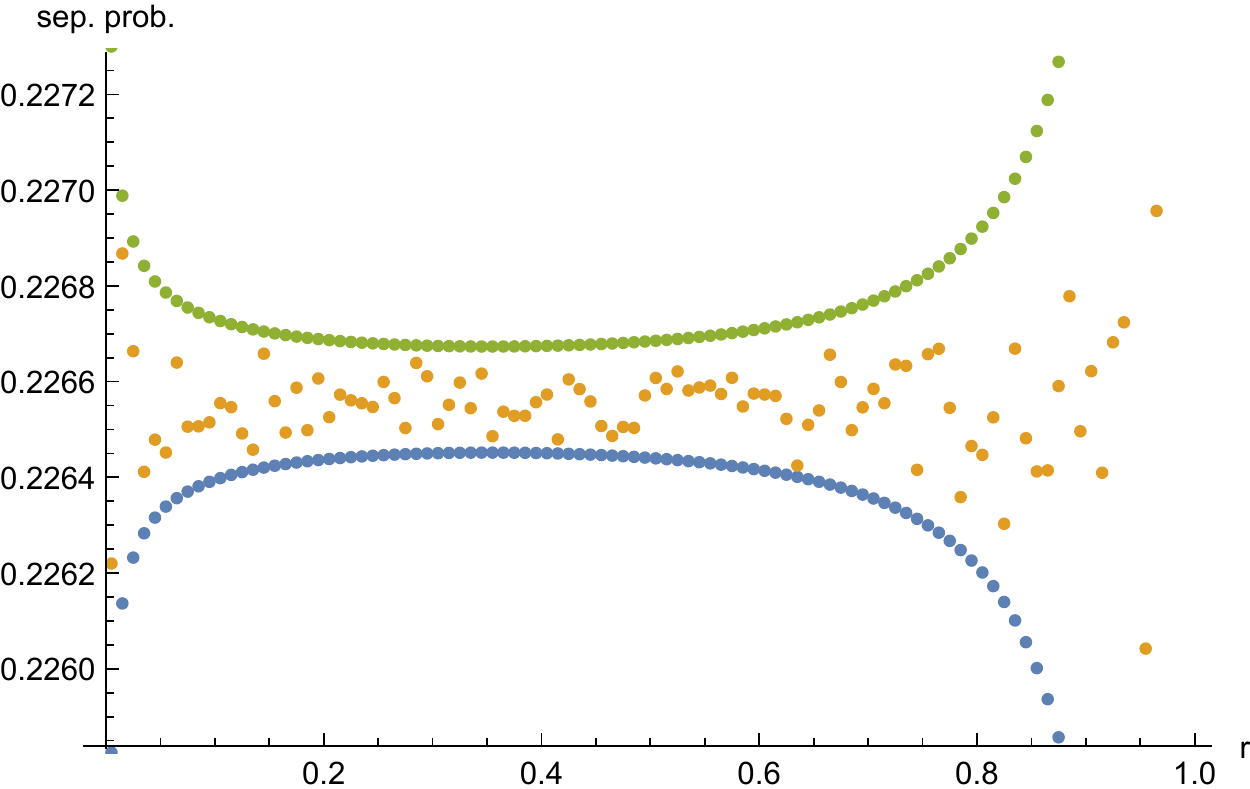}
\caption{\label{fig:RealconfidenceDivision}Two-{\it rebit} separability probability for the Hilbert-Schmidt case  associated with the inequality
$|\rho^{PT}| > |\rho|$, along with 95$\%$ confidence limits about $(\frac{1}{2}) (\frac{29}{64})=\frac{29}{128} \approx 0.2265625$.} 
\end{figure}
\begin{figure}
\includegraphics{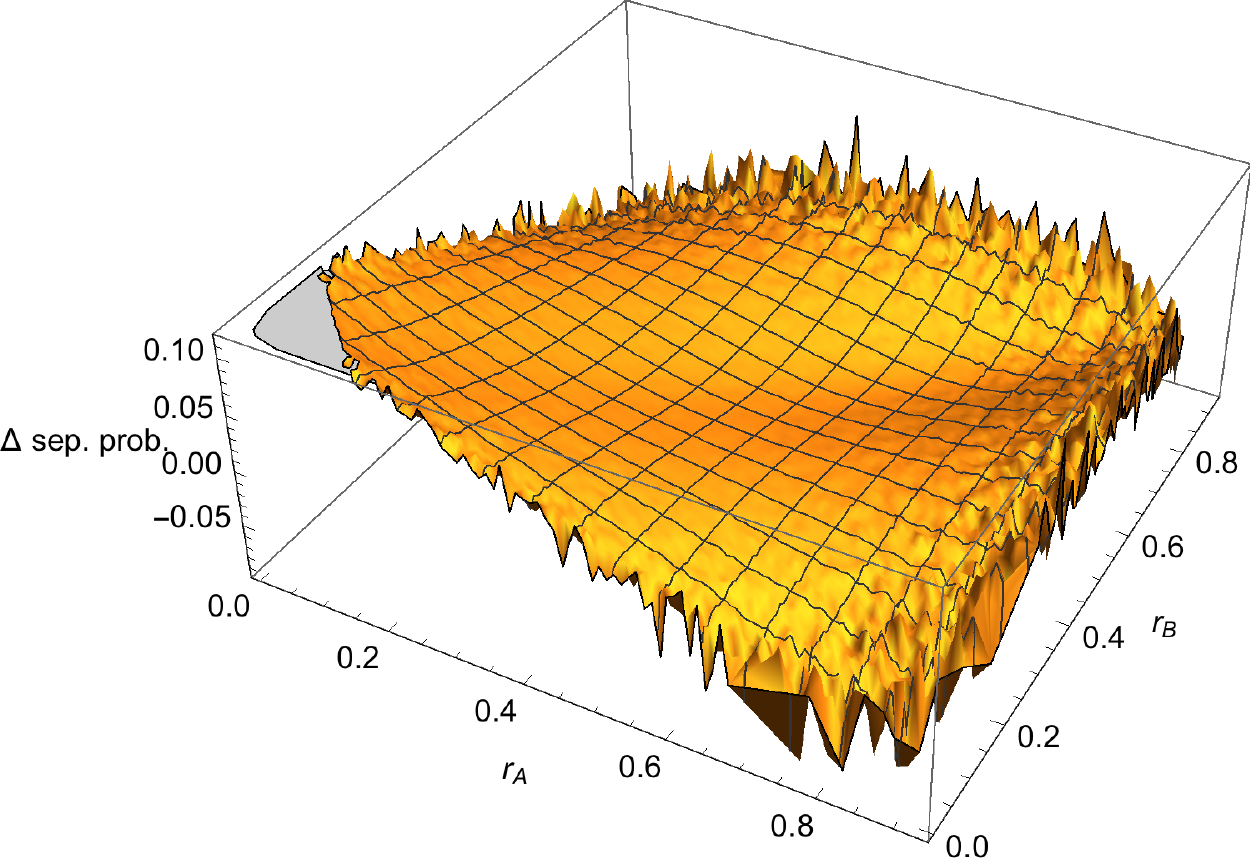}
\caption{\label{fig:MathStackResiduals} Residuals (discrepancies) of fit of $\frac{8 p(r_A,r_B)}{33}$--given by (\ref{tung})--to the estimated Hilbert-Schmidt two-qubit separability probabilities given in Fig.~\ref{fig:SepProbRatios}.}
\end{figure}
\newpage
\clearpage
\begin{figure}
\includegraphics{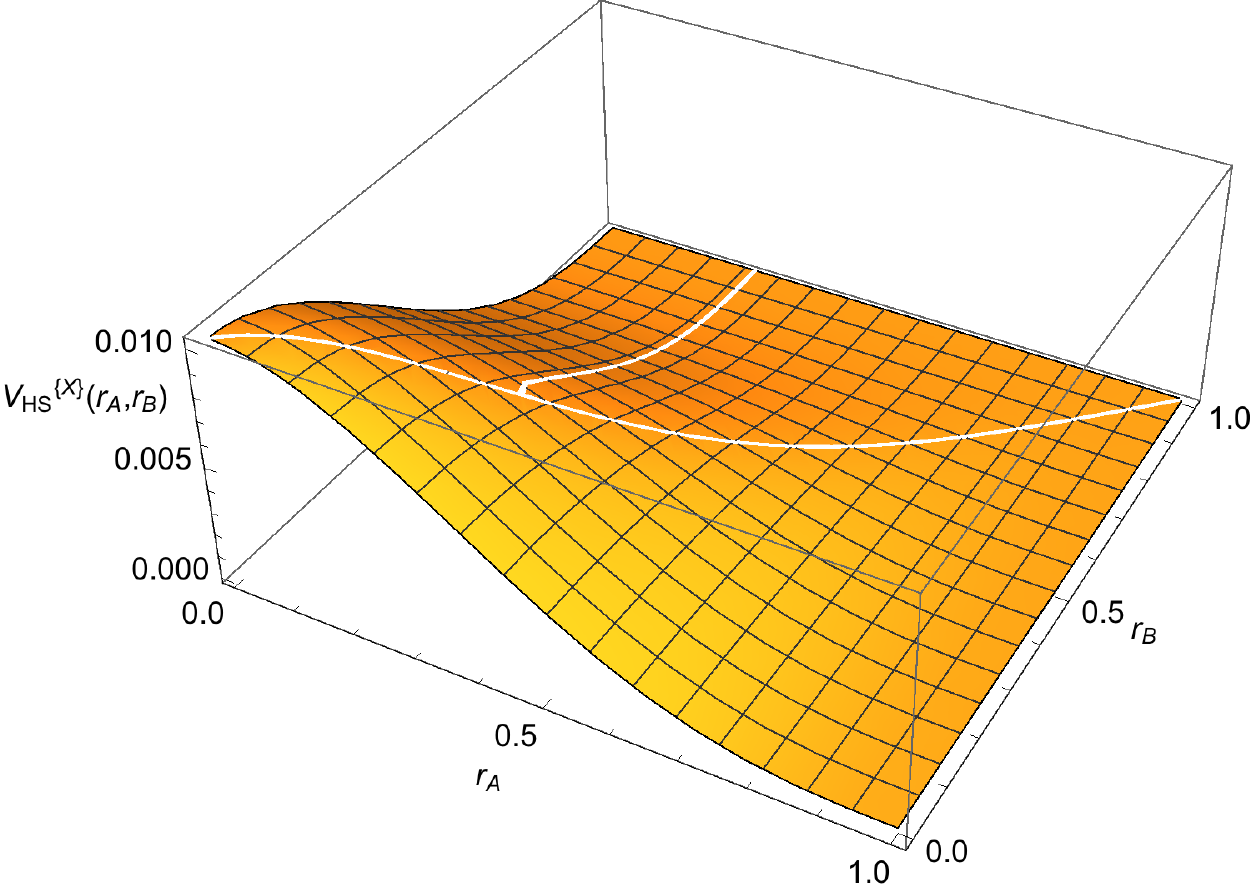}
\caption{\label{fig:Xbivariate}Bivariate Hilbert-Schmidt volume distribution (\ref{Xtotal}) for the $X$-state model}
\end{figure}
\begin{figure}
\includegraphics{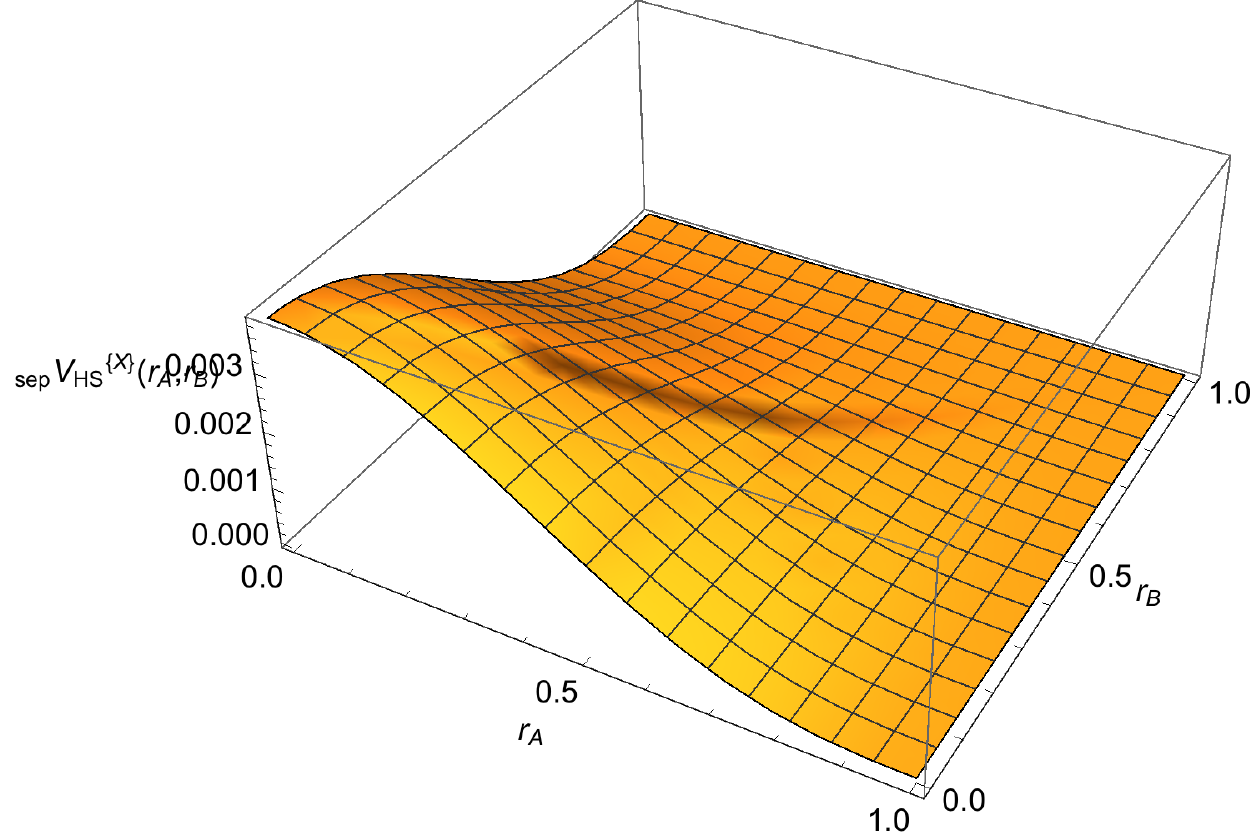}
\caption{\label{fig:XbivariateSep}Bivariate Hilbert-Schmidt 
{\it separable} volume distribution (\ref{Xsep}) for the $X$-state model}
\end{figure}
\begin{figure}
\includegraphics{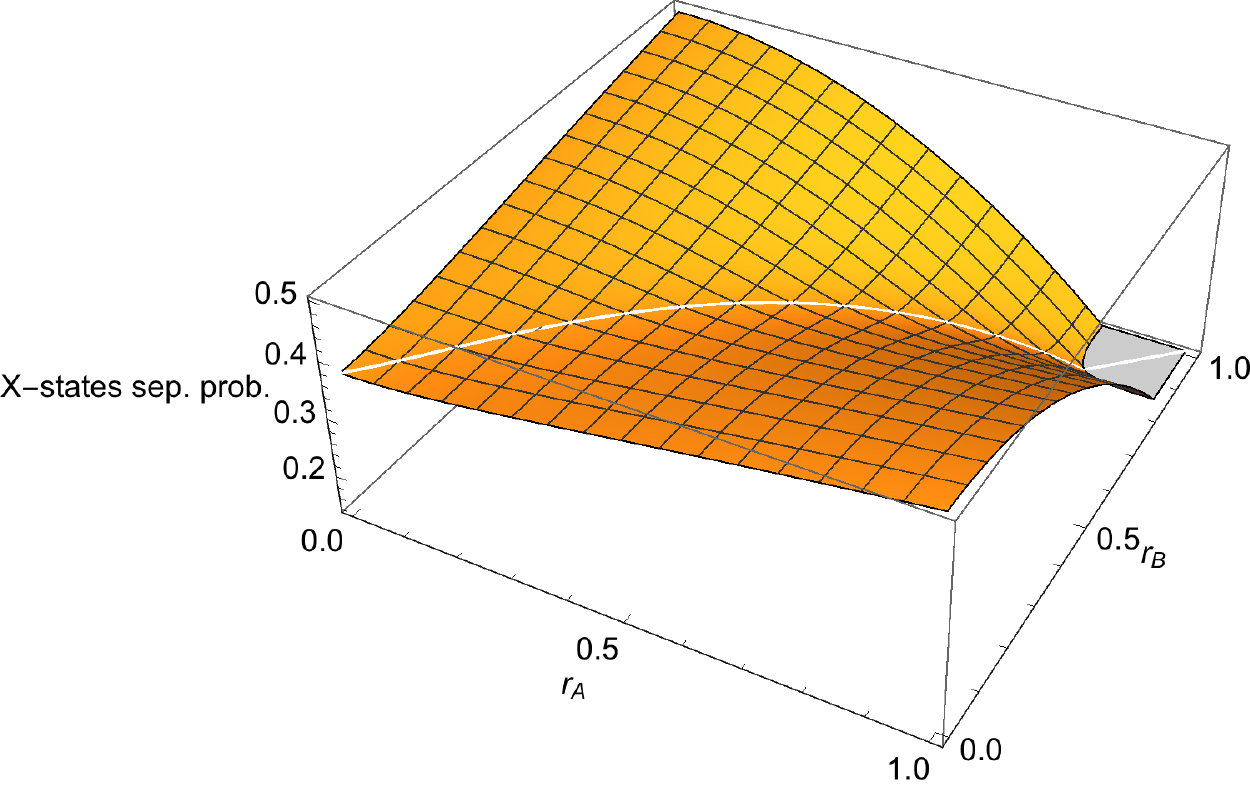}
\caption{\label{fig:XbivariateSepProb}Bivariate Hilbert-Schmidt 
separability probability distribution  (\ref{BivSepProb}) for the $X$-state model--the ratio of Fig.~\ref{fig:XbivariateSep} to Fig.~\ref{fig:Xbivariate}}
\end{figure}
\begin{figure}
\includegraphics{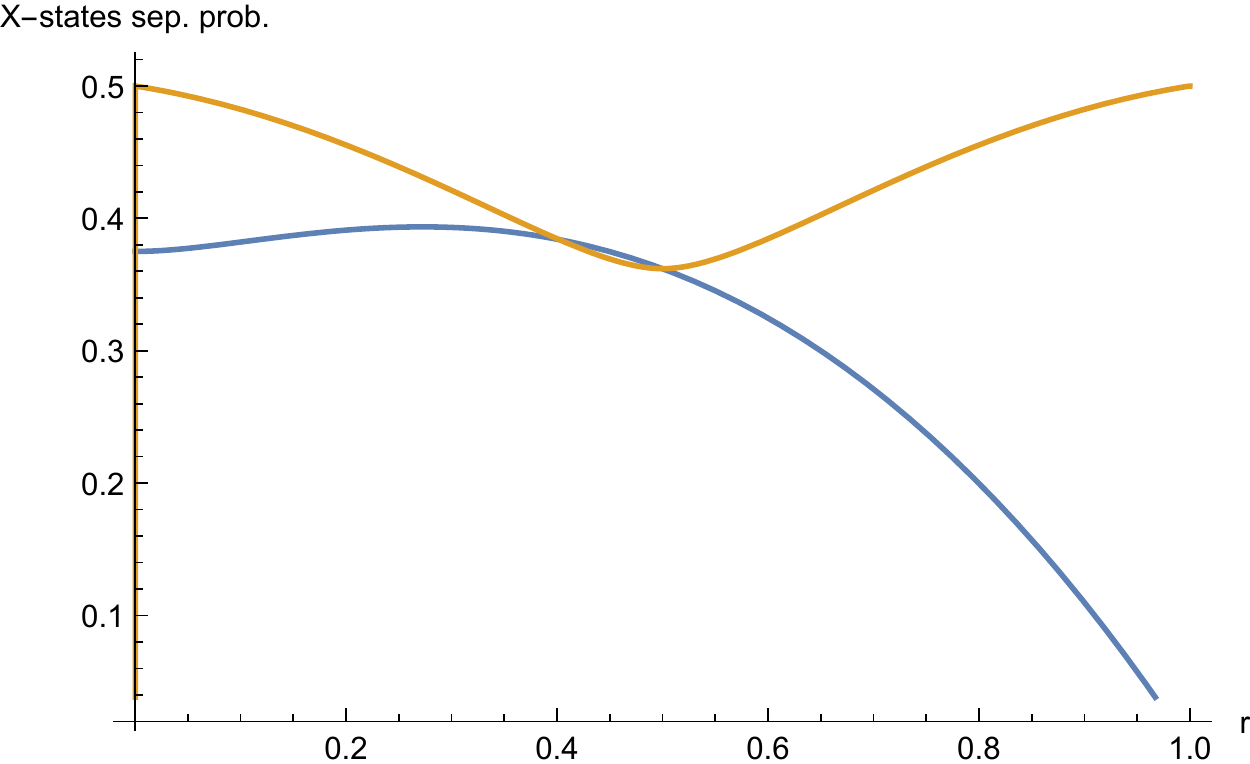}
\caption{\label{fig:XbivariateUpperLower}(Lower) $r_A=r_B$--given by (\ref{Lower})--and (upper) $r_A+r_B=1$ curves--given by (\ref{Upper})--for bivariate Hilbert-Schmidt 
$X$-states separability probability distribution. The minimum of the upper curve is at $r=\frac{1}{2}$, while the maximum of the lower curve is at 0.27227007. In the interval $r \in [0.40182804, \frac{1}{2}]$ the $p^{\{X-states\}}(r_A=1-r_B)$ curve is dominated by the $p^{\{X-states\}}(r_A=r_B)$ curve.}
\end{figure}
\begin{figure}
\includegraphics{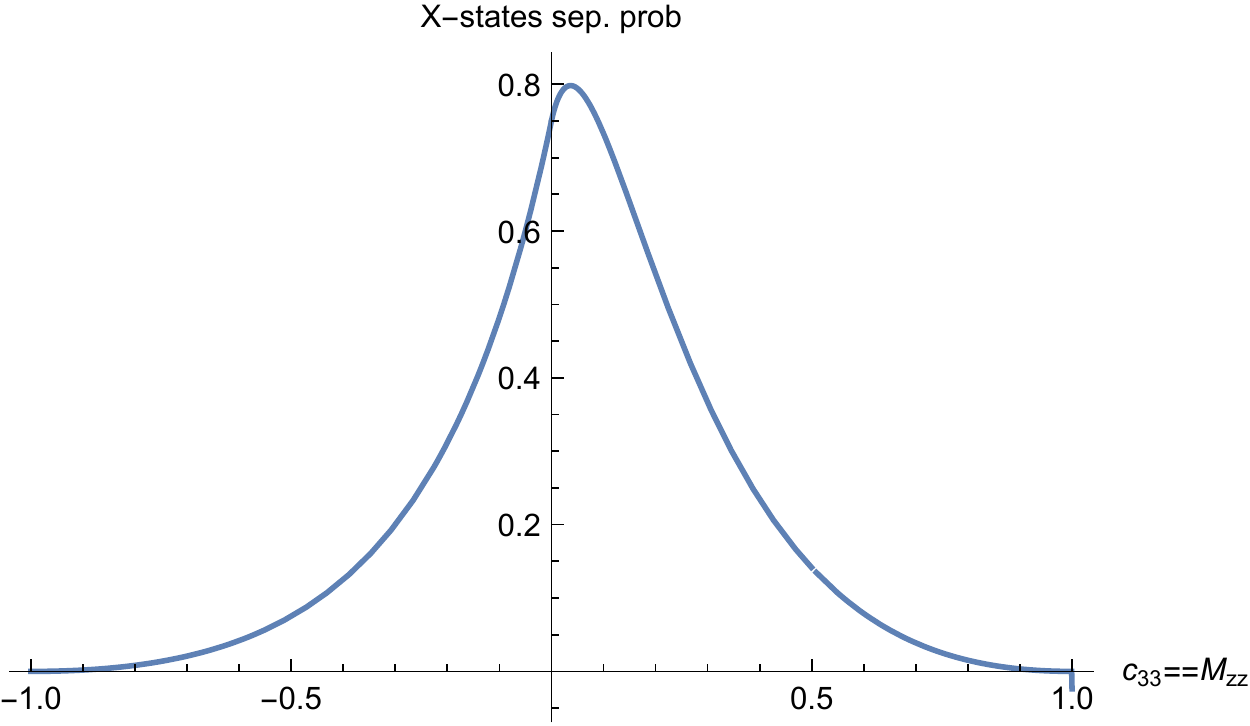}
\caption{\label{fig:MzzSepProb}$X$-states separability probability as a function of the Fano correlation parameter
$c_{33} \equiv M_{zz}$. For $-1 <M_{zz} < 0$, the curve is simply 
$\frac{3 \left(M_{\text{zz}}+1\right){}^2}{2 \left(M_{\text{zz}}-2\right) \left(2
   M_{\text{zz}}-1\right)}.$ The maximum ($\approx 0.79819147$) occurs near  $c_{33} \equiv M_{zz}= 0.03677089.$}
\end{figure}
\begin{figure}
\includegraphics{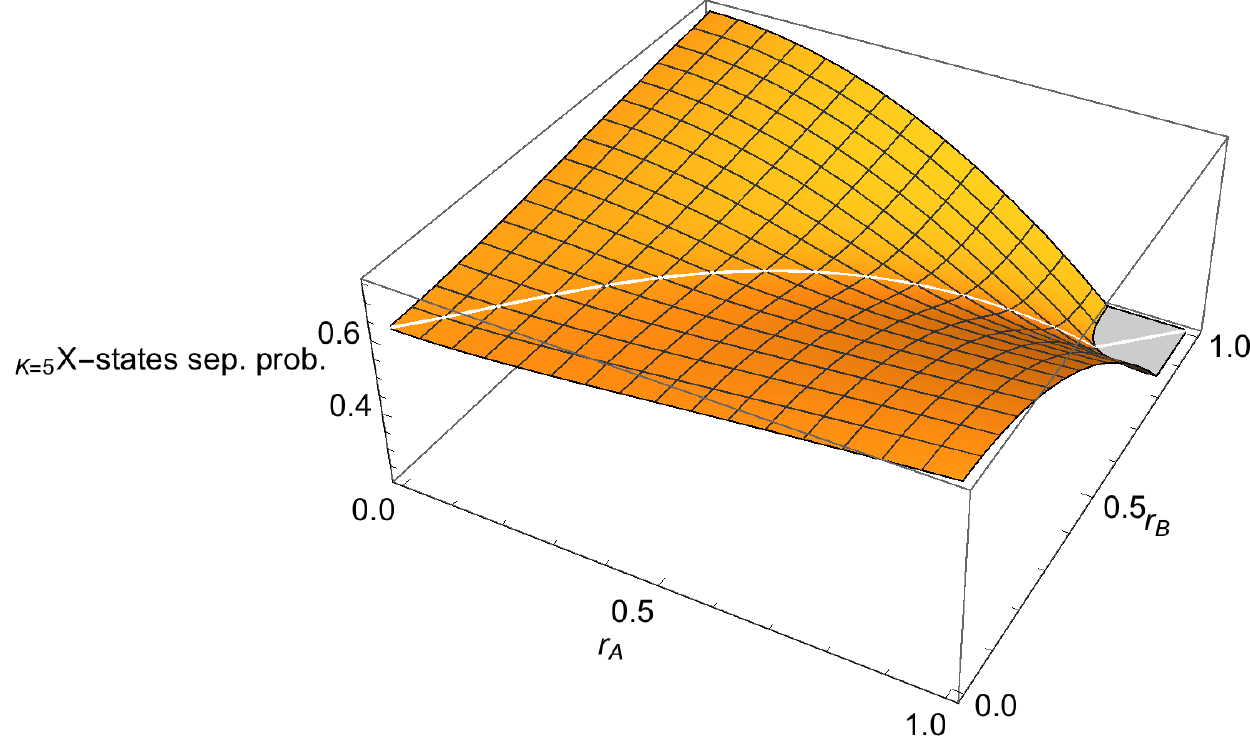}
\caption{\label{fig:XstatesK5}Bivariate separability probability function (eq. (\ref{BivSepProbK5}))
for the $K=5$ induced measure $X$-states model}
\end{figure}
\begin{figure}
\includegraphics{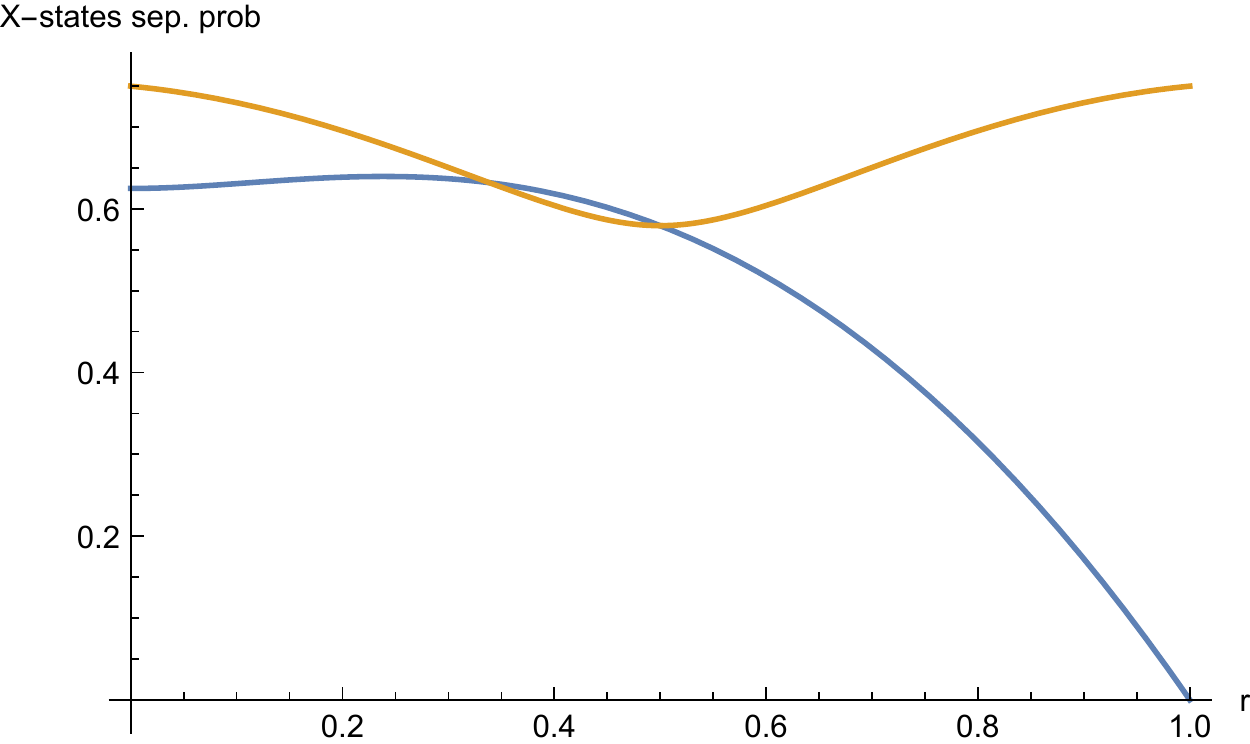}
\caption{\label{fig:XstatesK5joint}The $r_A=r_B$ and (largely dominant) $r_A=1-r_B$ sections of the $K=5$ 
$X$-states separability probability function (\ref{BivSepProbK5})}
\end{figure}
\begin{figure}
\includegraphics{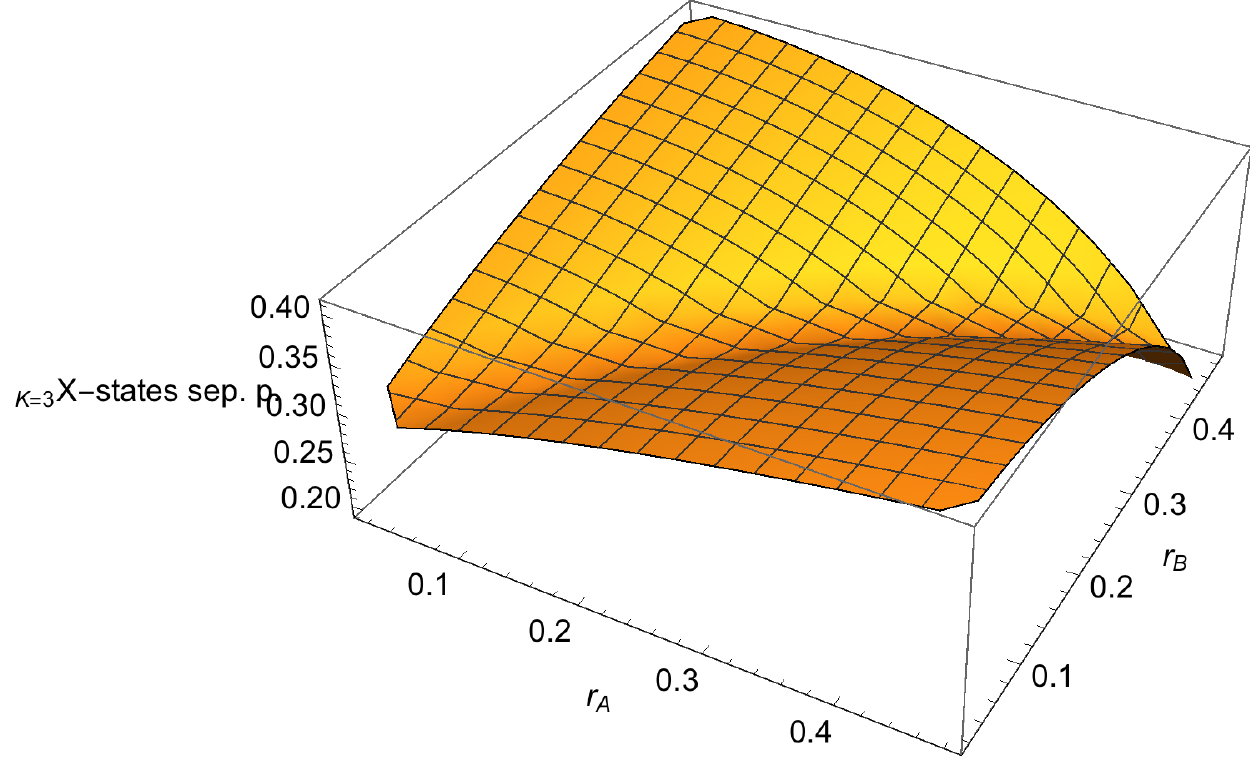}
\caption{\label{fig:K3XstatesPlot}Numerically-based estimate of bivariate separability probability function for the $K=3$ induced measure $X$-states model}
\end{figure}
\begin{figure}
\includegraphics{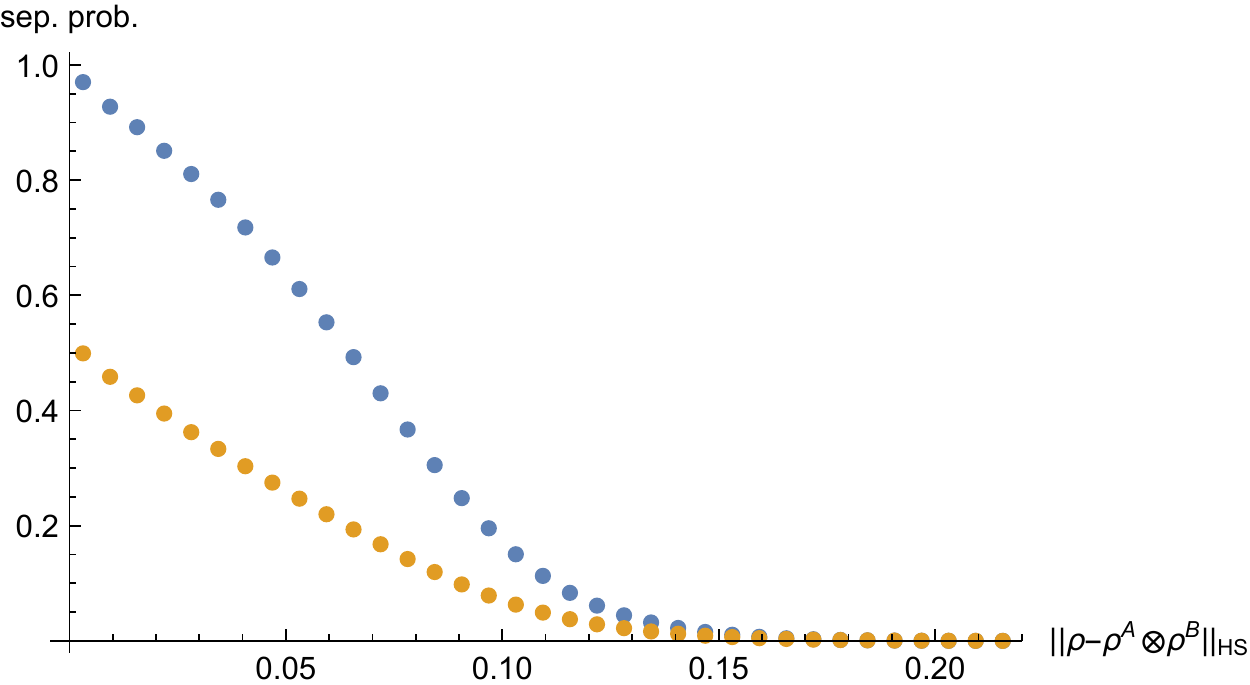}
\caption{\label{fig:HolikPlastino}Two-qubit separability probabilities as a function of
the entanglement expression $||\rho-\rho^A \otimes \rho^B||_{\it{HS}}$ for the Hilbert-Schmidt (dominant curve) and Bures measures}
\end{figure}

\end{document}